\documentclass[11pt]{article}

\usepackage[sort&compress,super,numbers,comma]{natbib}
\usepackage[margin=0.75in]{geometry}
\usepackage{amsmath,amssymb,amsfonts}
\usepackage{threeparttable,colortbl}

\usepackage{authblk}

\usepackage{xurl} 
\usepackage{hyperref}
\hypersetup{colorlinks,breaklinks,linkcolor=Blue,urlcolor=Blue,anchorcolor=Blue,citecolor=Blue}
\usepackage{color}
\definecolor{DarkBlue}{rgb}{0.0,0.08,0.45}
\definecolor{Blue}{rgb}{0.0,0.0,1.0}
\definecolor{Red}{rgb}{1.0,0.0,0.0}
\definecolor{RedOrange}{rgb}{0.9,0.0,0.2}
\definecolor{dgrn}{RGB}{0,150,0}
\definecolor{dgray}{gray}{0.3}
\definecolor{gray}{gray}{0.7}

\newcommand{\etal}{\textit{et~al.}}
\newcommand{\ie}{\textit{i.e.}}

\newcommand{\B}[1]{\textbf{#1}}
\newcommand{\mc}[3]{\multicolumn{#1}{#2}{#3}}
\newcommand{\pcite}[1]{\protect\cite{#1}}

\newcommand{\DOI}[1]{DOI: \href{https://doi.org/#1}{#1}}

\newcommand{\CaptionCite}[1]{\protect\captioncite{\cite{#1}}}

\makeatletter
\def\captioncite#1{%
     \begingroup
         \@fileswfalse
         #1
    \endgroup
}
\makeatother

\begin{document}
\title{\Large 
\textbf{Comment on ``Politicizing science funding undermines public trust in science, academic freedom, 
and the unbiased generation of knowledge''}
}
\author{
	John M. Herbert\footnote{\href{mailto:herbert@chemistry.ohio-state.edu}{herbert@chemistry.ohio-state.edu}} 
}
\affil{
\em	Department of Chemistry \& Biochemistry  \\
\em	The Ohio State University, Columbus, Ohio USA
}
\date{}\maketitle


\begin{abstract}\noindent
In a commentary published in mid-2024 (to which the present work is a direct response), a number of scientists 
argue that U.S.\ funding agencies have ``politicized'' the process
by which grants are awarded, in service of diversifying the scientific workforce.   The 
commentary in question, however, makes numerous 
unfounded assertions while recycling citations to a fusillade of opinion essays written by the same cabal of authors, 
in an effort to resemble a work of serious scholarship.   Basic 
fact-checking is provided here, demonstrating numerous claims that are unsupported by the source material
and readily debunked.   The present work also serves to document the reality of inclusion and diversity plans 
for scientific grant proposals to U.S. funding agencies, as they existed at the end of 2024.  
It is intended as a bulwark against retroactive false narratives, as the U.S. moves 
into a period of intense antagonism towards diversity, equity, and inclusion activities.
\end{abstract}

\section{Introduction}
A recent commentary by Efimov \etal\cite{EfiFliGeo24}
argues that U.S.\ funding agencies have ``politicized'' the process
by which grants are awarded, in service of diversifying the scientific workforce.   The commentary makes numerous 
unfounded assertions while recycling citations to a fusillade of opinion essays written by the same cabal of authors. 
A list of problematic claims, unsupported by the source material, is presented in Table~\ref{table}.
The most egregious errors are discussed below.

\begin{table*}[!hpt]\scriptsize
\centering
\caption{
	Summary of key errors and misrepresentations in Efimov \etal\CaptionCite{EfiFliGeo24} 
}\label{table}
\begin{threeparttable}
\renewcommand{\arraystretch}{1.3}
\begin{tabular}{p{0.45\textwidth} p{0.45\textwidth}}
\hline
\normalsize\B{Error or misrepresentation}$^\textit{a}$ & \normalsize\B{Explanation and rebuttal}
\\ \hline
\mc{2}{l}{\cellcolor{gray}\B{The scientific enterprise is strictly meritocratic}} \\
``Claims, such as `The presence of disparities is proof of systemic racism' and `Meritocracy is a myth' 
are propagated widely despite the vagueness of the claims and their lack of support by concrete data''.  Nevertheless, 
applicants for federal funding ``must profess the belief (one not itself supported by science) that certain systemic barriers exist''. 
& 
There is overwhelming, well-documented evidence for systemic bias and unmeritocratic hiring practices in academic 
science.\cite{Herbert:2023}
\\
\mc{2}{l}{\cellcolor{gray}\B{DEI activities are required for STEM funding}} \\
``[F]unding agencies are requiring applicants$\ldots$ to dedicate a part of the research budget to [DEI] implementation'', 
and this is
``[d]iverting funding from science into activities unrelated to the production of knowledge''.
&
Few (if any) program solicitations ever had explicit requirements for DEI activities.  ``Inclusion Plans'' and similar documents need not
have funds attached.  Furthermore, \textit{who} is allowed to serve as a scientific investigator is a crucial aspect of the scientific process.
\\
\mc{2}{l}{\cellcolor{gray}\B{Funding agencies are increasingly favoring DEI over scientific merit}} \\
The ``merit-based approach... is now being deemphasized''.
&
Some agencies do require a discussion of broadening participation but the review process continues to prioritize scientific merit.
\\
\mc{2}{l}{\cellcolor{gray}\B{Grant proposals require extravagant spending on DEI activities}} \\
Using rhetorical artifice, Efimov \etal\ imply that 5--10\% of a NSF budget, or perhaps \$50,000, constitutes 
typical spending for DEI activities.  They insinuate that this is mandated by funding agencies.
 &
The 5--10\% figure is taken out of context and the \$50,000 value comes from the website of an industry consulting firm, 
provided without context.   
%
%
Funding agencies seldom (if ever) micromanage budgets in the manner that is suggested, nor do peer reviewers.
 \\
\mc{2}{l}{\cellcolor{gray}\B{Applicants from ``inclusive'' institutions or research groups must prevaricate to obtain funding}} \\
``If an applicant's institution has already overcome any systemic barriers it may have had (or the applicant believes it has done so), 
then the applicant must lie or the proposal is doomed to fail. $\ldots$
[I]f access to an applicant's research team is already fair and non-discriminatory, 
why should an applicant be required to write an inclusion plan[?]''.
&
This is an intentional feint that equates ``treating everyone the same'' with ``removing systemic barriers'', 
a fallacy that is specifically addressed elsewhere.\pcite{Herbert:2025}
It is difficult to imagine 
that any institution has truly ``overcome'' such barriers; that's what makes them \textit{systemic}.  In any case, these comments overstate 
the role of Inclusion Plans and PIER plans in assessing a proposal's scientific merit.
\\
\mc{2}{l}{\cellcolor{gray}\B{DEI activities are unrelated to funding agencies' scientific mission}} \\
Funding agencies are 
``participating in activities that are arguably unrelated to their stated missions''.
&
Scientific workforce development is an important part of these agencies' mission.  Training a workforce whose demographics
reflect the U.S. population is one aspect of that pursuit.
\\
\mc{2}{l}{\cellcolor{gray}\B{Diversity initiatives by the funding agencies originate with U.S. President Biden}} \\
It is insinuated that no such initiatives existed until a recent pair of Executive Orders from President Biden.
&
The NSF's ``Broader Impacts'' mandate 
has existed since 1997 and persisted through presidential administrations from both political parties.\cite{Ren25}
Broader Impacts may (but need not) include activities related to broadening participation.\cite{Her24,Ren25}  
\\
\mc{2}{l}{\cellcolor{gray}\B{President Biden demands that scientific funding be distributed evenly according to race}} \\
A recent directive from President Biden\pcite{EO-14091} is purported to 
``mandate that research funding be distributed `equitably'---i.e., proportionally to demographic representation---among identity groups.''
&
The directive in question aims to ensure that federal programs are delivering equitable outcomes. 
It calls for creation of ``Equity Teams'' that will
``coordinate the implementation of equity initiatives and ensure 
that their respective agencies are delivering 
equitable outcomes for the American people".  Nowhere are funds allocated for any ``identity groups''.
\\
\mc{2}{l}{\cellcolor{gray}\B{Demographic data are requested by funding agencies for nefarious purposes}} \\
Investigators are obligated to fulfill ``\textit{de facto} DEI quotas'' as a condition of grant renewal, so that awards are 
``allocated according to identity metrics''.
&
Submission of this information to the NSF is voluntary,\pcite{Pli20} and awards are not made using identity metrics.
In support of this assertion, the only citation that is provided 
pertains to a NIH program specifically targeted at inclusion and diversity,\pcite{NIH-PRIDE}
for which demographic data are a natural part of the reporting requirements.
\\
\mc{2}{l}{\cellcolor{gray}\B{DEI requirements amount to illegal compelled speech}} \\
``The First Amendment of the Constitution of the United States strictly forbids compelling people to say things they do not believe are true.''
&
This is a gross misrepresentation of the First Amendment.  There is no Constitutional right to federal support for research.
\\
\mc{2}{l}{\cellcolor{gray}\B{Funding agencies circumvent discrimination laws through subterfuge}} \\
``Funding agencies attempt to circumvent the laws prohibiting them from basing funding decisions on race or ethnicity by cloaking 
DEI requirements in nebulous language [citation omitted] and by disguising racial preferences".   
& 
The NASA slide deck that is cited\pcite{NahWat23} actually warns that 
``[t]okenizing diverse team members'' is a common \textit{weakness} of Inclusion Plans.  
\\
\hline
\end{tabular}
\begin{tablenotes}[flushleft]\footnotesize\item
	$^a$All text in quotation marks in the left column is from Ref.~\protect\citenum{EfiFliGeo24}. 
\end{tablenotes}
\end{threeparttable}
\end{table*}

\section{Fact-Checking}  
The fundamental claim in Efimov \etal\cite{EfiFliGeo24} 
is that U.S. funding agencies are prioritizing diversity, equity, and inclusion (DEI) 
over scientific merit in the review process for research funding.   This is debunked in detail elsewhere.\cite{Her24}
In fact, few (if any) program solicitations require DEI activities.
Activities that are not mandated by the solicitation are not required elements for the purpose of merit review.  
If DEI activities are not required, then no portion of the budget needs to be set aside for them.

To support the accusation that DEI activities are ``[d]iverting funding from science'', all  
the authors can muster is a slide deck\cite{Ren23}
from a principal investigator (PI) whose work focuses specifically on public outreach.\cite{ARIS-team}
This is cited to support a blanket assertion that 5--10\% of the budget for National Science Foundation (NSF) awards must be 
spent on DEI activities.   In reality, budgets for research awards are seldom (if ever) micromanaged at that level.

At the heart of the hyperbolic outrage in Efimov \etal\ 
is the fact that funding agencies do encourage would-be investigators 
to foster inclusive research environments.  This takes various forms at different agencies.  At the NSF, it falls under the
``Broader Impacts'' review criterion.   Broadening participation by underrepresented
groups is one way that an investigator may satisfy this criterion, but other possibilities have nothing to do with 
diversity.\cite{Her24,Ren25}
The National Aeronautics and Space Administration (NASA) and the U.S. Department of Energy (DOE) 
mandate an ``Inclusion Plan'' and a ``Promoting Inclusive and Equitable Research'' (PIER) plan, respectively.   These documents are 
intended to stimulate activities that reduce systemic barriers, in the interest of 
``[a]dvancing scientific discovery by harnessing a diverse range of views, expertise, and experiences''.\cite{DOE-PIER}
In the DOE's list of review criteria, the quality of the PIER plan is placed well below the scientific and
technical merit of the proposal.\cite{DOE-PIER}

To stoke outrage over the cost of DEI activities, Efimov \etal\
quote a NASA slide deck suggesting that PIs \textit{could} choose to hire paid diversity 
consultants, then they provide an estimate (from an industry consulting firm's website) that this might cost \$8,000--\$50,000. 
Those context-free numbers notwithstanding, Efimov \etal\ neglect to mention that the same NASA presentation states  
that Inclusion Plans will \textit{not} be used to rank proposals.\cite{NahWat23}   
It also warns against 
``[t]okenizing diverse team members'', \ie, it actively \textit{discourages} the type of ``DEI quotas'' feared by Efimov \etal\
The quality of the DOE PIER plan is placed well below scientific and
technical merit in the list of review criteria.\cite{DOE-PIER}

The authors repeatedly conflate DEI activities with race-based quotas, going so far as to imply that 
a recent Executive Order by the Biden administration mandates that federal funding be distributed 
to ``identity groups'' in proportion to demographics. 
There is literally nothing in the Order to support this allegation.  It does direct agencies to 
``coordinate the implementation of [existing] equity initiatives'',\cite{EO-14091} but 
to equate this with racial quotas is a paranoid delusion of anti-DEI zealots.

That paranoia is evident in concern over ``extensive collection of demographic information by the funding 
agencies''.\cite{EfiFliGeo24}   That data collection is often voluntary,\cite{Pli20}    
Furthermore the authors, as scientists, 
should recognize that the first step in problem solving is data collection to analyze the problem.  
Therefore, this attack feels like a means to obscure (and thus avoid) the diversity problem itself, preserving the 
unearned advantage of legacy elites.\cite{McI89}


\section{Discussion}  

Efimov \etal\cite{EfiFliGeo24} repeatedly deny there is evidence for systemic barriers 
to participation for underrepresented individuals in science, technology, engineering, and mathematics (STEM).
In support of that claim, the authors cite their own opinion commentaries along with several papers
(on the topic of gender bias in faculty hiring) by 
Ceci and Williams, whose methods have long been 
criticized.\cite{WilSmi15,Gro15,Fra15a,Zev15,Kro23}
Evidence abounds that such barriers \textit{do} 
exist.\cite{Aza20,McG20a,GosKwoMur21,SetJonBuc21,McGBotNap22,BlaCec22,Her23a,Herbert:2025} 
To argue otherwise is to assert that the wildly skewed demographics of STEM faculty,\cite{BakKoe24}
who are most often the PIs on federal grants, is an equilibrium situation that reflects a deliberate choice 
by underrepresented individuals.  
Given that 13\% of U.S. undergraduates are Black and 22\% are Hispanic,\cite{DavFry19,NCES}
to accept this argument one must believe that some conscious decision explains why 
fewer than 2\% of Chemistry and Biology faculty in the U.S. are Black, and only 3\% are 
Hispanic,\cite{Nel17,LiKoe17,DowWid20} with similarly low representation in other STEM disciplines.\cite{BakKoe24}
Sufficient STEM Ph.D.s are awarded to Black and Hispanic Americans 
to infer that ``pipelines'' are open,\cite{Nel17,MatLewHop22}
yet these individuals are not hired as faculty at rates that will achieve demographic parity
within this century.\cite{Her23c}  
Even beyond higher education, underrepresentation plagues the STEM 
workforce more generally.\cite{NSF-diversity}

To invoke a mantra of ``treating every student the same'', as Efimov \etal\ do,\cite{EfiFliGeo24} is to deny 
the reality of these systemic barriers and, in effect, to treat every student as a privileged White man.\cite{Herbert:2025}
It is similarly absurd to deny that systemic barriers cannot explain why 
the fraction of female faculty hired by U.S. universities has been unchanged or has decreased over the past decade, in nearly
every discipline.\cite{WapZhaCla22}   The manner in which 
academic careers interfere with family considerations is certainly a factor,\cite{Her23a} and that's a systemic barrier.
One way to foster an inclusive environment is to reduce the hurdles that mothers face in STEM workplaces.

It is difficult to imagine that demographics so far out-of-equilibrium with the U.S. population can be consistent
with ``meritocratic'' hiring, but we don't have to imagine.   The data show that 20\% of U.S. universities produce 80\% of its
faculty, with five institutions responsible for 1/8 of all faculty at doctoral institutions.\cite{WapZhaCla22}
At elite institutions, faculty of ``lesser pedigree'' are no less productive,\cite{WayMorLar19} yet they are rarely hired. 
Meanwhile, STEM Ph.D.\ candidates from underrepresented groups abound,\cite{MatLewHop22} meaning that 
``pipelines'' are open but these individuals are not being hired at sufficient rates to move the needle on faculty demographics.
Although the conceit that science is strictly meritocratic is deeply instilled in American culture,\cite{KluSmi86}
and also in STEM education,\cite{CecBla10,Cec13b,SerSilCec18}
it simply does not stand up to actual data, nor to the lived experience of minoritized STEM scientists who cope with systemic
racism on a daily basis.\cite{WooCamMcG16,McGGriHou19,McG20a,GosKwoMur21}

Efimov \etal\ 
object that DEI activities ``[d]ivert funding from science into activities unrelated to the production of knowledge'', 
but this ignores agencies' obligation to develop the scientific workforce.  \textit{Who} 
is serving as PI affects the questions that are asked and the methods used to answer them, as the National Institutes of Health (NIH)
have acknowledged in recent initiatives.\cite{NIH-PRIDE,RicCraNga21} 
Although Efimov \etal\ 
worry that DEI initiatives ``undermine$\ldots$ the unbiased generation of knowledge'', in fact the opposite is true.
If the scientific workforce does not reflect the population as a whole, then the research produced by that community will be 
impacted.\cite{McG20b}  Therefore, efforts to broaden participation should be viewed as attempts to solicit research
that reflects America's strength as a multicultural and multiracial society. 

\section{Epilogue} 
Given that the Trump administration has now ended PIER plans,\cite{PIER-dead}  
Inclusion Plans,\cite{AIP-Trump} and DEI activities in general,\cite{EO-14173,Quinn:2025,Knox:2025}
certain aspects of this debate now seem outdated.
Nevertheless, it is important to correct the record.   At a minimum, we need to 
preserve the memory of the way that the U.S. scientific funding system 
actually operated in 2024, lest anti-DEI crusaders be allowed to memory-hole that reality and 
retroactively replace it with the false narrative presented by  Efimov \etal\cite{EfiFliGeo24}
(As a bulwark against this, source material from U.S. government websites is included here as Supplementary Material, 
lest these sources disappear entirely.)
Consistent with the assessment above, that anti-DEI provocateurs would prefer to hide from the clear
statistical reality of While male domination of STEM, one of the Trump administration's early moves was to shutter the National
Center for Education Statistics (NCES),\cite{NCES-dead1} 
one of the most important sources of diversity statistics regarding both students (at all levels) and faculty.\cite{NCES}
Dissolution of NCES 
will negatively impact not just higher education but K--12 education as well.\cite{NCES-dead2}

Most of this Comment (save for this Epilogue) was written in early September 2024,
shortly after Ref.~\citenum{EfiFliGeo24} was published.  After a lengthy but unsuccessful attempt to
procure co-authors, which I ultimately abandoned following the U.S. presidential election in November 2024, I submitted this work
to \textit{Frontiers in Research Metrics and Analysis}, as a ``General Commentary''  
on Ref.~\citenum{EfiFliGeo24}.  
On January 18, 2025, I received a single review calling it a ``strong commentary'' and asking only that I remove the phrase 
``paranoid delusion''.  I did so (although I've restored it here), 
and immediately resubmitted.   After some time, I received a message that the review had been
withdrawn, at which point I noticed that the Associate Editor had been reassigned.  Upon contacting the journal, 
I was informed that both the reviewer and the original Associate Editor had withdrawn ``due to a lack of time\slash busy schedules''.
Notably, this took place shortly after the Trump administration paused all
funding agency decisions for scientific grants.  Because \textit{Frontiers} journals publish both the editor and reviewer names 
on each paper, I suspect that these individuals stepped away for reasons other than 
being too busy to approve a two-word change to a 1,000-word Comment.

A few days later, the manuscript was rejected without further review. I was told, on behalf of the Editorial Board, that it 
``does not meet the standards of scientific quality or rigor required by the journal to be considered for peer review'', in part because it 
``spend[s] too much time and space examining each claim instead of making [a] point''.  
(General Commentaries are 
limited to 1,000 words and one figure or table.\cite{Frontiers-rules})
After a lengthy back-and-forth with multiple editors, I was invited to write a longer, stand-alone opinion article although it was intimated that
a point-by-point rebuttal of Ref.~\citenum{EfiFliGeo24} would not be tolerated, and that the requirement to publish the editor and 
reviewer names could not be waived.  Given the contemporary political climate in the U.S., specifically with regard to higher education
and DEI programs,\cite{Quinn:2025,Knox:2025,Dutton:2025,Blake:2025,Stanley:2025}
I believe that lack of anonymity will skew the pool of willing reviewers.  Pro-DEI sentiments are unlikely to be evaluated fairly in the
absence of blind review.



\begin{thebibliography}{10}

\bibitem{EfiFliGeo24}
I.~R. Efimov, J.~S. Flier, R.~P. George, A.~I. Krylov, L.~S. Maroja,
  J.~Schaletzky, J.~Tanzman, and A.~Thompson, ``Politicizing science funding
  undermines public trust in science, academic freedom, and the unbiased
  generation of knowledge'', {\em Front.\ Res.\ Metr.\ Anal.}, {\bf 9},
  1418065 (2024).  \DOI{10.3389/frma.2024.1418065}

\bibitem{Her24}
J.~M. Herbert, ``Fact-checking the `politicization' of scientific funding'',
  {\em The Chronicle of Higher Education} (October 24, 2024).
  \url{https://www.chronicle.com/blogs/letters/fact-checking-the-politicization-of-scientific-funding}
  (accessed 2024-10-24).

\bibitem{Ren23}
S.~D. Renoe, ``Let's talk broader impacts'', {\em U.S. National Science
  Foundation, Division of Molecular and Cellular Biosciences} (June 7, 2023).
  \url{https://www.nsf.gov/bio/mcb/June7_2023_BIwSusanRenoe_ARIS.pdf}
  (accessed 2024-09-11).

\bibitem{ARIS-team}
Leadership Team, 
{\em Advancing Research Impact in Society} (2024).
\url{https://researchinsociety.org/about/leadership-team} (accessed 2024-09-12).

\bibitem{Herbert:2023}
J.~M. Herbert, ``Tackling pedigree bias in faculty hiring'',
{\em Inside Higher Ed} (September 27, 2023).
\url{https://www.insidehighered.com/opinion/views/2023/09/27/tackling-pedigree-bias-tenure-track-hiring-opinion}
(accessed 2025-03-20).

\bibitem{Herbert:2025}
J.~M. Herbert and N.~M. Dickson-Karn,
``The fallacy of `treating all students the same'\,'', 
{\em Inside Higher Ed} (March 20, 2025).
\url{https://www.insidehighered.com/opinion/views/2025/03/20/fallacy-treating-every-student-same-opinion}
(accessed 2025-03-20).  

\bibitem{Ren25}
S.~D. Renoe, ``An insider perspective on broader impacts'',
{\em BioScience}, {\bf 75}, 207--211.
\DOI{10.1093/biosci/biaf004}

\bibitem{EO-14091}
Executive Order 14091, ``Further Advancing Racial Equity and Support
  for Underserved Communities Through the Federal Government'', {\em Fed.\
  Reg.}, {\bf 88}, 10825--10833 (2023).
  \url{https://www.federalregister.gov/documents/2023/02/22/2023-03779/further-advancing-racial-equity-and-support-for-underserved-communities-through-the-federal}
  (accessed 2024-09-11).

\bibitem{Pli20}
S.~H. Plimpton, ``Privacy act and public burden statements'', 
{\em National Science Foundation} (June 1, 2020).
\url{https://www.nsf.gov/pubs/policydocs/pappg20_1/privacy_burden.jsp} (accessed 2024-09-12).

\bibitem{NIH-PRIDE}
``Part 1.  Overview Information.  Programs for Inclusion and Diversity Among Individuals Engaged in Health-Related-Research
(PRIDE) (R25 Clinical Trial Not Allowed)'',  {\em National Institutes of Health (NIH); National Heart, Lung, and Blood Institute (NHLBI)}
(2023).
\url{https://grants.nih.gov/grants/guide/rfa-files/RFA-HL-24-004.html} (accessed 2024-09-15).

\bibitem{NahWat23}
A.~L. Nahm and R.~N. Watkins, ``Inclusion plans in research proposals'', {\em
  National Aeronautics and Space Administration} (June 22, 2023).
  \url{https://smd-cms.nasa.gov/wp-content/uploads/2023/08/12-inclusionplans-nahm-watkins.pdf}
  (accessed 2024-09-11).

\bibitem{DOE-PIER}
``Promoting Inclusive and Equitable Research (PIER) Plans''. 
\newblock U.S. Department of Energy (September 3, 2024).
\url{https://science.osti.gov/-/media/Office-Hours/pdf/2024/DOE-SC-Office-Hours--PIER-Plans---Sept3-2024-final.pdf}
(accessed 2024-09-11).

\bibitem{McI89}
P.~{McIntosh}, ``White privilege: {Unpacking} the invisible knapsack'', {\em
  Peace and Freedom}, pp.~10--12 (1989).
  (\url{https://nationalseedproject.org/Key-SEED-Texts/white-privilege-unpacking-the-invisible-knapsack},
  accessed 2024-09-12).

\bibitem{WilSmi15}
J.~C. Williams and J.~L. Smith, ``The myth that academic science isn't biased
  against women'', {\em The Chronicle of Higher Education} (July 8, 2015).
  \url{https://www.chronicle.com/article/the-myth-that-academic-science-isnt-biased-against-women}
  (accessed 2024-09-11).

\bibitem{Gro15}
L.~Grossman, ``Claiming sexism in science is over just wishful thinking'', {\em
  New Scientist} (April 17, 2015).
  \url{https://www.newscientist.com/article/dn27367-claiming-sexism-in-science-is-over-is-just-wishful-thinking}
  (accessed 2024-09-11).

\bibitem{Fra15a}
M.~R. Francis, ``\,`{A} surprisingly welcome atmosphere'\,'', {\em Slate} (April
  20, 2015).
  \url{https://slate.com/human-interest/2015/04/no-sexist-hiring-in-stem-fields-a-vaunted-new-study-makes-that-claim-unconvincingly.html}
  (accessed 2024-09-11).

\bibitem{Zev15}
Z.~Zevallos, ``The myth about women in science? {Bias} in the study of gender
  inequality in {STEM}'', {\em Other Sociologist} (April 16, 2015).
  \url{https://othersociologist.com/2015/04/16/myth-about-women-in-science}
  (accessed 2024-09-11).

\bibitem{Kro23}
M.~C. Kropinski, ``A misleading portrayal of women's equality in science'',
  {\em Inside Higher Ed} (May 13, 2023).
  \url{https://www.insidehighered.com/opinion/letters/2023/05/13/misleading-portrayal-womens-equality-science}
  (accessed 2024-09-11).

\bibitem{Aza20} 
S.~Azad, Ed.
\newblock {\em Addressing Gender Bias in Science \& Technology}, Vol.  1354 of
  {\em ACS Symposium Series}, Washington D.C., (2020). American Chemical
  Society. \DOI{10.1021/bk-2019-1328}

\bibitem{McG20a}
E.~O. {McGee}, ``Interrogating structural racism in {STEM} higher education'',
  {\em Educ.\ Researcher}, {\bf 49}, 633--644 (2020).  \DOI{10.3102/0013189X20972718}

\bibitem{GosKwoMur21}
M.~L. Gosztyla, L.~Kwong, N.~A. Murray, C.~E. Williams, N.~Behnke, P.~Curry,
  K.~D. Corbett, K.~N. {DSouza}, J.~G. {de~Pablo}, J.~Gicobi, M.~Javidnia,
  N.~Lotay, S.~M. Prescott, J.~P. Quinn, Z.~M.~G. Rivera, M.~A. Smith, K.~T.~Y.
  Tang, A.~Venkat, and M.~A. Yamoah, ``Responses to 10 common criticisms of
  anti-racism action in {STEMM}'', {\em PLoS Comput.\ Biol.}, {\bf 17},
  e1009141 (2021). \DOI{10.1371/journal.pcbi.1009141}

\bibitem{SetJonBuc21}
I.~H. Settles, M.~K. Jones, N.~T. Buchanan, and K.~Dotson, ``Epistemic
  exclusion: {Scholar(ly)} devaluation that marginalizes faculty of color'',
  {\em J.~Divers.\ High.\ Educ.}, {\bf 14}, 493--507 (2021). \DOI{10.1037/dhe0000174}

\bibitem{McGBotNap22}
E.~O. {McGee}, P.~K. Botchway, D.~E. {Naphan-Kingery}, A.~J. Brockman,
  S.~{Houston~II}, and D.~T. White, ``Racism camouflaged as imposterism and the
  impact on {Black} {STEM} doctoral students'', {\em Race Ethn.\ Educ.}, {\bf
  25}, 487--507 (2022).  \DOI{10.1080/13613324.2021.1924137}

\bibitem{BlaCec22}
M.~{Blair-Loy} and E.~A. Cech, {\em Misconceiving Merit: Paradoxes of
  Excellence and Devotion in Academic Science and Engineering}, University of
  Chicago Press: Chicago, 2022.

\bibitem{Her23a}
J.~M. Herbert, ``Academic free speech or right-wing grievance?'', {\em Digital
  Discov.}, {\bf 2}, 260--297 (2023). \DOI{10.1039/D2DD00111J}
  
\bibitem{BakKoe24}
S.~P. Baker and C.~Koedel, ``Diversity trends among faculty in {STEM} and
  non-{STEM} fields at selective public universities in the {U.S.} from 2016 to
  2023'', {\em Humanit.\ Soc.\ Sci.\ Commun.}, {\bf 11}, 1149:1--9 (2024).
  \DOI{10.1057/s41599-024-03687-x}

\bibitem{DavFry19}
L.~Davis and R.~Fry, ``College faculty have become more racially and ethnically
  diverse, but remain far less so than students'', {\em Pew Research Center}
  (July 31, 2019).
  \url{https://www.pewresearch.org/fact-tank/2019/07/31/us-college-faculty-student-diversity}
  (accessed 2024-09-11).

\bibitem{NCES}
``Characteristics of postsecondary students''.
\newblock National Center for Education Statistics, U.S. Department of
  Education, Institute of Education Sciences (May 2022).
\url{https://nces.ed.gov/programs/coe/indicator/csb} (accessed 2023-01-21) .

\bibitem{Nel17}
D.~J. Nelson, ``Diversity of science and engineering faculty at research universities'',
in {\em Diversity in the Scientific Community Volume~1: Quantifying Diversity and Formulating Success}, 
D.~J. Nelson and H.~N. Cheng, Ed., Vol.~1255 of {\em ACS Symposium Series;}
\newblock American Chemical Society: Washington D.C., 2017;
\newblock pp.~15--86. \DOI{10.1021/bk-2017-1255.ch002}

\bibitem{LiKoe17}
D.~Li and C.~Koedel, ``Representation and salary gaps by race-ethnicity and
  gender at selective public universities'', {\em Educ.\ Researcher}, {\bf 46},
  343--354 (2017). 
  \DOI{10.3102/0013189X17726535}

\bibitem{DowWid20}
A.~{Downey-Mavromatis} and A.~Widener, ``Racial and ethnic diversity of {US}
  chemistry faculty has changed little since 2011'', {\em Chemical \&
  Engineering News} (November 8, 2020).
  \url{https://cen.acs.org/education/Racial-ethnic-diversity-US-chemistry/98/i43}
  (accessed 2024-09-11).
  
\bibitem{BakKoe24}
S.~P. Baker and C.~Koedel,
``Diversity trends among faculty in {STEM} and non-{STEM} fields at selective public universities in the {U.S.} from 2016 to 2023'',
{\em Humanit.\ Soc.\ Sci.\ Commun.}, {\bf 11}, 1149 (2024).
\DOI{10.1057/s41599-024-03687-x}

\bibitem{MatLewHop22}
J.~N. Matias, N.~A. Lewis, and E.~C. Hope, ``{US} universities not succeeding
  in diversifying faculty'', {\em Nat.\ Hum.\ Behav.}, {\bf 6}, 1606--1608
  (2022). \DOI{10.1038/s41562-022-01495-4}

\bibitem{Her23c}
R.~Hernandez, ``Discipline-based diversity research in chemistry'', {\em Acc.\
  Chem.\ Res.}, {\bf 56}, 787--797 (2023). \DOI{10.1021/acs.accounts.2c00797}
  
\bibitem{NSF-diversity}
National Center for Science and Engineering Statistics,
{\em Diversity and STEM: Women, Minorities, and Persons with Disabilities 2023}.
Special Report NSF 23-315, National Science Foundation: Alexandria, 2023.
\url{https://www.nsf.gov/reports/statistics/diversity-stem-women-minorities-persons-disabilities-2023}
(accessed 2025-03-20)

\bibitem{WapZhaCla22}
K.~H. Wapman, S.~Zhang, A.~Clauset, and D.~B. Larremore, ``Quantifying
  hierarchy and dynamics in {US} faculty hiring and retention'', {\em Nature},
  {\bf 610}, 120--127 (2022). \DOI{10.1038/s41586-022-05222-x}

\bibitem{WayMorLar19}
S.~F. Way, A.~C. Morgan, D.~B. Larremore, and A.~Clauset, ``Productivity,
  prominence, and the effects of academic environment'', {\em Proc.\ Natl.\
  Acad.\ Sci.\ USA}, {\bf 116}, 10729--10733 (2019). \DOI{10.1073/pnas.1817431116}

\bibitem{KluSmi86}
J.~R. Kluegel and E.~R. Smith,
{\em Beliefs about Inequality: Americans' Views of What Is and What Ought to Be}, 
Routledge: New York: 1986.  \DOI{10.4324/9781351329002}

\bibitem{CecBla10}
E.~A. Cech and M. Blair-Loy,
``Perceiving a glass ceiling?  Meritocratic versus structural explanations of gender inequality among women in science and technology'',
{\em Soc.\ Probl.}, {\bf 57}, 371--397 (2010). \DOI{10.1525/sp.2010.57.3.371}

\bibitem{Cec13b}
E.~A. Cech, ``The (mis)framing of social justice: {Why} ideologies of
  depoliticization and meritocracy hinder engineers' ability to think about
  social injustices'', in {\em Engineering Education for Social Justice},
  J.~Lucena, Ed., Vol. ~10 of {\em Philosophy of Engineering and Technology;}
\newblock Springer Science+Business Media: Dordrecht, 2013;
\newblock pp.~67--84. \DOI{10.1007/978-94-007-6350-0\_4}

\bibitem{SerSilCec18}
C.~Seron, S.~Silbey, E.~Cech, and B. Rubineau,
``{I} am not a feminist, but$\ldots$'': Hegemony of a meritocratic ideology and the limits of critique among women in engineering'',
{\em Work Occup.}, {\bf 45}, 131--167 (2018). \DOI{10.1177/0730888418759774}

\bibitem{WooCamMcG16}
C.~V. Wood, P.~B. Campbell, and R.~{McGee}, ``\,`{An} incredibly steep hill':
  {How} gender, race, and class shape perspectives on academic careers among
  beginning biomedical {PhD} students'', {\em J.~Women Minor.\ Sci.\ Eng.},
  {\bf 22}, 159--181 (2016). \DOI{10.1615/JWomenMinorScienEng.2016014000}

\bibitem{McGGriHou19}
E.~O. {McGee}, D.~M. Griffith, and S.~L. {Houston, II}, ``\,`{I} know {I} have
  to work twice as hard and hope that makes me good enough': {Exploring} the
  stress and strain of black doctoral students in engineering and computing'',
  {\em Teach.\ Coll.\ Rec.}, {\bf 121}, 1--38 (2019). \DOI{10.1177/016146811912100407}

\bibitem{RicCraNga21}
R.~R. Richardson, D.~C. Crawford, J.~Ngai, and A.~C. {Beckel-Mitchner},
  ``Advancing scientific excellence through inclusivity in the {NIH} {BRAIN}
  initiative'', {\em Neuron}, {\bf 109}, 3361--3364 (2021). \DOI{10.1016/j.neuron.2021.10.021}

\bibitem{McG20b}
E.~O. {McGee}, {\em Black, Brown, Bruised: How Racialized {STEM} Education
  Stifles Innovation}, Harvard Education Press: Cambridge, 2020.
  
\bibitem{PIER-dead}
``Executive Order Update to PIER Plan Requirement''
\newblock U.S. Department of Energy (January 28, 2025).
\url{https://www.energy.gov/science/articles/executive-order-update-pier-plan-requirement}
(accessed 2025-03-19).

\bibitem{AIP-Trump}
L.~McKenzie, ``Science agencies disband DEI initiatives in response to Trump orders''
(January 24, 2025).
\url{https://ww2.aip.org/fyi/science-agencies-disband-dei-initiatives-in-response-to-trump-orders}
(accessed 2025-03-19).

\bibitem{EO-14173}
Executive Order 14173, 
``Ending Illegal Discrimination and Restoring Merit-Based Opportunity''', 
  {\em Fed.\ Reg.}, {\bf 90}, 8633--8636 (2025).
  \url{https://www.federalregister.gov/documents/2025/01/31/2025-02097/ending-illegal-discrimination-and-restoring-merit-based-opportunity}
  (accessed 2025-05-17).

\bibitem{Quinn:2025}
R.~Quinn, 
``Trump is targeting DEI in Higher Ed.  But what does he mean?" (February 27, 2025).
\url{https://www.insidehighered.com/news/diversity/2025/02/27/trump-targeting-dei-higher-ed-what-does-he-mean}
(accessed 2025-03-20).

\bibitem{Knox:2025}
L.~Knox and S.~Weissman,
``Wary colleges scramble to meet DEI deadline'',
{\em Inside Higher Ed} (February 28, 2025).
\url{https://www.insidehighered.com/news/diversity/race-ethnicity/2025/02/28/colleges-scramble-meet-federal-anti-dei-deadline}
(accessed 2025-03-19).

\bibitem{NCES-dead1}
J.~Mehta and C.~Turner,
``Trump administration targets Education Department research arm in latest cuts'',
{\em NPR} (February 10, 2025).
\url{https://www.npr.org/2025/02/10/nx-s1-5292444/trump-musk-education-department-schools-students-research-cuts}
(accessed 2025-03-20).

\bibitem{NCES-dead2}
N.~Modan,
``What will NCES layoffs mean for the Nation's Report Card?",
{\em K--12 Dive} (March 18, 2025).
\url{https://www.k12dive.com/news/Education-Department-nces-layoffs-leaves-naep-assessments-nations-report-card-barebones/742837}
(accessed 2025-03-20).

\bibitem{Frontiers-rules}
``Article types'', {\em Front.\ Res.\ Metr.\ Anal.}.
\url{https://www.frontiersin.org/journals/research-metrics-and-analytics/for-authors/article-types}
(accessed 2025-03-19).

\bibitem{Dutton:2025}
C.~Dutton, ``Changing the course'',
  {\em The Chronicle of Higher Education} (March 5, 2025).
\url{https://www.chronicle.com/article/changing-the-course} (accessed 2025-03-19).

\bibitem{Blake:2025}
J.~Blake and K.~Knott,
``Trump's demands to Columbia reflect assault on Higher Ed, experts say'',
{\em Inside Higher Ed} (March 14, 2025).
\url{https://www.insidehighered.com/news/government/politics-elections/2025/03/14/trump-escalates-attack-columbia-his-latest-demands}
(accessed 2025-03-19).

\bibitem{Stanley:2025}
J.~Stanley, ``Trump is setting the {US} on a path to educational authoritarianism'',
{\em The Guardian} (March 17, 2025).
\url{https://www.theguardian.com/commentisfree/2025/mar/17/trump-us-path-educational-authoritarianism}
(accessed 2025-03-17).

\end{thebibliography}

\end{document}


\title{\Large 
\textbf{
	\underline{Supplementary Material for}: \\[10pt]
	Comment on ``Politicizing science funding undermines public trust in science, academic freedom, 
	and the unbiased generation of knowledge''
}}
\author{
	John M. Herbert\footnote{\href{mailto:herbert@chemistry.ohio-state.edu}{herbert@chemistry.ohio-state.edu}} 
}
\affil{
\em	Department of Chemistry \& Biochemistry  \\
\em	The Ohio State University, Columbus, Ohio USA
}
\date{}\maketitle


\noindent
This document contains version of the source material that were downloaded in late 2024, lest they disappear 
from government websites during the Trump administration.

\tableofcontents\pagebreak
\section{Slide Deck about Broader Impacts (\texorpdfstring{Ref.~\protect\citenum{Ren23}}{})}
\includepdf[pages=-]{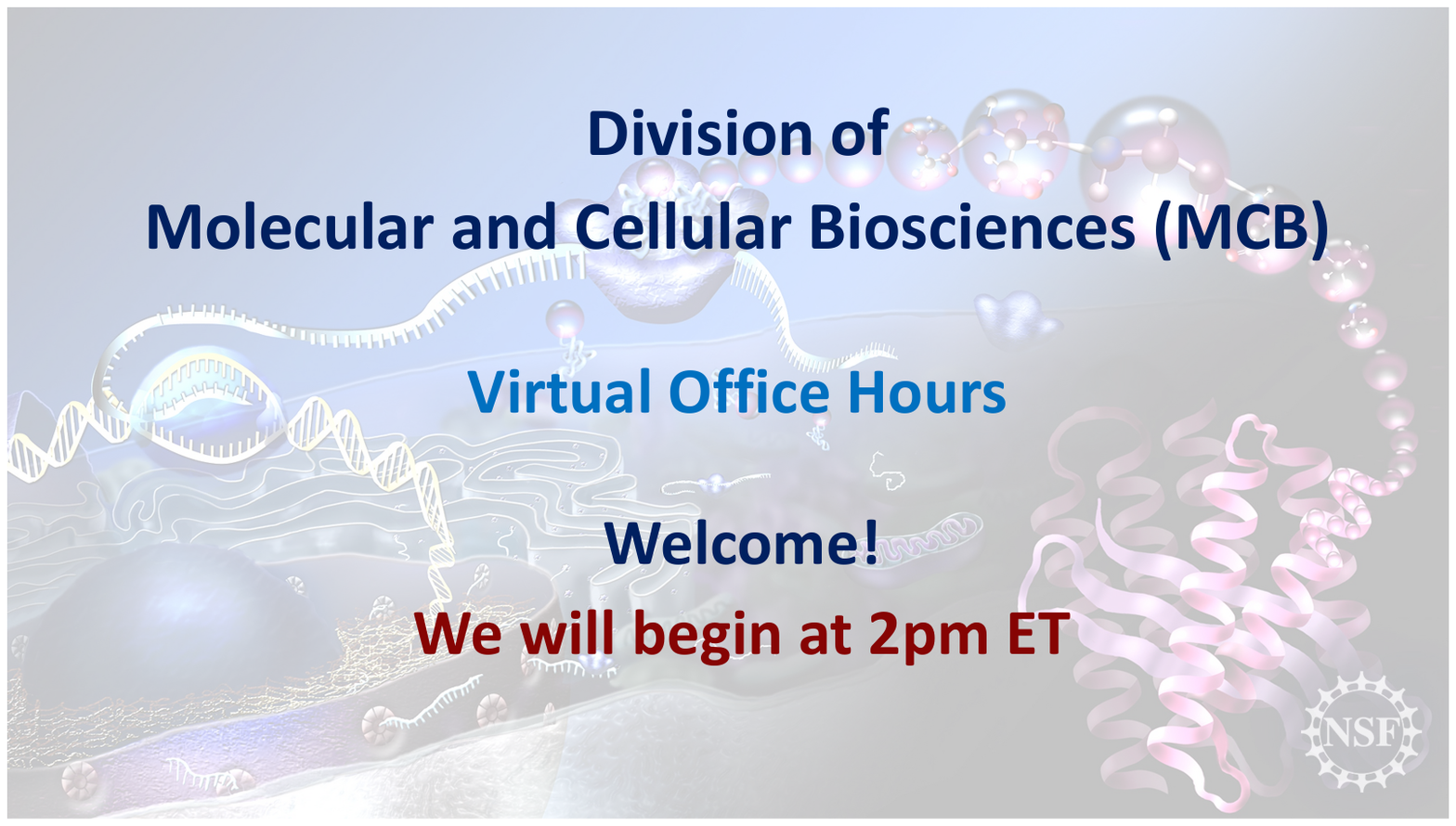}

\section{ARIS Leadership Team (\texorpdfstring{Ref.~\protect\citenum{ARIS-team}}{})}
\includepdf[pages=-]{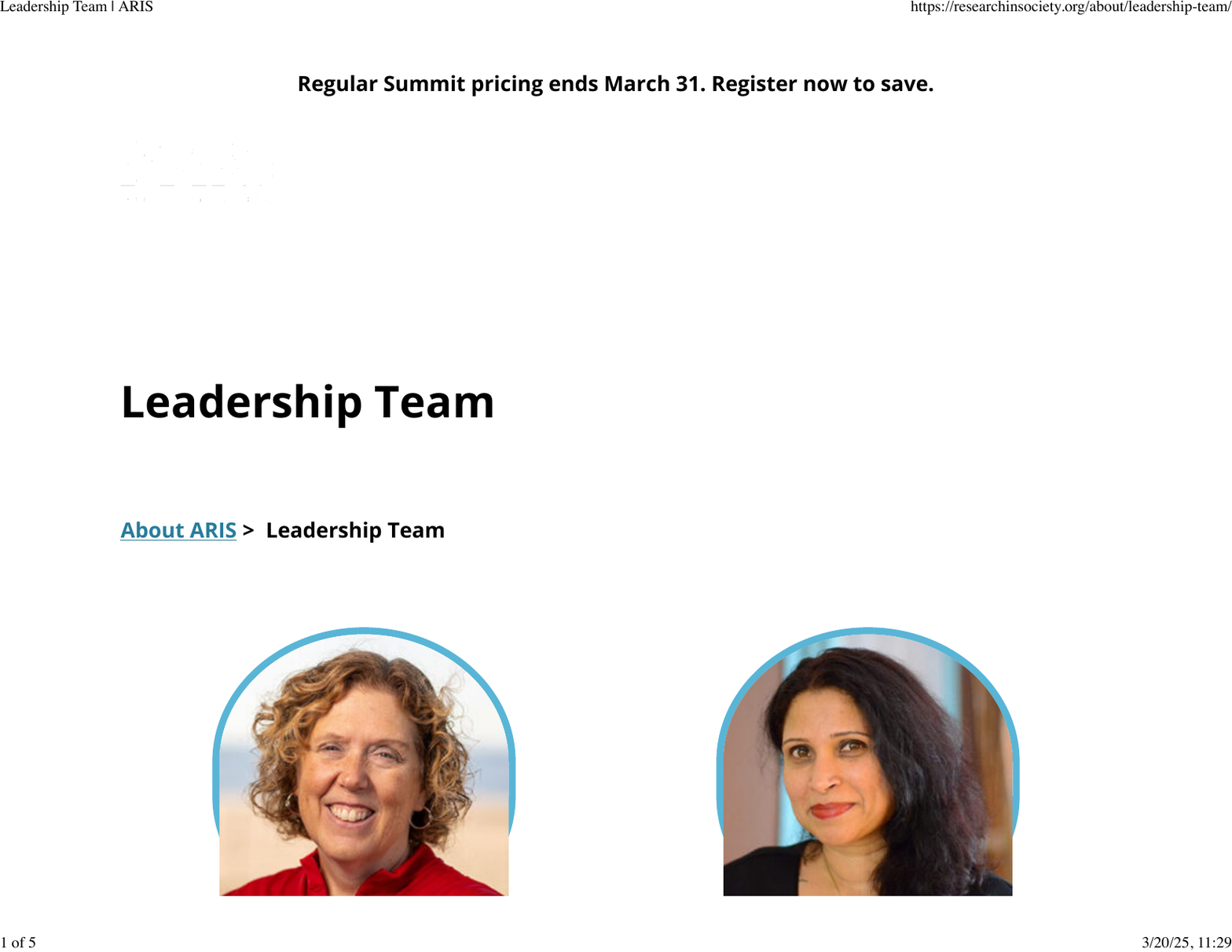}

\section{NIH-PRIDE Funding Opportunity Announcement (\texorpdfstring{Ref.~\protect\citenum{NIH-PRIDE}}{})}
\includepdf[pages=-]{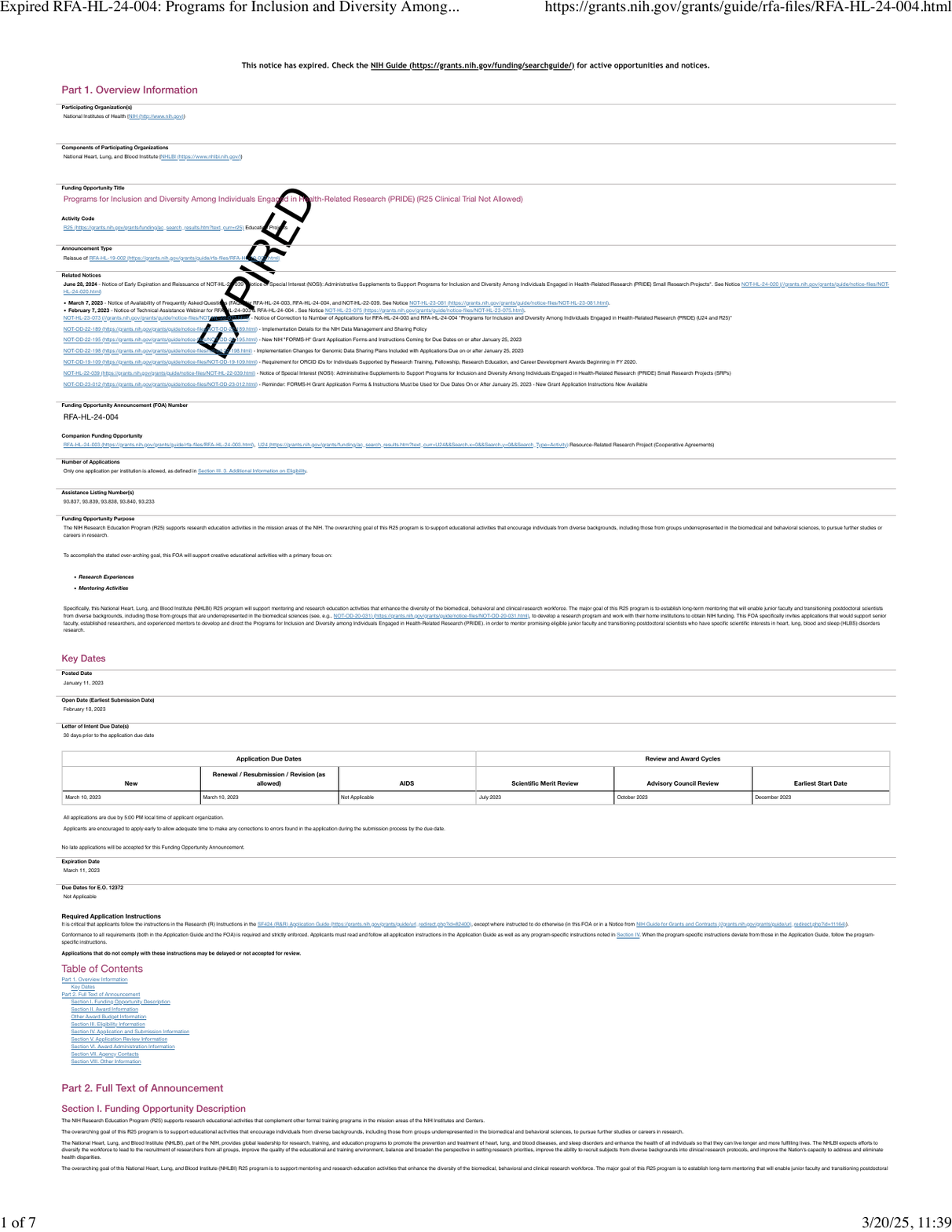}

\section{NSF Privacy Act and Public Burden Statement (\texorpdfstring{Ref.~\protect\citenum{Pli20}}{})}
\includepdf[pages=-]{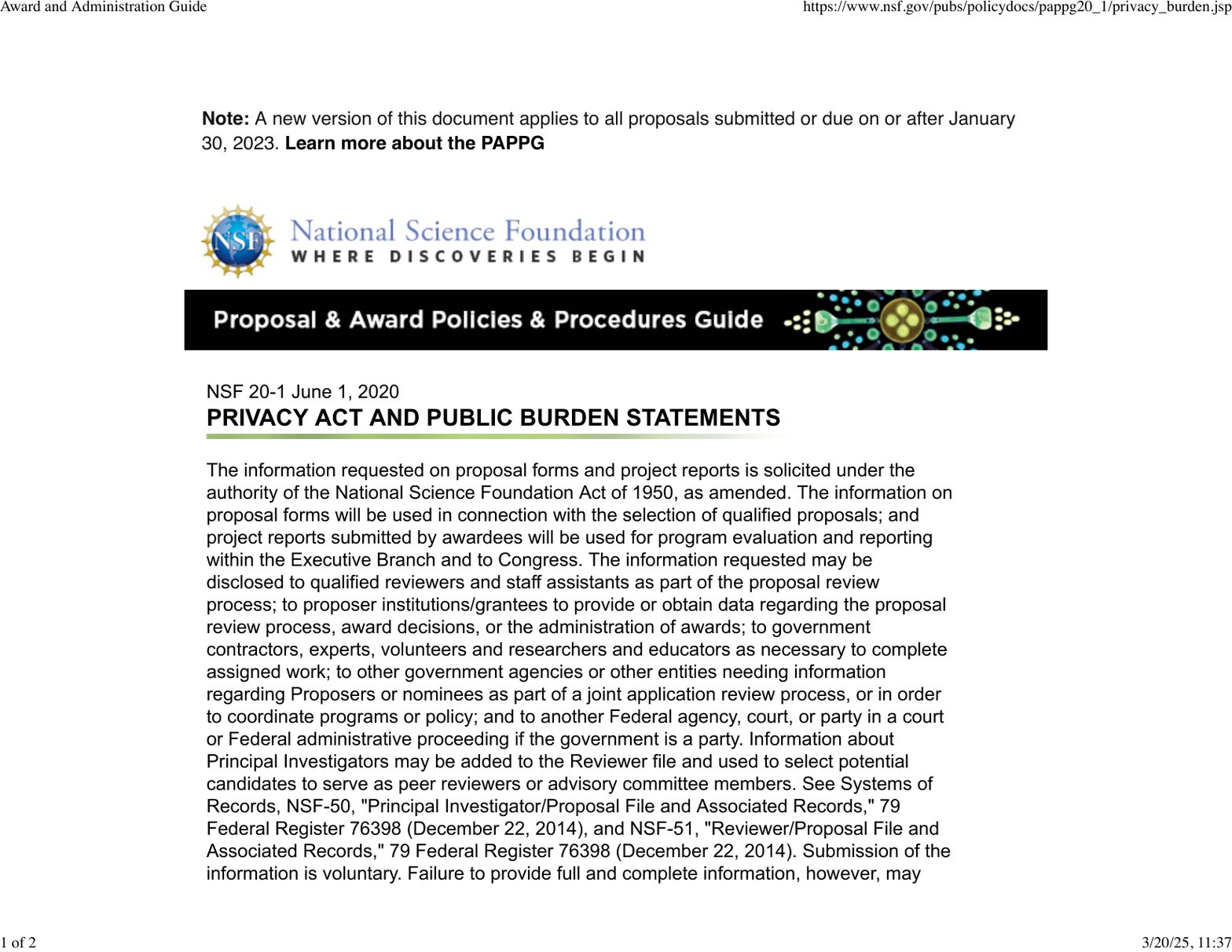}

\section{Slide Deck about NASA Inclusion Plans (\texorpdfstring{Ref.~\protect\citenum{NahWat23}}{})}
\includepdf[pages=-]{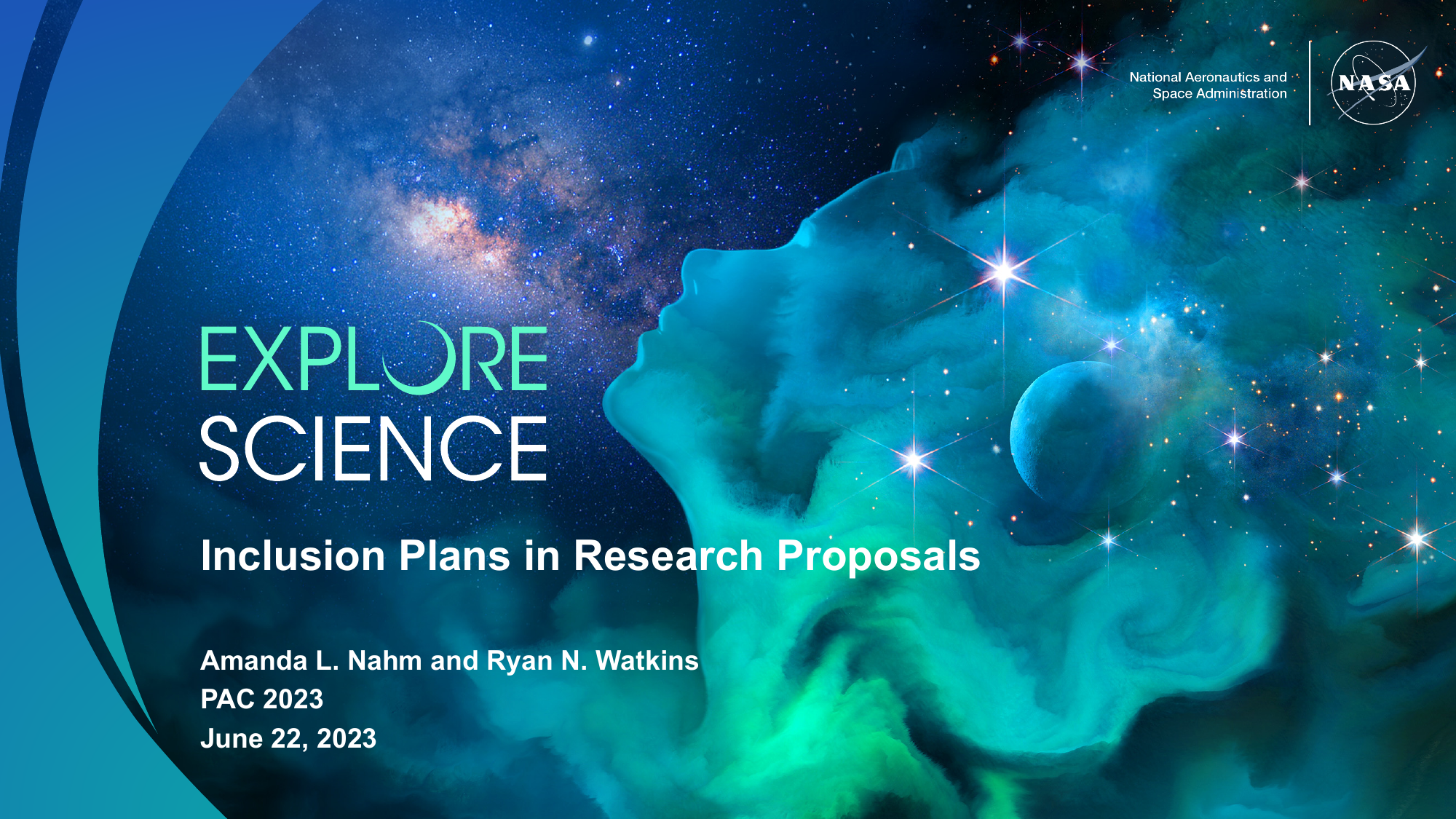}

\section{Slide Deck about DOE PIER Plans (\texorpdfstring{Ref.~\protect\citenum{DOE-PIER}}{})}
\includepdf[pages=-]{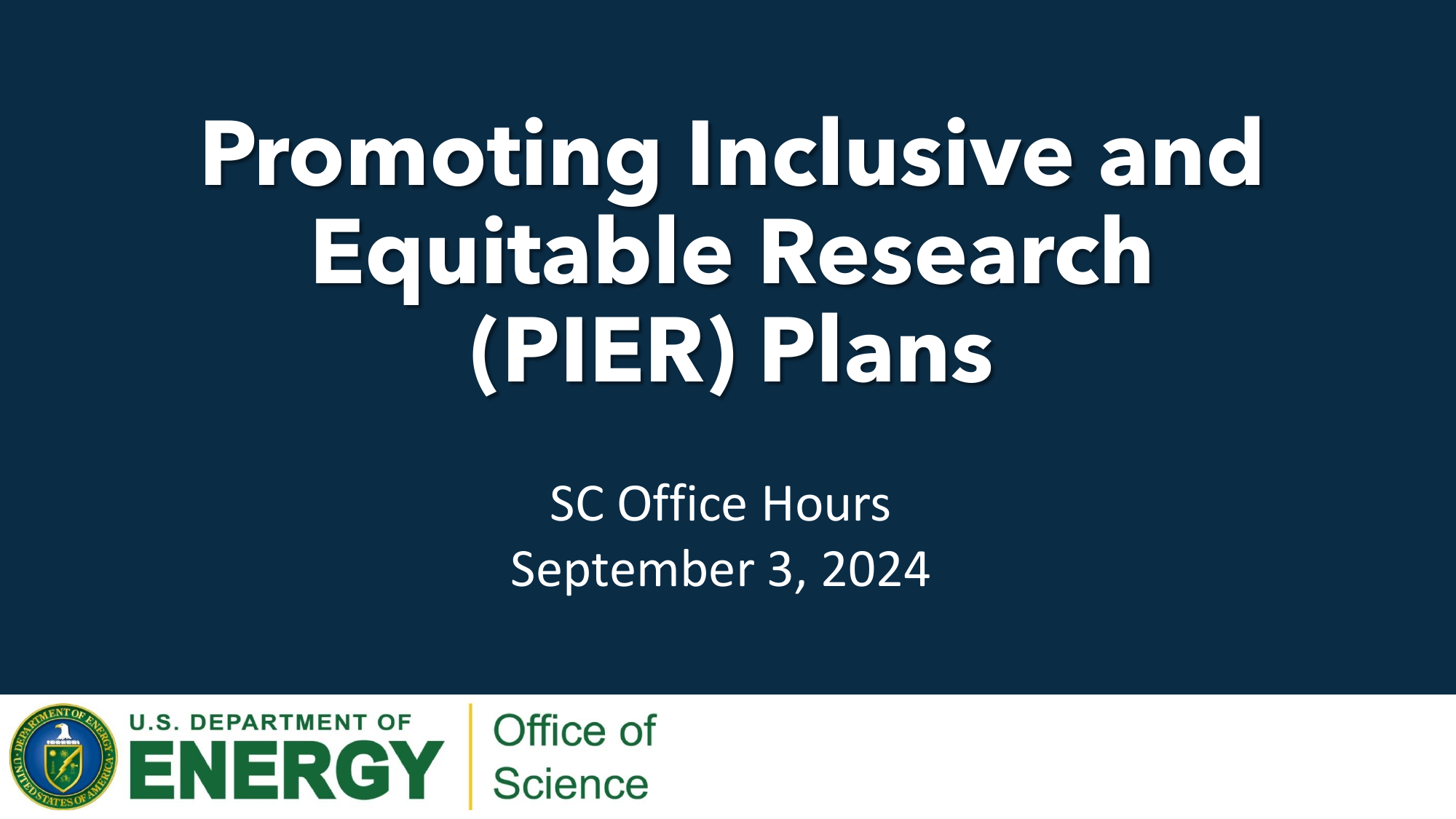}

\section{NCES Statistical Characterization of Post-Secondary Students (\texorpdfstring{Ref.~\protect\citenum{NCES}}{})}
\includepdf[pages=-]{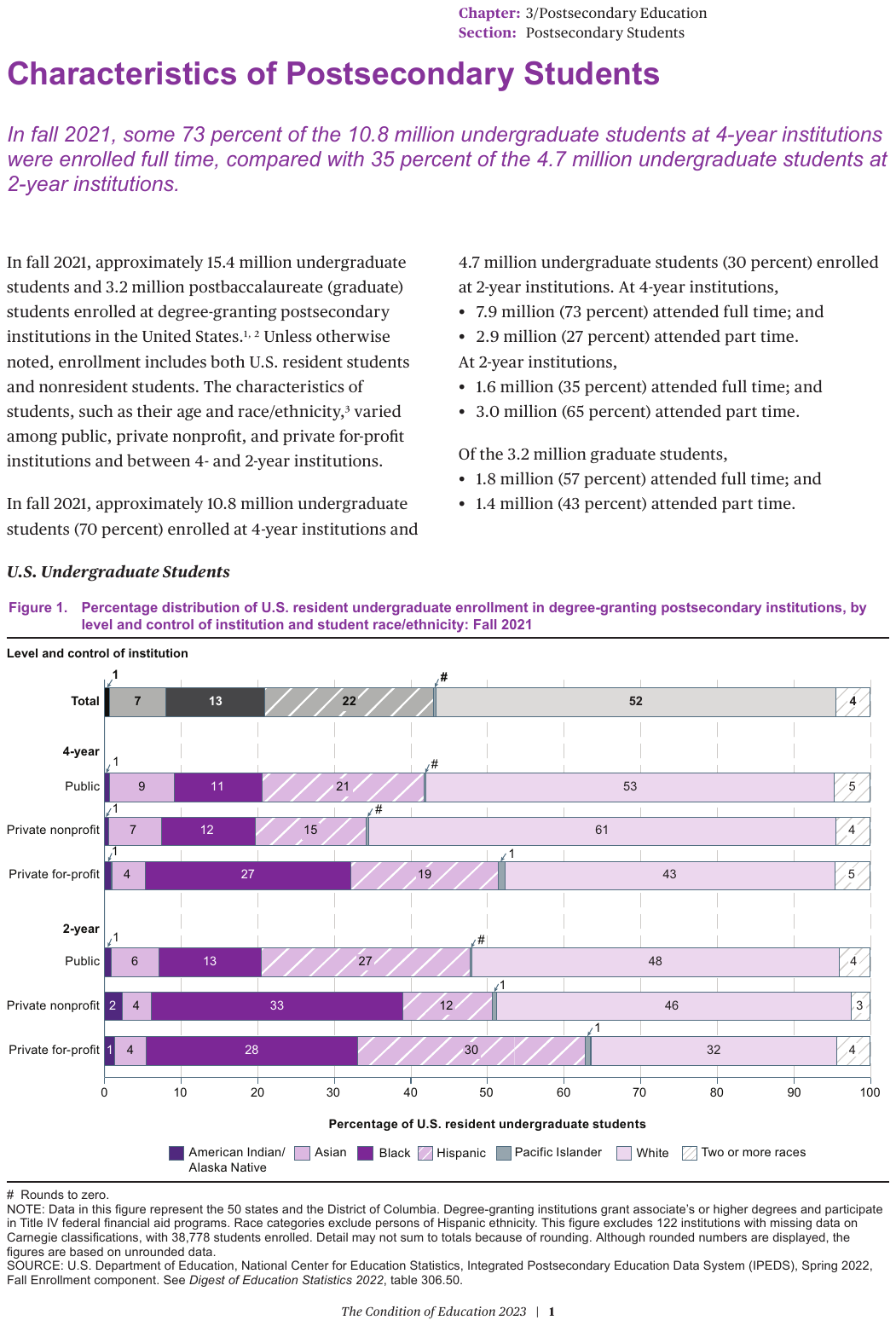}

\section{Report on STEM Workforce Diversity (\texorpdfstring{Ref.~\protect\citenum{NSF-diversity}}{})}


\section*{Supplementary material}

# Diversity and STEM

Women, Minorities, and Persons with Disabilities  2023

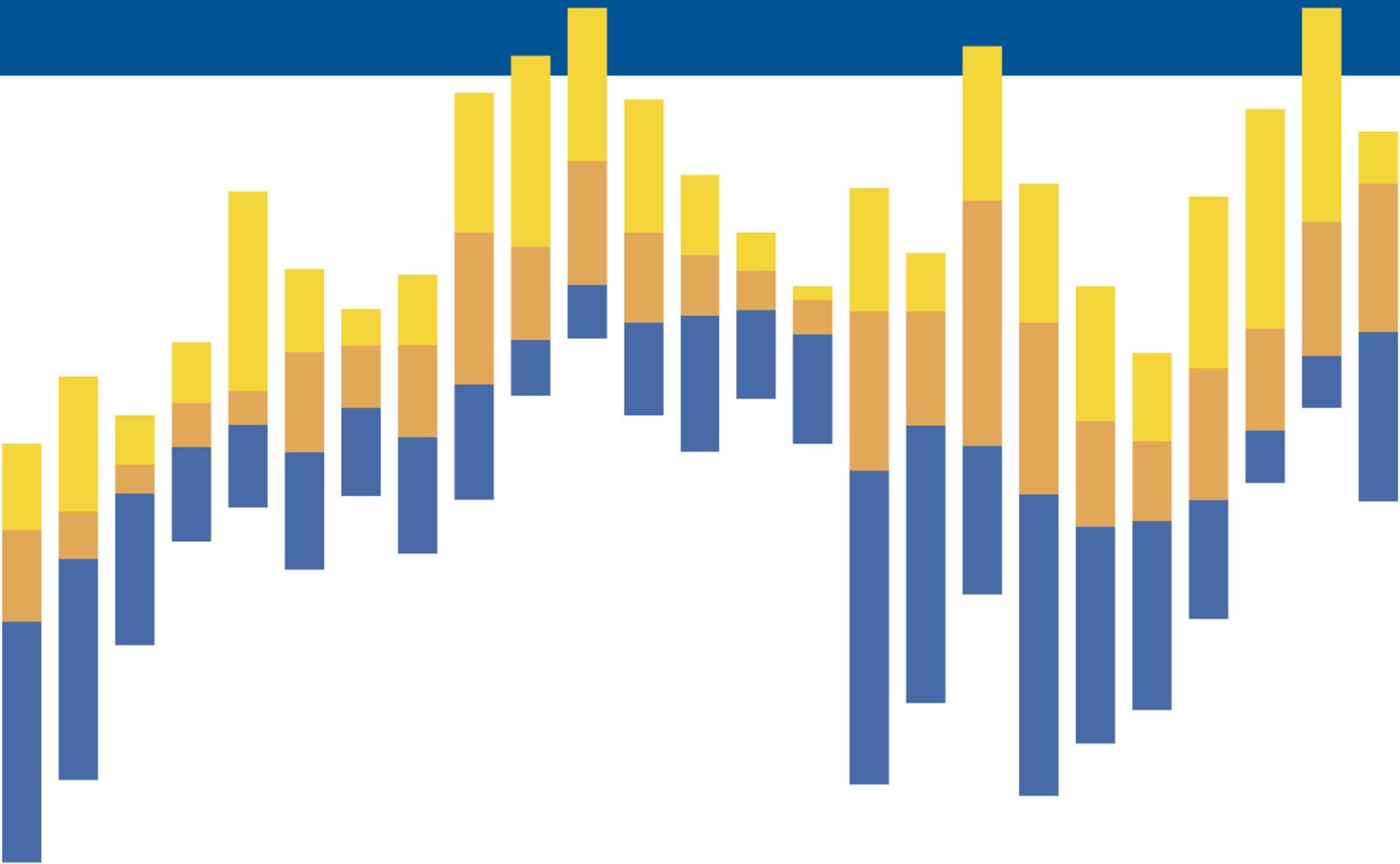

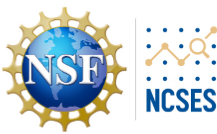

National Center for Science and Engineering Statistics
Directorate for Social, Behavioral and Economic Sciences
National Science Foundation

# Table of Contents

# Executive Summary

A diverse workforce provides the potential for innovation by leveraging different backgrounds, experiences, and points of view. Innovation and creativity, along with technical skills relying on expertise in science, technology, engineering, and mathematics (STEM), contribute to a robust STEM enterprise. Furthermore, STEM workers have higher median earnings and lower rates of unemployment compared with non-STEM workers. This report provides high-level insights from multiple data sources into the diversity of the STEM workforce in the United States.

## Key Takeaways

### STEM Workforce

- The U.S. STEM workforce gradually diversified between 2011 and 2021, with increased representation of women and underrepresented minorities—Hispanics or Latinos, Blacks or African Americans, and American Indians or Alaska Natives.
- In 2021, among people ages 18 to 74 years, women made up half (51%) of the total U.S. population and about a third (35%) of people employed in STEM occupations.
- In 2021, nearly a quarter (24%) of individuals in the U.S. workforce were employed in STEM occupations.
- Hispanic workers represented 15% of the total STEM workforce in 2021, and Asian and Black workers were 10% and 9%, respectively. American Indians and Alaska Natives together made up less than 1% of the U.S. population and STEM workforce in 2021.
- In 2021, among female STEM workers, 68% had science and engineering (S&E)–related jobs (health care workers, S&E managers, S&E precollege teachers, and technologists and technicians); women represented nearly two-thirds (65%) of workers in S&E-related occupations.
- In 2021, nearly two-thirds (63%) of Hispanic individuals in STEM jobs worked in middle-skill occupations (jobs that require considerable STEM skills and expertise but do not typically require a bachelor's degree for entry); among those in middle-skill occupations, 24% were Hispanic.
- In 2020, women had lower median earnings than men in S&E, S&E-related, and middle-skill occupations.
- In 2020, Hispanic, Black, and American Indian or Alaska Native STEM workers had lower median earnings than White or Asian STEM workers.
- Persons with a disability who worked part time in an S&E-related occupation in 2021 reported wanting to work full time at about twice the rate of those without a disability (28% vs. 15%).

### STEM Workforce and Education

- Nearly two-thirds (65%) of women working in STEM jobs in 2021 had at least a bachelor's degree education, compared with less than half (43%) of men in STEM jobs.
- Among the college-educated workforce in S&E occupations in 2021, women's representation ranged from 61% of social and related scientists to 16% of engineers.
- In 2021, about two-thirds (65%) of STEM workers with at least one disability had less than a bachelor's degree education.

- Underrepresented minorities—Hispanic, Black, and American Indian or Alaska Native individuals—made up a higher share of the skilled technical workforce (32%) in 2021 than of workers who were employed in STEM occupations with at least a bachelor's degree (16%).

## S&E Degrees

- Hispanic, Black, and American Indian or Alaska Native persons collectively account for 37% of the U.S. population ages 18–34 years in 2021 and 26% of S&E bachelor's, 24% of S&E master's, and 16% of S&E doctoral degrees earned by U.S. citizens and permanent residents in 2020.
- At the associate's level only, Hispanic students earned a higher share of S&E degrees among U.S. citizens and permanent residents in 2020 (32%), relative to the Hispanic share of the college-age population (22%).
- Black students had higher representation among the U.S. citizens and permanent residents in the social and behavioral sciences, earning 12% of bachelor's degrees in these fields in 2020, relative to 5% of bachelor's degrees in engineering.
- The number of S&E degrees earned by women between 2011 and 2020 increased by 63% at the associate's level, 34% at the bachelor's level, 45% at the master's degree level, and 18% at the doctorate level.
- In 2020, women earned 66% of bachelor's, 67% of master's, and 60% of doctoral degrees in the social and behavioral sciences.
- In 2020, women were underrepresented among degree recipients at all degree levels in physical and earth sciences, mathematics and computer sciences, and engineering.
- Among S&E doctorate recipients in 2021, individuals earning degrees in psychology and social sciences had the highest rate of disability (13%) and those in engineering had the lowest rate (8%).

# Introduction

## Overview

The U.S. science, technology, engineering, and mathematics (STEM) workforce fuels innovation and provides important contributions to the nation. New advancements and discoveries in science and technology are rapidly changing the world of work and increasing the demand for technically skilled employees. As this demand has increased, so has the number of STEM workers. In 2021, 34.9 million people worked in STEM occupations, up from 29.0 million in 2011. Today, nearly a quarter (24%) of the U.S. workforce is employed in STEM occupations.

Representation of different groups based on sex, race or ethnicity, and disability status varies within the STEM workforce, and representation in STEM occupations is uneven relative to the distribution of these groups in the working age population. Women, persons with disabilities, and persons from some racial and ethnic minority groups—Hispanic or Latino, Black or African American, and American Indian or Alaska Native—are underrepresented in the STEM workforce when compared to their share of the total population. In 2021, women made up half (51%) of the total population ages 18 to 74 years and about a third (35%) of those employed in a STEM occupation (figure 1-1). Although 9% of the population had one or more disabilities at that time, 3% of those who work in STEM occupations did. When combined, Hispanics, Blacks, and American Indians or Alaska Natives—collectively referred to as underrepresented minorities—made up 31% of the total population and 24% of STEM workers in 2021.

Women and some racial and ethnic minority groups are also underrepresented in postsecondary science and engineering (S&E) education—which may be indicative of their future participation in the STEM workforce. Women earned approximately half of the S&E degrees at the associate's and bachelor's degree levels in 2020, which was similar to their share of the population ages 18 to 34 years (figure 1-2). This age group, referred to here as the college-age population, includes most students completing degrees at the associate through doctoral levels.[1] Compared with the proportion of women earning associate's degrees, women accounted for lower shares of advanced S&E degree recipients, earning 46% of S&E master's degrees and 41% of S&E doctoral degrees in 2020.

Racial and ethnic groups also vary in their representation among S&E degree recipients relative to their share of 18-to-34-year-olds. Compared with their proportions of the college-age population, Whites and Asians account for a disproportionately high share of S&E degree recipients at the bachelor's level and above, whereas Hispanics, Blacks, and American Indians or Alaska Natives account for a disproportionately low share of these degree recipients. The gap is increasingly pronounced at higher degree levels. Underrepresented minorities collectively accounted for 37% of the college-age population in 2021 and 26% of S&E bachelor's, 24% of S&E master's, and 16% of S&E doctoral degree recipients in 2020.

### Figure 1-1

**Characteristics of the U.S. population ages 18–74, by labor force status: 2021**

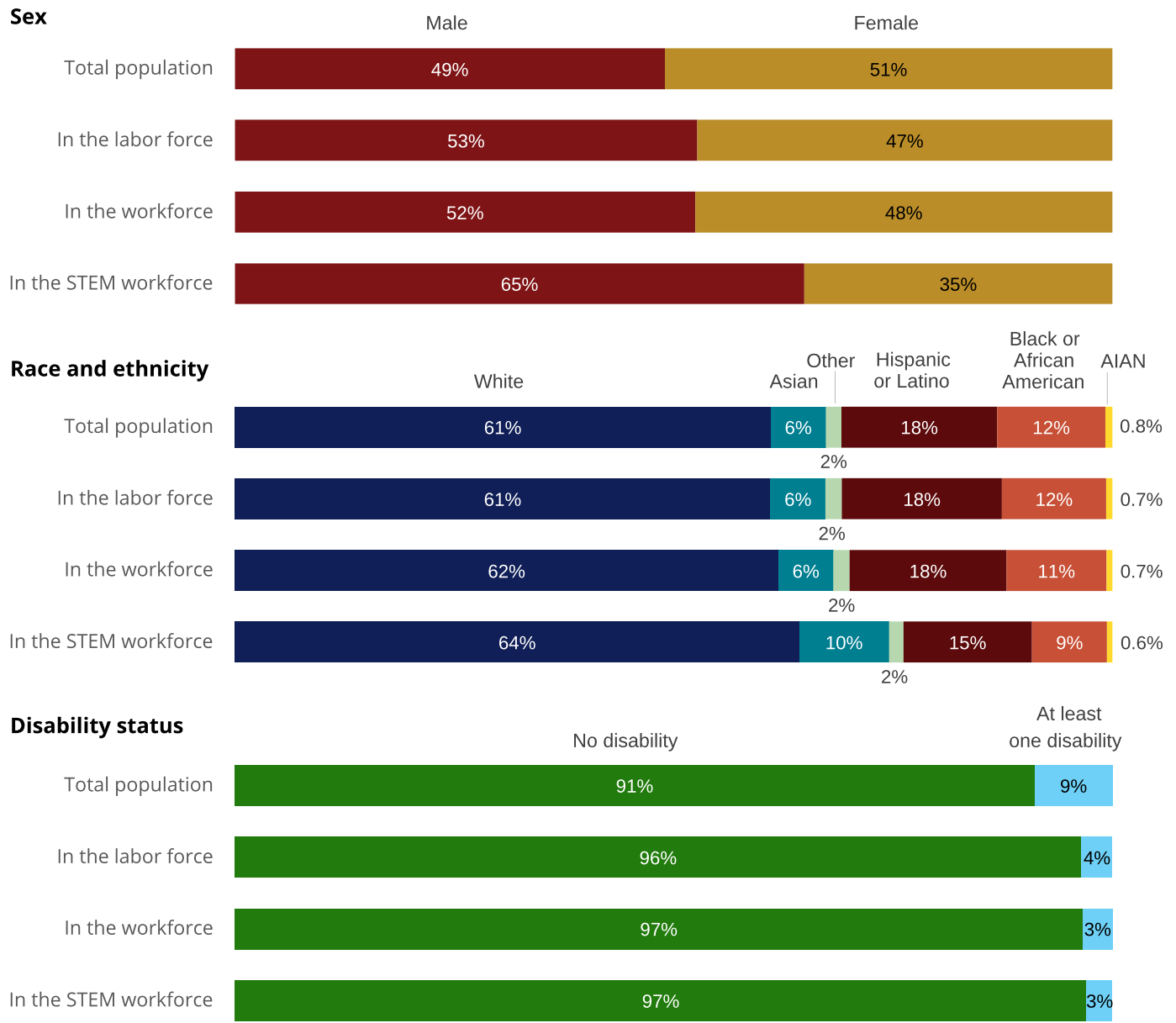

AIAN = American Indian or Alaska Native; STEM = science, technology, engineering, and mathematics.

**Note(s):**
Civilian noninstitutionalized population plus armed forces living off post or with their families on post. Hispanic or Latino may be any race; race categories exclude Hispanic origin. Other includes Native Hawaiian and Other Pacific Islander and more than one race. Respondents can report more than one disability. Those who reported difficulty with one or more functionalities were classified as having a disability. The labor force includes those who are employed and those who are not working but actively seeking work (unemployed). The workforce includes only employed individuals. Due to rounding, percentages may not sum to 100 or subgroup totals.

**Source(s):**
Census Bureau, Current Population Survey, Annual Social and Economic Supplement, 2021.

**Figure 1-2**

**Characteristics of S&E degree recipients, by degree level: 2020**

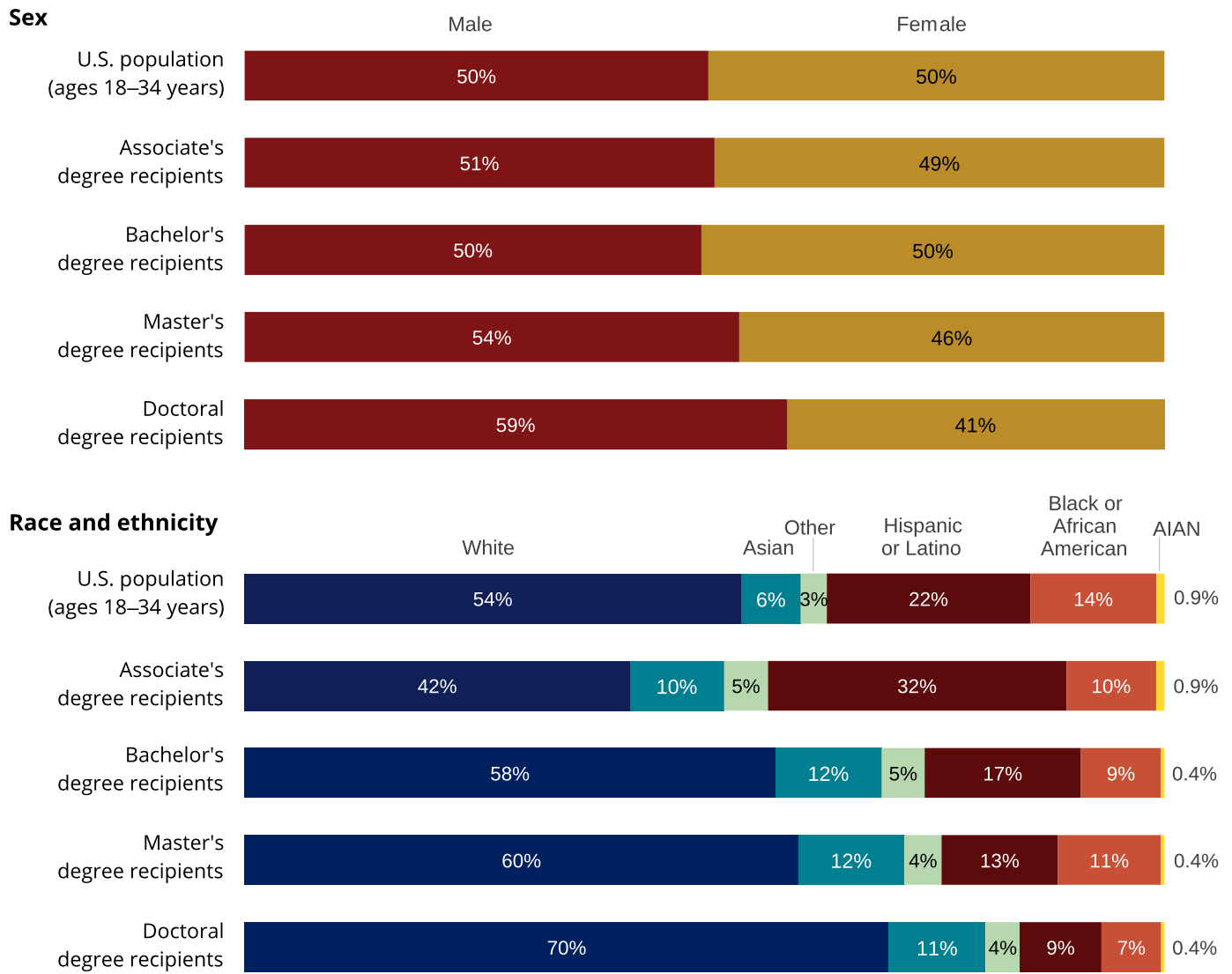

AIAN = American Indian or Alaska Native; S&E = science and engineering.

**Note(s):**
Hispanic or Latino may be of any race; race categories exclude Hispanic origin. Other includes Native Hawaiian and Other Pacific Islander and more than one race. Race and ethnicity data for S&E degree recipients are available only for U.S. citizens and permanent residents.

**Source(s):**
National Center for Education Statistics, Integrated Postsecondary Education Data System, Completions Survey, 2020. U.S. population data from Census Bureau, Current Population Survey, Annual Social and Economic Supplement, 2021.

## Purpose of Report

The purpose of this report, *Diversity and STEM: Women, Minorities, and Persons with Disabilities 2023,* is to provide statistical information about these three groups, all of whom have historically been underrepresented in the S&E enterprise. As mandated by the Science and Engineering Equal Opportunities Act (Public Law 96-516), the National Science Foundation through the National Center for Science and Engineering Statistics (NCSES) publishes a biennial report to assess the current standing of women, persons from racial or ethnic minority groups, and persons with disabilities in STEM employment and S&E education in relation to other groups. The analyses presented here are relevant to policymakers, program managers, and researchers interested in diversity and representation in the U.S. science and engineering enterprise.

## Structure of Report

This report begins by describing the representation of women, persons from underrepresented minority groups, and persons with disabilities in the STEM workforce as a whole and in three types of STEM occupations (S&E, S&E-related, and middle-skill occupations) using data from 2011 and 2021. Analyses of comparative wage and salary earnings and the impact of educational attainment are presented using the survey data from 2021. The section on unemployment rates across these groups highlights the effects of the COVID-19 pandemic, focusing on the differences between 2019 and 2021. The section on the STEM workforce with at least a bachelor's degree describes the representation of these groups in specific STEM occupations in 2021 and how part-time versus full-time employment varies across the groups.

Two sections of this report focus on S&E education. The section on degrees earned shows the data by broad S&E field from 2011 to 2020, with an emphasis on bachelor's degree data for underrepresented minority groups. Representation of women among recipients of S&E technologies associate's degrees, a typical entry point into the skilled technical workforce, is described in a sidebar. The section on enrollment addresses the effect of the COVID-19 pandemic on graduate enrollment, exploring differences based on sex, race, and ethnicity. Collectively, these sections provide insight into the likely representation of these groups in the future STEM labor force.

## Defining the STEM Workforce, Racial and Ethnic Categories, and Disability Status

When discussing the STEM workforce, this report uses the definition of the STEM labor force as outlined in the *Science and Engineering Indicators 2022* report The STEM Labor Force of Today: Scientists, Engineers, and Skilled Technical Workforce (NSB, NSF 2021) and the infographic *Workforce Statistics* (NCSES 2022)*.* This new definition of the STEM labor force includes workers in S&E, S&E-related, and middle-skill occupations (see sidebar The STEM Workforce of the United States and the "Glossary" section for definitions).

**SIDEBAR**

**The STEM Workforce of the United States**

The science, technology, engineering, and mathematics (STEM) workforce is made up of individuals at all education levels who work in a wide variety of occupations. This workforce is defined by broad occupation type—science and engineering (S&E), S&E-related, and middle-skill occupations—and by educational attainment—either having at least a bachelor's degree or not.

*S&E occupations* typically require a bachelor's degree for entry and are broadly composed of workers who are computer and mathematical scientists; biological, agricultural, and environmental life scientists; physical scientists; social scientists; and engineers.

*S&E-related occupations* require STEM skills and expertise, but they do not fall into the five main S&E occupational categories listed above. The main occupational categories and positions that make up this group include health care workers, S&E managers, S&E precollege teachers, and technologists and technicians.

*Middle-skill occupations* require considerable STEM skills and expertise but do not typically require a bachelor's degree for entry. These positions are primarily in the areas of construction trades, installation, maintenance, and production.

STEM workers with a bachelor's degree or higher and those without such a degree are employed in all three broad occupation types (figure 1-A). However, their distribution tends to reflect the degree or training requirements of the occupations in each broad group. S&E occupations often require advanced education, so most of those workers have at least a bachelor's degree (82%). In contrast, middle-skill occupations tend to require certification, licensing, or on-the-job training and thus have a high percentage of workers without a bachelor's degree (85%).

Grouping STEM workers by educational attainment creates two workforce groups. The STEM workforce with at least a bachelor's degree includes individuals who have attained a bachelor's degree or higher and who work in S&E, S&E-related, or middle-skill occupations. These workers include engineers; software developers; physicians; registered nurses; industrial production managers; and farmers, ranchers, and other agricultural managers.

The STEM workforce without a bachelor's degree, referred to as the skilled technical workforce, is comprised of workers in S&E, S&E-related, and middle-skill occupations that require a high-level of knowledge in a technical domain but do not require a bachelor's degree. These workers hold positions such as computer support specialists, industrial engineers, licensed nurses, pharmacy technicians, carpenters, and electricians.

**Figure 1-A**

STEM workforce ages 18–74, by education: 2021

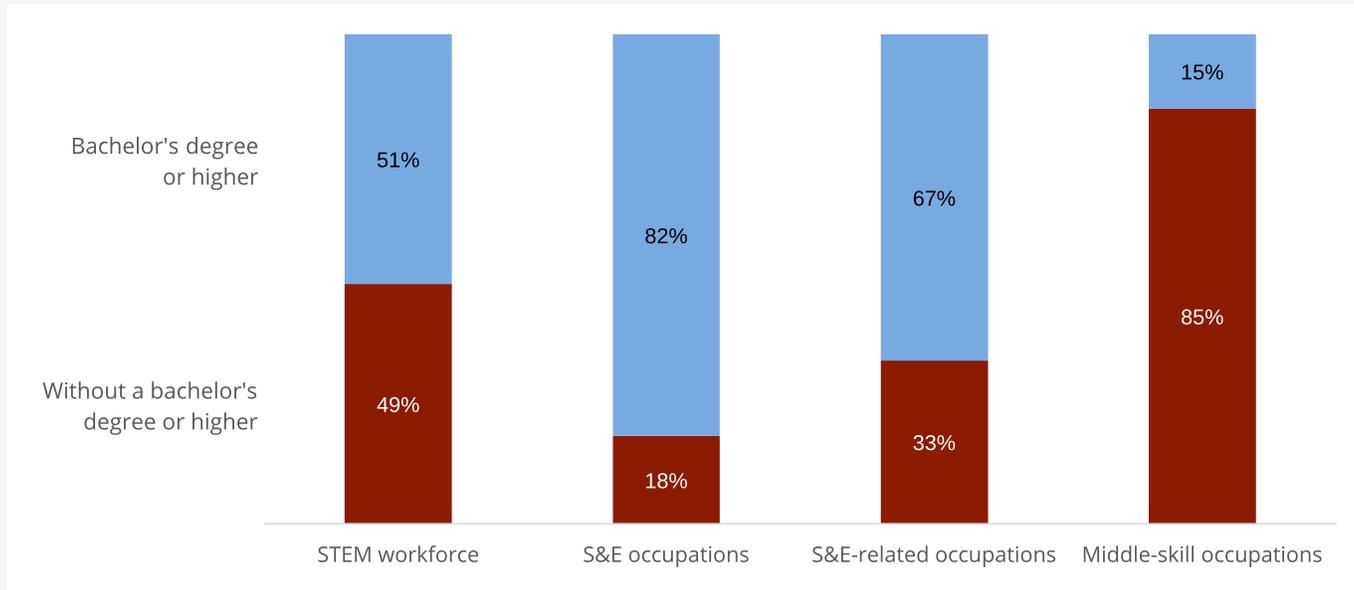

S&E = science and engineering; STEM = science, technology, engineering, and mathematics.

**Note(s):**
Civilian noninstitutionalized population plus armed forces living off post or with their families on post.

**Source(s):**
Census Bureau, Current Population Survey, Annual Social and Economic Supplement, 2021.

---

For more information about the STEM workforce, see the infographic *Workforce Statistics*\* and the *Science and Engineering Indicators 2022* report The STEM Labor Force of Today: Scientists, Engineers, and Skilled Technical Workers.[†]

\* National Center for Science and Engineering Statistics (NCSES). 2022. *Workforce Statistics*. Infographic NCSES 22-203. Alexandria, VA. Available at https://ncses.nsf.gov/136/assets/0/files/ncses_workforcestatistics_onepager.pdf.

[†] National Science Board, National Science Foundation (NSB, NSF). 2021. The STEM Labor Force of Today: Scientists, Engineers and Skilled Technical Workers. *Science and Engineering Indicators 2022*. NSB-2021-2. Alexandria, VA. Available at https://ncses.nsf.gov/pubs/nsb20212.

Where possible, data for all three underrepresented minority groups—Hispanics, Blacks, and American Indians or Alaska Natives—are presented separately. However, the group of American Indians or Alaska Natives had a sample size that was too small to produce statistically reliable estimates in some cases, and those data are not shown. The sidebar Defining Race and Ethnicity provides details about how racial and ethnic groups are defined by the data sources used in this report.

> **SIDEBAR**
>
> **Defining Race and Ethnicity**
>
> *Current Population Survey*
>
> The Census Bureau's Current Population Survey (CPS) uses two separate questions to ask respondents about their race and Hispanic origin. With respect to race, a respondent can be White, Black or African American, American Indian or Alaska Native, Asian, Native Hawaiian or Other Pacific Islander, or a combination of two or more of these groups. To determine Hispanic origin, respondents were asked, Are you Spanish, Hispanic, or Latino? A respondent's ethnicity can be Hispanic or not Hispanic, regardless of race.
>
> CPS data on race and Hispanic origin were combined to create the six nonoverlapping racial and ethnic groups used in this report:
>
> - Hispanic: Hispanic, any race
> - White: single race only, not Hispanic
> - Black: single race only, not Hispanic
> - American Indian or Alaska Native: single race only, not Hispanic
> - Asian: single race only, not Hispanic
> - Other: Native Hawaiian or Other Pacific Islander, single race only, not Hispanic; more than one race, not Hispanic
>
> *Integrated Postsecondary Education Data System*
>
> For this report, data on race and ethnicity from the National Center for Education Statistics' Integrated Postsecondary Education Data System (IPEDS) were combined to align with the categories used in the CPS. The CPS collects data on race and ethnicity for all survey respondents, regardless of citizenship status. By comparison, the race and ethnicity data in IPEDS are available only for U.S. citizens and permanent residents; these details are not available for temporary residents. All race and ethnicity comparisons in this report that use IPEDS degree award data are limited to U.S. citizens and permanent residents with reported race and ethnicity.
>
> *Surveys from the National Center for Science and Engineering Statistics*
>
> Data on race and ethnicity from the National Center for Science and Engineering Statistics' (NCSES's) Survey of Graduate Students and Postdoctorates in Science and Engineering (GSS) are collected similarly to those in IPEDS and were aligned to match CPS data. All race and ethnicity comparisons in this report that use GSS data are limited to U.S. citizens and permanent residents with reported race and ethnicity.
>
> For the National Survey of College Graduates (NSCG) and the Survey of Doctorate Recipients (SDR), both surveys from NCSES, the racial and ethnic groups correspond to those used in the CPS. All race and ethnicity comparisons in this report that use NSCG and SDR data include U.S. citizens, foreign citizens on permanent visas, and individuals on temporary visas.

The data sources used in this report estimate the number of persons with at least one disability in two different ways (see sidebar Defining Persons with at Least One Disability). The Current Population Survey (CPS) uses a set of six yes-or-no questions to assess disability. The NCSES surveys, including the National Survey of College Graduates, the Survey of Earned Doctorates, and the Survey of Doctoral Recipients, use a set of five scale questions to determine functional limitations. Although the two different classification criteria for disability measure the same broad concept, caution should be taken when comparing the estimates derived from the CPS and NCSES sources.

**SIDEBAR**

**Defining Persons with at Least One Disability**

The Census Bureau's Current Population Survey (CPS) uses a series of yes-or-no questions to ask about six disability types: hearing difficulty, vision difficulty, cognitive difficulty, ambulatory difficulty, self-care difficulty, and independent living difficulty. The questions include the following:

- Do you have difficulty dressing or bathing?
- Are you deaf or do you have serious difficulty hearing?
- Are you blind or do you have serious difficulty seeing even when wearing glasses?
- Because of a physical, mental, or emotional condition, do you have difficulty doing errands alone such as visiting a doctor's office or shopping?
- Do you have serious difficulty walking or climbing stairs?
- Because of a physical, mental, or emotional condition, do you have serious difficulty concentrating, remembering, or making decisions?

Respondents who report having any one of the six disability types are considered to have a disability.

For more information about how disability data are collected from the CPS, see Guidance for Disability Data Users and How Disability Data Are Collected from the Current Population Survey.

The six-item set of questions used on the CPS is one way to gauge disability. This report also presents disability data from the National Center for Science and Engineering Statistics (NCSES) surveys: the National Survey of College Graduates (NSCG), the Survey of Earned Doctorates (SED), and the Survey of Doctoral Recipients (SDR). These NCSES surveys assess functional limitations by asking respondents to indicate the usual degree of difficulty they have with the following five activities:

- Seeing words or letters in ordinary newsprint (with glasses or contact lenses, if you usually wear them)
- Hearing what is normally said in conversation with another person (with a hearing aid, if you usually use one)
- Walking without human or mechanical assistance or using stairs
- Lifting or carrying something as heavy as 10 pounds, such as a bag of groceries
- Concentrating, remembering, or making decisions because of physical, mental, or emotional condition

To allow for consistent use of the data on disabilities across multiple data sources, this report considered an individual with a disability to have answered yes to any of the six CPS disability questions or to have responded with "moderate," "severe," or "unable to do" degree of difficulty to the activities listed on the NCSES surveys. Despite this attempt to provide comparable definitions, caution should be used when comparing the data on disabilities presented from CPS and NCSES sources.

# The STEM Workforce

## Overview

The science, technology, engineering, and mathematics (STEM) workforce is made up of individuals at all education levels who work in science and engineering (S&E), S&E-related, and middle-skill occupations (see sidebar The STEM Workforce of the United States). Using data from the Census Bureau's Current Population Survey (CPS), this section describes (1) the size of this workforce, (2) the representation of people according to sex, race, ethnicity, and disability status, and (3) how representation of these groups has changed between 2011 and 2021. Data show that although men and Whites still make up the largest share, the STEM workforce has been gradually diversifying over the past 10 years, with increased representation of women and underrepresented minorities—that is, Hispanics or Latinos, Blacks or African Americans, and American Indians or Alaska Natives.

## Representation in the STEM Workforce

### About a quarter of the U.S. workforce is employed in STEM occupations.

Of the 146.4 million people ages 18 to 74 in the workforce, 34.9 million (24%) were employed in STEM occupations in 2021 (figure 2-1). Although men and women represented similar proportions of the total workforce—52% men and 48% women (figure 1-1)—a greater share of men (29%) than women (18%) worked in STEM occupations. Among racial or ethnic groups, Asian workers had the highest share employed in STEM (39%), whereas the lowest share was among Black workers (18%). Within the other racial and ethnic groups, 20% to 25% worked in STEM. Workers with one or more disabilities represent a small proportion (3%) of the total workforce. Among workers with at least one disability, 21% worked in STEM occupations, which is slightly less than the 24% of nondisabled workers in STEM occupations.

In terms of age—and contrary to what is typically believed—STEM workers with at least one disability are not overwhelmingly concentrated in the older age groups. According to the 2021 CPS, among the STEM workforce ages 18 to 74 with one or more disabilities, 18% were in the early stages of their career (ages 18 to 34), 39% were midcareer (ages 35 to 54), and 43% were in the later stages of their career (age 55 and older).

**Figure 2-1**

**Occupations of the workforce ages 18–74, by sex, ethnicity, race, and disability status: 2021**

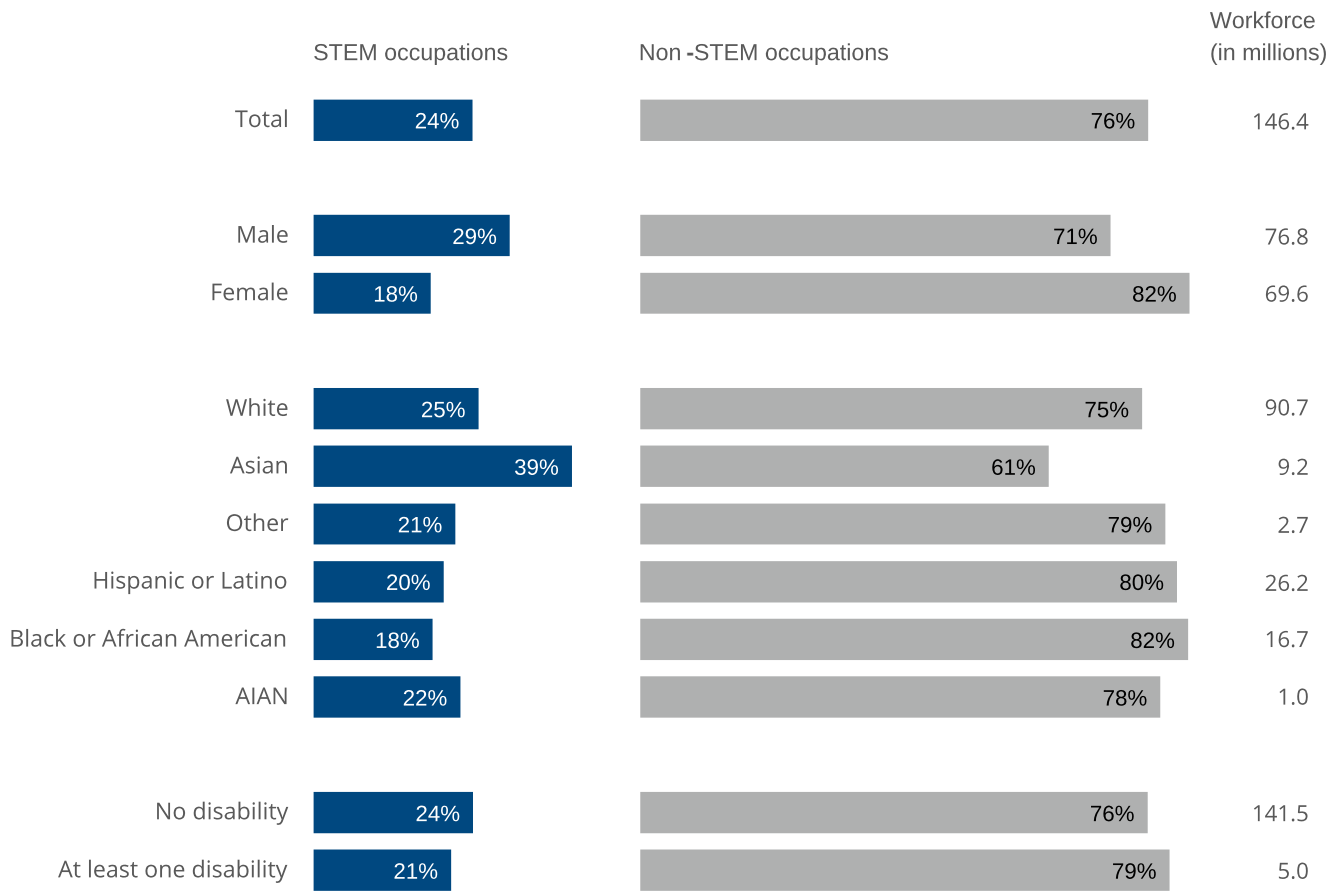

AIAN = American Indian or Alaska Native; STEM = science, technology, engineering, and mathematics.

**Note(s):**
Civilian noninstitutionalized population plus armed forces living off post or with their families on post. Hispanic or Latino may be any race; race categories exclude Hispanic origin. Other includes Native Hawaiian and Other Pacific Islander and more than one race. Respondents can report more than one disability. Those who reported difficulty with one or more functionalities were classified as having a disability.

**Source(s):**
Census Bureau, Current Population Survey, Annual Social and Economic Supplement, 2021.

## Growth in the STEM Workforce between 2011 and 2021

### The size of the STEM workforce grew between 2011 and 2021 for all groups.

Between 2011 and 2021, the STEM workforce grew by 5.9 million, from 29.0 million to 34.9 million, representing a 20% increase (figure 2-2). STEM workers as a percentage of the total workforce also increased, from 21% in 2011 to 24% in 2021.

Although fewer women than men work in STEM occupations, their share of the STEM workforce grew at a faster rate. Between 2011 and 2021, the number of women in the STEM workforce increased 31%, from 9.4 million to 12.3 million. For men, 22.6 million were employed in STEM occupations in 2021, up 15% from the 19.7 million employed in 2011.

In 2021, White workers—at 22.4 million—represented the largest race and ethnic group in the STEM workforce, followed by workers who were Hispanic (5.1 million), Asian (3.6 million), Black (3.0 million), and American Indian or Alaska Native (216,000). Hispanic STEM workers experienced the greatest numeric growth over this period—2.0 million—increasing from 3.1 million to 5.1 million. Although the increase in STEM workers was mostly for those without disabilities, the number of STEM workers with at least one disability also increased, reaching about 1.0 million in 2021.

Figure 2-2

**STEM workforce ages 18–74, by sex, ethnicity, race, and disability status: 2011 and 2021**

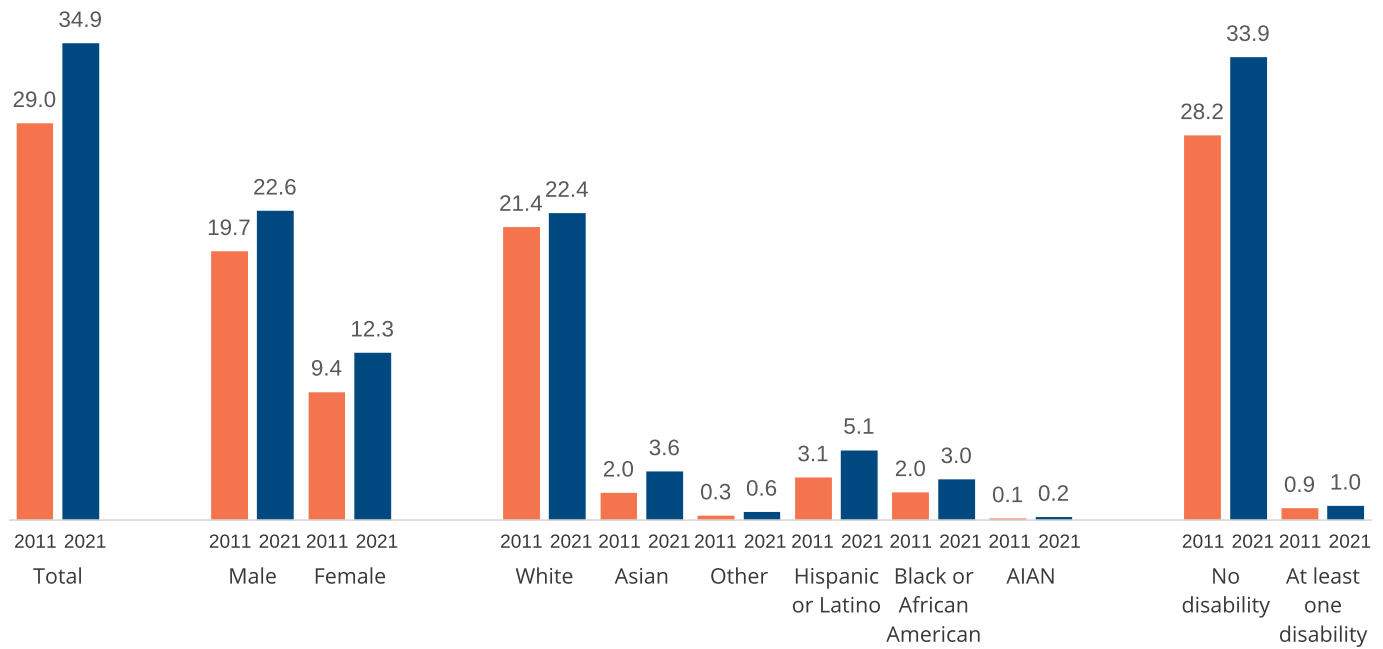

(Numbers in millions)

AIAN = American Indian or Alaska Native; STEM = science, technology, engineering, and mathematics.

**Note(s):**
Civilian noninstitutionalized population plus armed forces living off post or with their families on post. Hispanic or Latino may be any race; race categories exclude Hispanic origin. Other includes Native Hawaiian and Other Pacific Islander and more than one race. Respondents can report more than one disability. Those who reported difficulty with one or more functionalities were classified as having a disability.

**Source(s):**
Census Bureau, Current Population Survey, Annual Social and Economic Supplement.

## The share of women and underrepresented minorities in the STEM workforce increased between 2011 and 2021.

Compared with women, men make up the greater share of the STEM workforce. In 2021, about two-thirds (65%) of those employed in STEM occupations were men and about one-third (35%) were women (figure 2-3). Consistent with women's faster growth than men's in the STEM workforce, the proportion of the STEM workforce that were women increased by 3 percentage points from 2011 to 2021.

Collectively, underrepresented minorities—Hispanics, Blacks, and American Indians or Alaska Natives—represented nearly a quarter (24%) of the STEM workforce in 2021, up from 18% in 2011. Of these three groups, the share of Hispanics increased the most, from 11% in 2011 to 15% in 2021. As the proportion of other racial and ethnic groups increased, the proportion of White STEM workers decreased from 74% in 2011 to 64% in 2021. These data show increasing diversity within the STEM workforce over this 10-year period.

Despite the increase in the number of STEM workers with a disability, the proportion of these workers in the STEM workforce was unchanged from 2011 to 2021.

**Figure 2-3**

**Characteristics of the STEM workforce ages 18–74: 2011 and 2021**

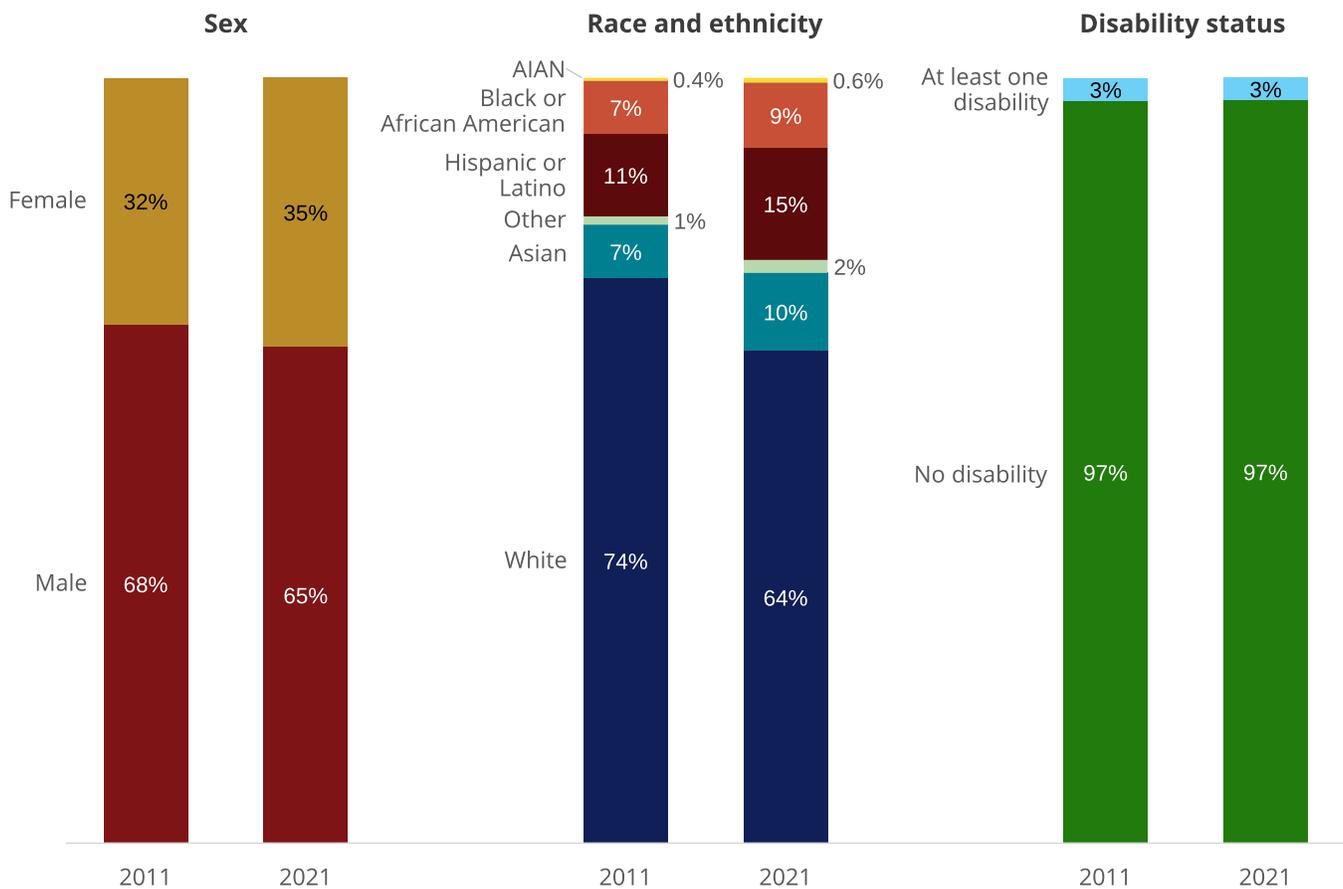

AIAN = American Indian or Alaska Native; STEM = science, technology, engineering, and mathematics.

**Note(s):**
Civilian noninstitutionalized population plus armed forces living off post or with their families on post. Hispanic or Latino may be any race; race categories exclude Hispanic origin. Other includes Native Hawaiian and Other Pacific Islander and more than one race. Respondents can report more than one disability. Those who reported difficulty with one or more functionalities were classified as having a disability. Due to rounding, percentages may not sum to 100 or subgroup totals.

**Source(s):**
Census Bureau, Current Population Survey, Annual Social and Economic Supplement.

417# STEM Occupations

## Overview

There are two ways the science, technology, engineering, and mathematics (STEM) workforce is divided for analysis in this section. One is based on the type of occupation, which falls into three broad categories: science and engineering (S&E), S&E-related, and middle-skill occupations (see sidebar The STEM Workforce of the United States). The other is based on education, which divides this workforce into two categories: (1) the STEM workforce, including workers with at least a bachelor's degree, and (2) the skilled technical workforce, including workers with a high level of technical knowledge but without a bachelor's degree. Most middle-skill positions do not require a bachelor's degree; thus, most workers in middle-skill occupations are part of the skilled technical workforce. In contrast, most S&E occupations require at least a bachelor's degree. S&E-related positions are a mix of those requiring one or more degrees and those that often require certification. Evaluation of the data from the Census Bureau's 2021 Current Population Survey reveals intriguing differences in the representation of women, people of different racial or ethnic groups, or people with disabilities among these categories of the STEM workforce. Asian STEM workers are found most commonly in S&E occupations. Women in the STEM workforce are found most commonly in S&E-related occupations. Men in the STEM workforce, as well as Hispanic or Latino and American Indian or Alaska Native STEM workers, are found most commonly in middle-skill occupations.

## Representation in S&E, S&E-Related, and Middle-Skill Occupations

### Within groups of STEM workers organized by sex, race, ethnicity, or disability status, the distribution by broad occupation type varies.

When the STEM workforce is divided into broad occupation types, 38% worked in middle-skill occupations, 37% in S&E-related occupations, and 25% in S&E occupations in 2021 (figure 3-1).

Dividing the STEM workforce by sex reveals striking differences in occupational distribution and concentration. Among men, half (52%) worked in middle-skill occupations, whereas over two-thirds (68%) of women worked in S&E-related occupations. The proportions were closer for S&E occupations than for the other two broad STEM occupation types: 28% of men and 20% of women were in S&E occupations.

Notable differences also exist when the distributions within the racial and ethnic groups are compared. Among Asian STEM workers, about half (52%) were employed in S&E occupations, with the smallest share employed in middle-skill occupations (13%). By comparison, among Hispanics and among American Indians or Alaska Natives in the STEM workforce, the greatest proportions worked in middle-skill occupations (63% and 52%, respectively) and the smallest proportions worked in S&E occupations (14% each). The highest share of Black or African American STEM workers was in S&E-related occupations (44%).

Among those in the STEM workforce, 46% of those with at least one disability worked in middle-skill occupations, compared with 38% of those with no disability. The share of STEM workers with at least one disability working in S&E-related occupations was lower than that of STEM workers without a disability (29% vs. 37%). A similar proportion of STEM workers with a disability and without a disability had positions in S&E occupations (24% and 25%, respectively).

**Figure 3-1**

**Occupations of the STEM workforce ages 18–74, by sex, ethnicity, race, and disability status: 2021**

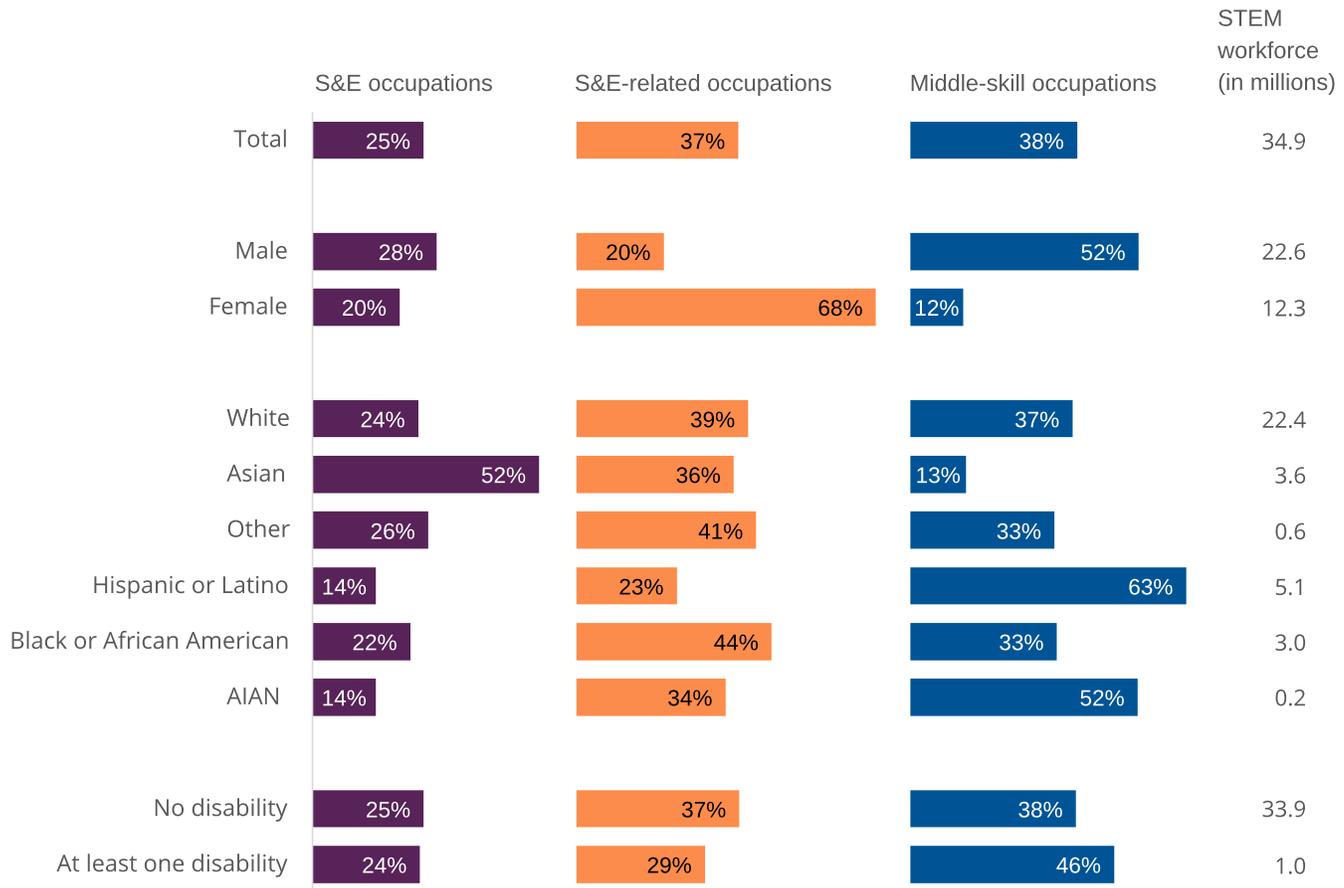

AIAN = American Indian or Alaska Native; S&E = science and engineering; STEM = science, technology, engineering, and mathematics.

**Note(s):**
Civilian noninstitutionalized population plus armed forces living off post or with their families on post. Hispanic or Latino may be any race; race categories exclude Hispanic origin. Other includes Native Hawaiian and Other Pacific Islander and more than one race. Respondents can report more than one disability. Those who reported difficulty with one or more functionalities were classified as having a disability. Due to rounding, percentages may not sum to 100.

**Source(s):**
Census Bureau, Current Population Survey, Annual Social and Economic Supplement, 2021.

## Characteristics of the STEM Workforce

### About two-thirds employed in S&E-related occupations are women; about a quarter employed in middle-skill occupations are Hispanic; about a fifth employed in S&E occupations are Asian.

In 2021, employment within each broad occupation type was dominated by either men or women (figure 3-2). Middle-skill occupations had the greatest concentration, where nearly 9 out of 10 (89%) workers were men. At 72%, men also had most of the S&E jobs. Conversely, among S&E-related occupations, nearly two-thirds (65%) were women.

**Figure 3-2**

**Characteristics of the STEM workforce ages 18–74, by occupation: 2021**

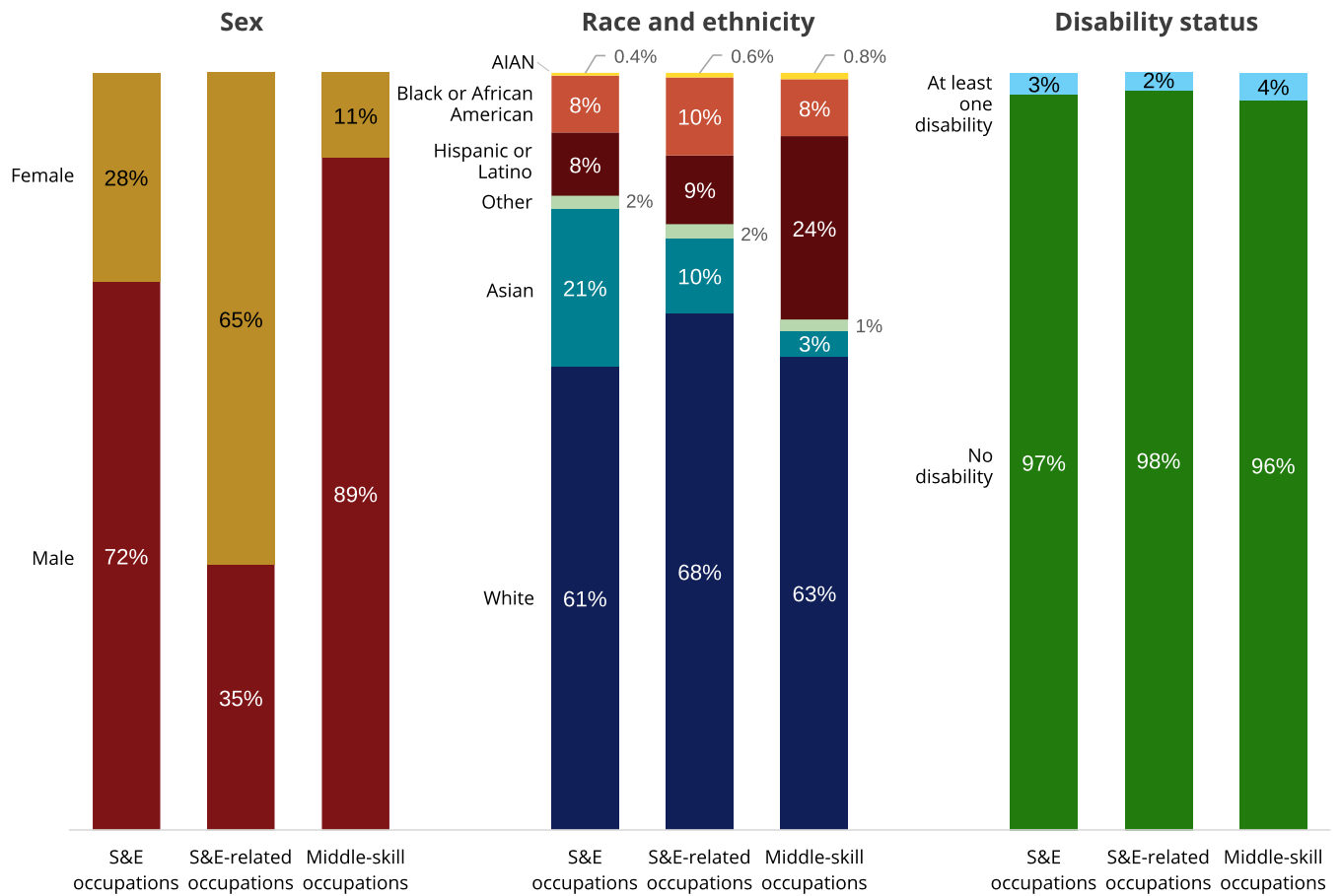

AIAN = American Indian or Alaska Native; S&E = science and engineering; STEM = science, technology, engineering, and mathematics.

**Note(s):**
Civilian noninstitutionalized population plus armed forces living off post or with their families on post. Hispanic or Latino may be any race; race categories exclude Hispanic origin. Other includes Native Hawaiian and Other Pacific Islander and more than one race. Respondents can report more than one disability. Those who reported difficulty with one or more functionalities were classified as having a disability.

**Source(s):**
Census Bureau, Current Population Survey, Annual Social and Economic Supplement, 2021.

When combined, underrepresented minorities—Hispanics, Blacks, and American Indians or Alaska Natives—made up about a quarter (24%) of all STEM workers (figure 1-1). When divided by broad occupation type, compared with their share of STEM workers, underrepresented minorities made up a lower share of S&E occupations (16%) and S&E-related occupations (20%) but a higher share of middle-skill occupations (33%).

Although Hispanics made up 15% of the STEM workforce (figure 2-3), they represented nearly a quarter (24%) of those working in middle-skill occupations (figure 3-2). Similarly, Asian STEM workers made up 10% of the STEM workforce but 21% of those in S&E occupations.

The share of workers with at least one disability was similarly low (2%–4%) across the three STEM occupation types, which is consistent with their low proportion in the overall STEM workforce (3%) (figure 2-3).

## Educational Attainment of the STEM Workforce

### Within groups of STEM workers organized by sex, race, ethnicity, or disability status, representation in the skilled technical workforce varies.

Dividing the STEM workforce by educational attainment yields two large groups (see sidebar The STEM Workforce of the United States). One is STEM workers who do not have a bachelor's degree, referred to as the skilled technical workforce, and the other is STEM workers with a bachelor's degree or higher. Overall, the STEM workforce was about equally split between the skilled technical workforce (49%) and workers with at least a bachelor's degree (51%) in 2021 (figure 3-3).

**Figure 3-3**

**Education of the STEM workforce ages 18–74, by sex, ethnicity, race, and disability: 2021**

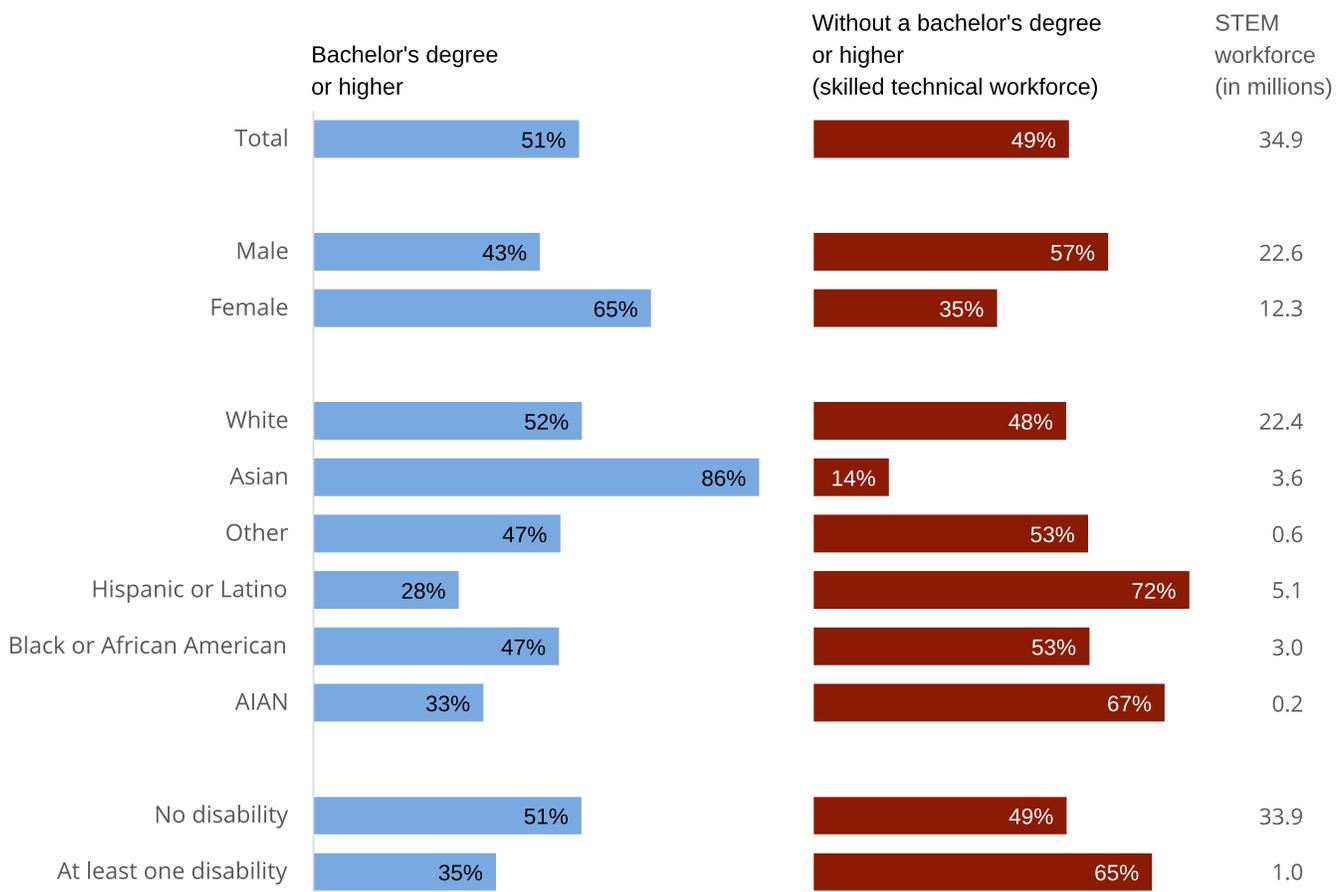

AIAN = American Indian or Alaska Native; STEM = science, technology, engineering, and mathematics.

**Note(s):**
Civilian noninstitutionalized population plus armed forces living off post or with their families on post. Hispanic or Latino may be any race; race categories exclude Hispanic origin. Other includes Native Hawaiian and Other Pacific Islander and more than one race. Respondents can report more than one disability. Those who reported difficulty with one or more functionalities were classified as having a disability.

**Source(s):**
Census Bureau, Current Population Survey, Annual Social and Economic Supplement, 2021.

Among male STEM workers, over half (57%) were employed in the skilled technical workforce, which includes workers in S&E, S&E-related, and middle-skill occupations. Conversely, nearly two-thirds (65%) of women with STEM jobs had a bachelor's degree or higher education.

About 7 out of 10 STEM workers in two racial or ethnic groups—Hispanic (72%) and American Indian or Alaska Native (67%)—worked in the skilled technical workforce. In contrast, 86% of Asian STEM workers had a bachelor's degree or higher, with just 14% of that group employed in the skilled technical workforce.

Among STEM workers with at least one disability, about two-thirds (65%) were employed in the skilled technical workforce, whereas STEM workers without a disability were evenly divided between the skilled technical workforce (49%) and those with at least a bachelor's degree (51%).

## Men make up three-fourths of the skilled technical workforce.

Among STEM workers in the skilled technical workforce, the share of men was three times that of women (75% vs. 25%) (figure 3-4). This pattern is similar to that of the male-female distribution of middle-skill occupations (figure 3-2). The higher share of women in the skilled technical workforce relates to the inclusion of workers across all three broad categories of STEM occupations—S&E, S&E-related, and middle-skill occupations—and the high share of women (65%) in S&E-related occupations (figure 3-2). Compared with the skilled technical workforce, the distribution by sex in STEM occupations with a bachelor's degree or higher was more evenly split: 55% male and 45% female.

The share of underrepresented minorities—Hispanics, Blacks, and American Indians or Alaska Natives—in the skilled technical workforce (32%) was twice the share of underrepresented minorities employed in STEM occupations with a bachelor's degree or higher (16%).

Over one in five (22%) employed in the skilled technical workforce were Hispanic, notably higher than the share of Hispanic workers employed in STEM jobs with at least a bachelor's degree (8%). Conversely, Asian workers made up 17% of those employed in STEM with a bachelor's degree or higher, about five times their share in the skilled technical workforce (3%).

The share of workers with at least one disability was similarly low across the two education levels (2% with a bachelor's degree or higher and 4% in the skilled technical workforce).

**Figure 3-4**

**Characteristics of the STEM workforce ages 18–74, by education: 2021**

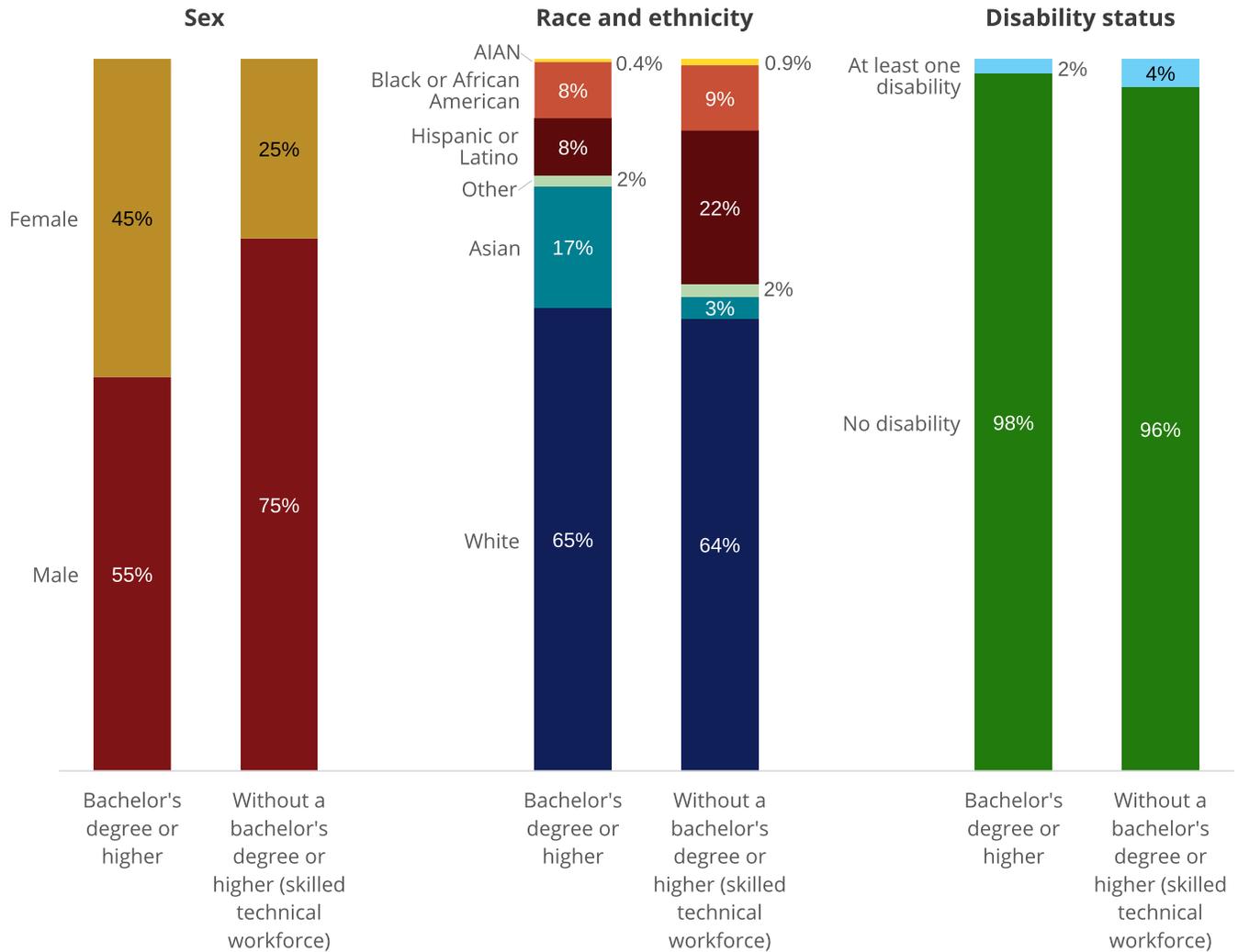

AIAN = American Indian or Alaska Native; STEM = science, technology, engineering, and mathematics.

**Note(s):**
Civilian noninstitutionalized population plus armed forces living off post or with their families on post. Hispanic or Latino may be any race; race categories exclude Hispanic origin. Other includes Native Hawaiian and Other Pacific Islander and more than one race. Respondents can report more than one disability. Those who reported difficulty with one or more functionalities were classified as having a disability.

**Source(s):**
Census Bureau, Current Population Survey, Annual Social and Economic Supplement, 2021.

# STEM Median Wage and Salary Earnings

## Overview

Employment in science, technology, engineering, and mathematics (STEM) has a positive impact on the pocketbook. Median wage and salary earnings[2] are higher for those working in STEM than in non-STEM occupations, regardless of sex, race, ethnicity, or disability status. Additionally, within the STEM workforce, higher education translates into higher pay. For all demographic groups of STEM workers, those with a bachelor's degree or higher have higher median earnings than those without college degrees. However, pay disparities exist in the STEM workforce. Female STEM workers earn less than male STEM workers. Black or African American, Hispanic or Latino, and American Indian or Alaska Native STEM workers earn less than White and Asian STEM workers. STEM workers with disabilities earn less than those without disabilities.

Various factors contribute to earnings differences.[3] Sex, race, ethnicity, and disability status are examined here by using data from the Census Bureau's Current Population Survey[4] to explore how demographic characteristics interact with occupation and education to influence earnings. For all groups, median earnings will be compared between STEM and non-STEM occupations and across STEM occupation type (S&E, S&E-related, and middle-skill occupations) and according to educational attainment (with or without at least a bachelor's degree).

## Earnings of STEM and Non-STEM Workers

### STEM workers earn more than non-STEM workers, regardless of sex, race, ethnicity, or disability status.

STEM workers had median wage and salary earnings of about $64,000, higher than the $40,000 earned by those working in non-STEM occupations (figure 4-1).

When the workforce is divided by sex, men who worked in STEM occupations made more than men who had non-STEM jobs ($65,000 vs. $48,000). The same was true for women, who earned $60,000 in STEM occupations and $36,000 in non-STEM occupations. However, when men and women are compared, men had higher median earnings than women in both STEM and non-STEM occupations.

For all racial and ethnic groups, STEM workers had higher median wage and salary earnings than their counterparts who worked in non-STEM jobs. When considering just STEM occupations, Asian workers had the highest median earnings ($92,000), followed by White workers ($66,000), whereas Hispanic workers ($45,000) and American Indian or Alaska Native workers ($50,000) had the lowest.

The same pattern exists among those in the workforce with at least one disability. Those who work in STEM had higher median earnings than those who worked in non-STEM occupations ($57,000 vs. $30,000). However, in both STEM and non-STEM occupation, those without a disability had higher median earnings than those with a disability.

**Figure 4-1**

**Median wage and salary earnings of the workforce ages 18–74 in STEM and non-STEM occupations, by sex, ethnicity, race, and disability status: 2020**

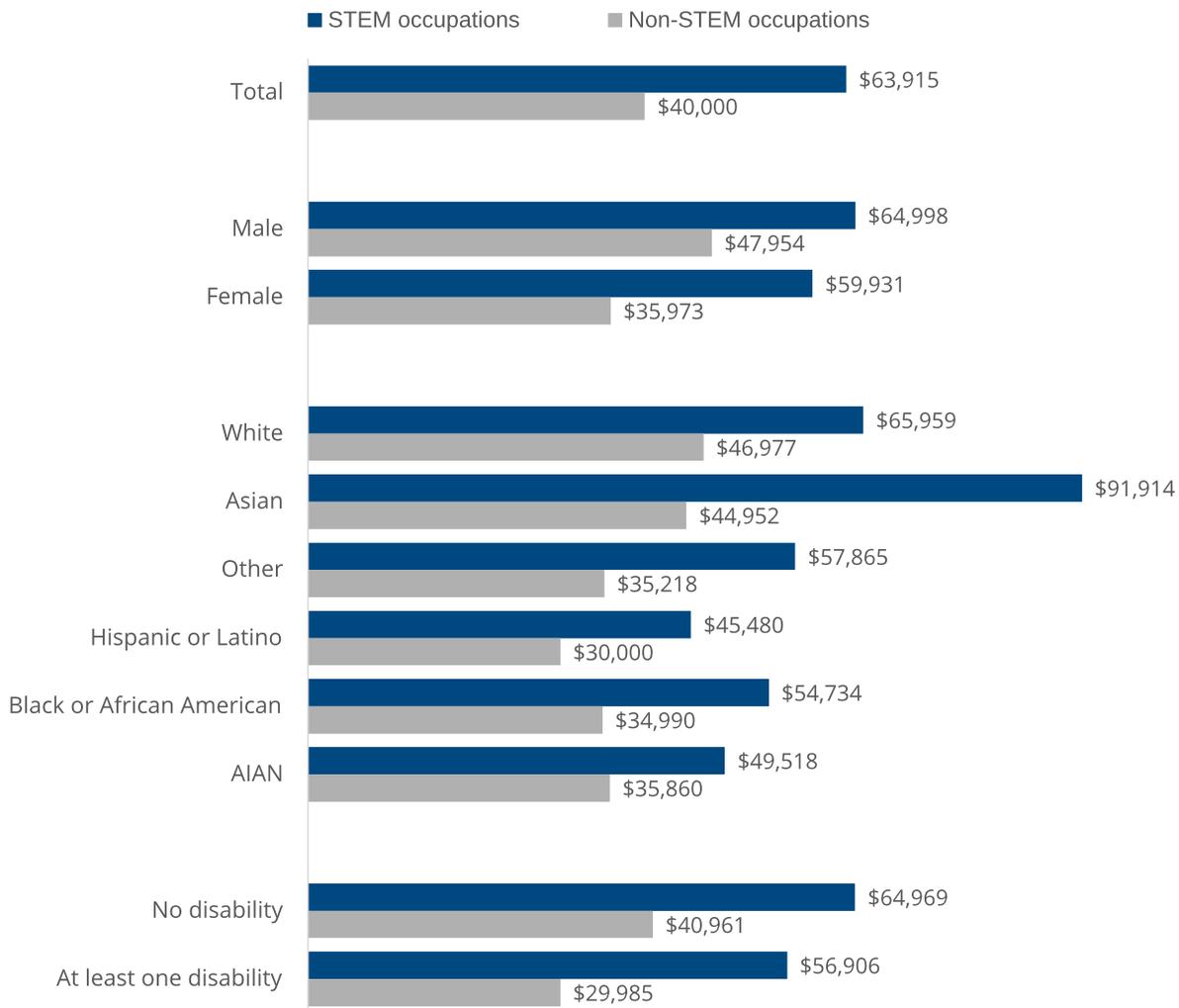

AIAN = American Indian or Alaska Native; STEM = science, technology, engineering, and mathematics.

**Note(s):**
Median wage and salary earnings for the previous year. Civilian noninstitutionalized population plus armed forces living off post or with their families on post. Hispanic or Latino may be any race; race categories exclude Hispanic origin. Other includes Native Hawaiian and Other Pacific Islander and more than one race. Respondents can report more than one disability. Those who reported difficulty with one or more functionalities were classified as having a disability.

**Source(s):**
Census Bureau, Current Population Survey, Annual Social and Economic Supplement, 2021.

## Earnings of Workers in S&E, S&E-Related, and Middle-Skill Occupations

### STEM workers in S&E occupations earn more than those in S&E-related or middle-skill occupations.

When STEM workers are divided by broad occupation type, median wage and salary earnings were highest for those in S&E occupations ($90,000), followed by S&E-related ($67,000), then middle-skill occupations ($50,000). This pattern—where earnings are highest for S&E occupations and lowest for middle-skill occupations—occurs regardless of sex, race, ethnicity, or disability status (figure 4-2).

Men had higher median earnings than women in all three broad occupation types. Among those with S&E jobs, men's median earnings were $100,000 in 2020, compared with $76,000 for women. In S&E-related occupations, the median earnings for men were $80,000, compared with $60,000 for women. Middle-skill earnings showed the smallest difference between men and women ($50,000 vs. $40,000).

In both S&E and S&E-related occupations, Asian workers had the highest median wage and salary earnings, followed by White workers and then by Hispanic and Black workers, whose earnings were comparable. The median earnings for Asian workers in S&E occupations were $107,000, which was higher than those for White ($90,000), Hispanic ($75,000), and Black ($73,000) workers in the same broad occupation type. Within middle-skill occupations, median earnings for White workers were higher than those for Black and Hispanic workers. There was no statistical difference in the median earnings of Asian, Hispanic, and Black middle-skill workers.

Among STEM workers with at least one disability, those employed in S&E occupations had the highest median wage and salary earnings ($89,000), followed by those in S&E-related ($53,000) and middle-skill ($45,000) occupations. Among workers in S&E-related occupations, those without a disability had higher median earnings than those with at least one disability ($68,000 vs. $53,000). Both disabled and nondisabled workers had comparable earnings if they work in S&E (about $90,000) or middle-skill (about $50,000) occupations.

**Figure 4-2**

**Median wage and salary earnings of the STEM workforce ages 18–74 by occupation and by sex, ethnicity, race, and disability status: 2020**

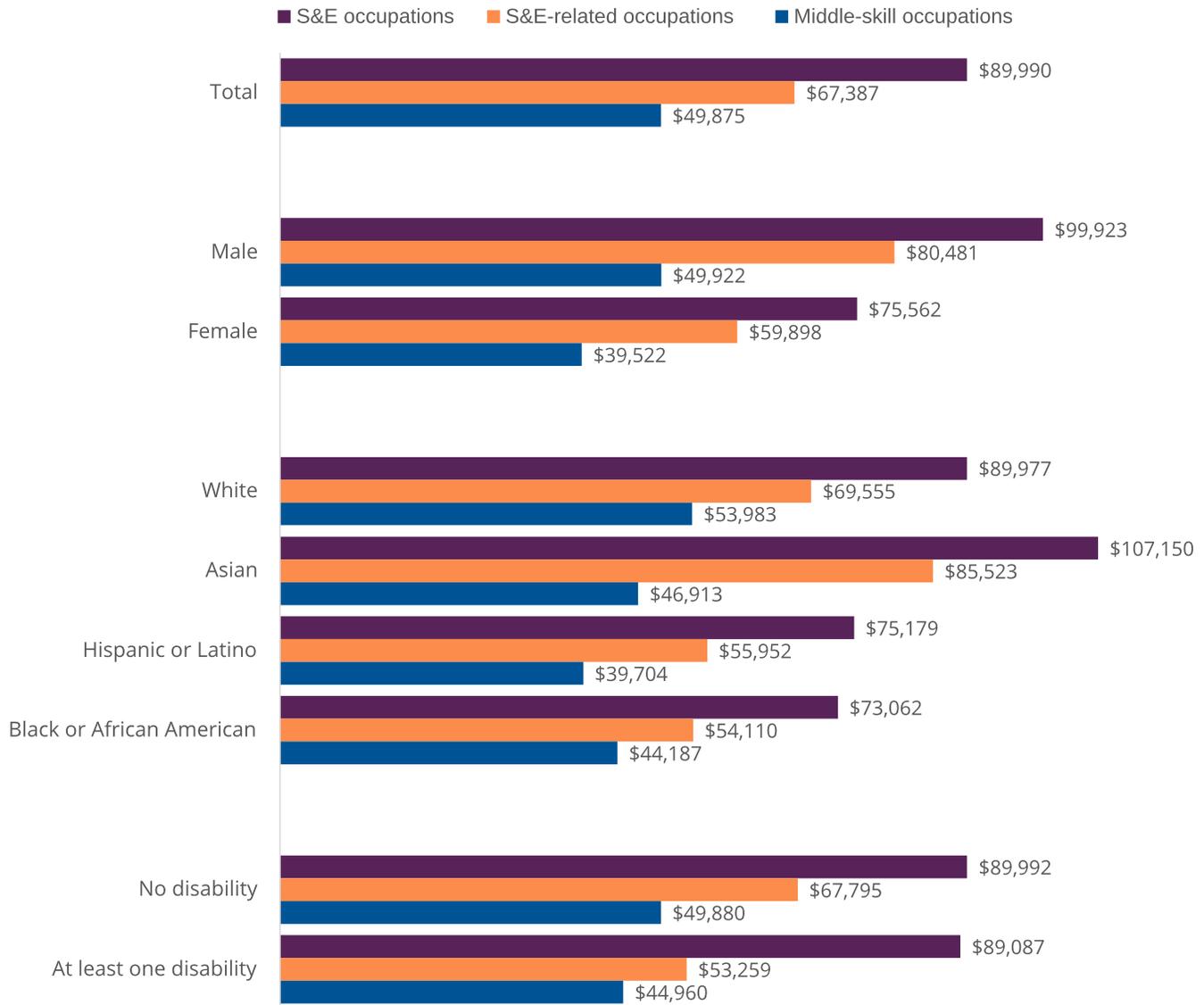

S&E = science and engineering; STEM = science, technology, engineering, and mathematics.

**Note(s):**
Median wage and salary earnings for the previous year. Civilian noninstitutionalized population plus armed forces living off post or with their families on post. Hispanic or Latino may be any race; race categories exclude Hispanic origin. Respondents can report more than one disability. Those who reported difficulty with one or more functionalities were classified as having a disability. Estimates suppressed for reliability for American Indian or Alaska Native and other (Native Hawaiian and Other Pacific Islander and more than one race); does not meet unweighted cell size requirements.

**Source(s):**
Census Bureau, Current Population Survey, Annual Social and Economic Supplement, 2021.

## Educational Attainment and Earnings in STEM Workforce

### In the STEM workforce, higher education means higher earnings.

For all demographic groups, STEM workers with a bachelor's degree or higher had higher median wage and salary earnings than those in the skilled technical workforce (figure 4-3).

**Figure 4-3**

**Median wage and salary earnings of the STEM workforce ages 18–74 by education and by sex, ethnicity, race, and disability status: 2020**

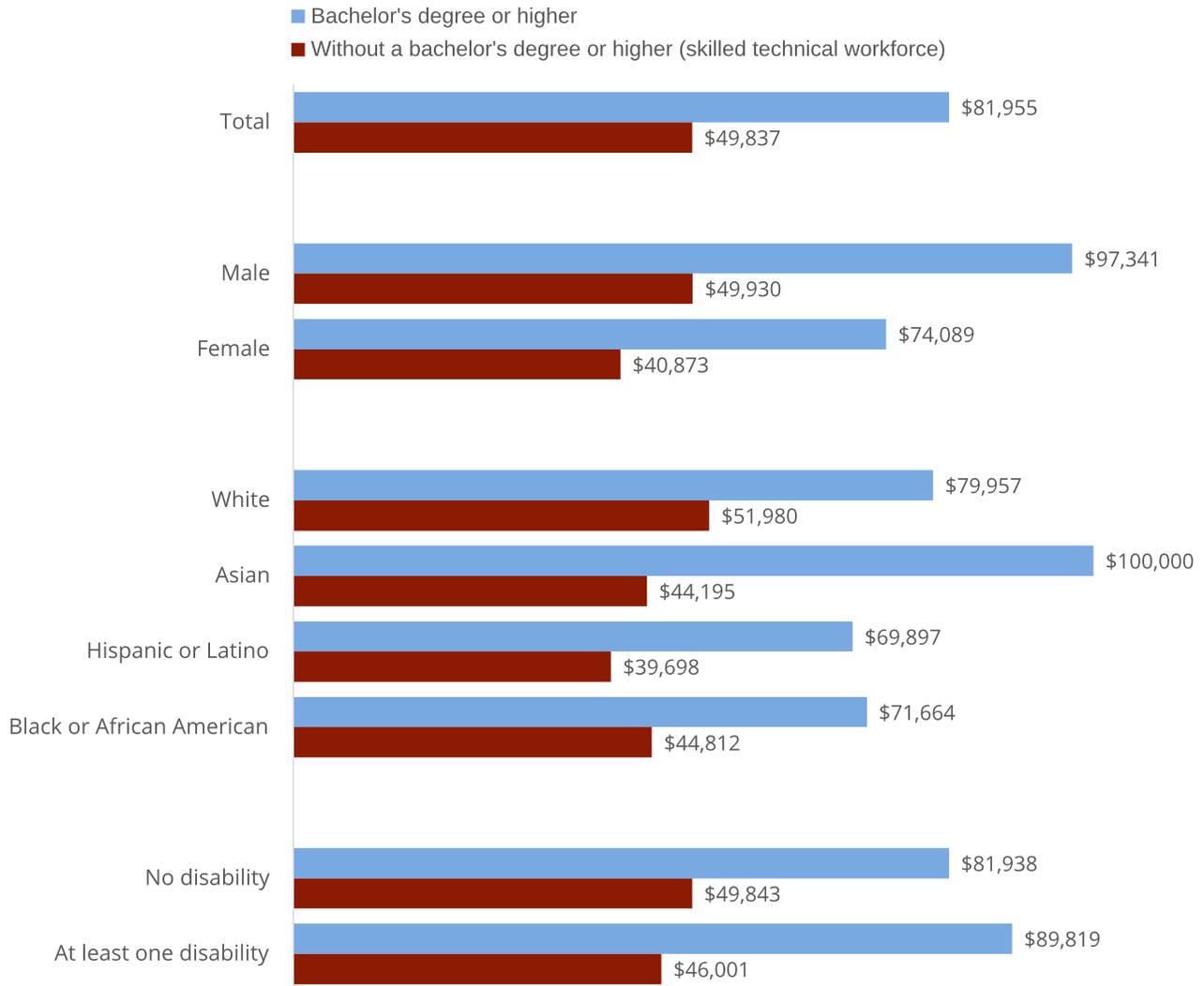

| Group | Bachelor's degree or higher | Without a bachelor's degree or higher (skilled technical workforce) |
|---|---|---|
| Total | $81,955 | $49,837 |
| Male | $97,341 | $49,930 |
| Female | $74,089 | $40,873 |
| White | $79,957 | $51,980 |
| Asian | $100,000 | $44,195 |
| Hispanic or Latino | $69,897 | $39,698 |
| Black or African American | $71,664 | $44,812 |
| No disability | $81,938 | $49,843 |
| At least one disability | $89,819 | $46,001 |

STEM = science, technology, engineering, and mathematics.

**Note(s):**
Median wage and salary earnings for the previous year. Civilian noninstitutionalized population plus armed forces living off post or with their families on post. Hispanic or Latino may be any race; race categories exclude Hispanic origin. Respondents can report more than one disability. Those who reported difficulty with one or more functionalities were classified as having a disability. Estimates suppressed for reliability for American Indian or Alaska Native and other (Native Hawaiian and Other Pacific Islander and more than one race); does not meet unweighted cell size requirements.

**Source(s):**
Census Bureau, Current Population Survey, Annual Social and Economic Supplement, 2021.

As seen in the gender pay disparity across broad STEM occupation types, male STEM workers typically make more than female STEM workers regardless of whether they have an advanced degree. The median wage and salary earnings for men with at least a bachelor's degree was $97,000 in 2020, compared with $74,000 for women in the same education category. In the skilled technical workforce, men's median earnings ($50,000) were higher than women's median earnings ($41,000).

In the STEM workforce with a bachelor's degree or higher, Asian workers ($100,000) had the highest median earnings, followed by White workers ($80,000) and then by Black ($72,000) and Hispanic ($70,000) workers (whose median earnings were comparable). In the skilled technical workforce, White workers had higher median earnings ($52,000) than Black ($45,000), Asian ($44,000), or Hispanic ($40,000) workers.

Earnings were not significantly different by disability status for STEM workers with or without a bachelor's degree.

# STEM Unemployment

## Overview

The unemployment rate measures the share of workers in the labor force who do not currently have a job but are available and actively looking for work. In general, underrepresented minorities—Hispanics or Latinos, Blacks or African Americans, and American Indians or Alaska Natives—experience higher rates of unemployment than their White and Asian counterparts.[5] Data from the Census Bureau's Current Population Survey show that during the COVID-19 pandemic, the unemployment rates of the labor force in science, technology, engineering, and mathematics (STEM) and non-STEM occupations increased between 2019 and 2021 for most demographic groups. The 2021 unemployment rates of these groups are also examined in STEM occupations according to broad occupational type—science and engineering (S&E), S&E-related, and middle-skill occupations—and according to educational attainment—with or without at least a bachelor's degree. Although the COVID-19 pandemic affected unemployment rates of individuals in STEM occupations less than it affected the rates of those in non-STEM occupations, persistent disparities among racial and ethnic groups exist. Black STEM workers experienced the highest rate of unemployment in 2021, followed by Hispanic STEM workers. White and Asian STEM workers experienced the lowest rates of unemployment.

## Unemployment Rates for STEM and Non-STEM Occupations in 2019 and 2021

### Unemployment rates for both STEM and non-STEM occupations increased between 2019 and 2021.

The COVID-19 pandemic had a profoundly negative impact on employment across the U.S. workforce. However, the impact was less for workers in STEM occupations than it was for those in non-STEM occupations (figure 5-1). Unemployment rates for STEM workers increased 1.5 percentage points between 2019 and 2021 (from 2.1% to 3.6%), whereas unemployment for non-STEM workers increased 2.8 percentage points (from 3.9% to 6.7%) during this period.

Unemployment rates reflect both employer and employee sides of employment. Unemployment rates are increased by job loss (employer side effect) and are decreased by workers leaving the workforce, either voluntarily or due to layoffs or job loss and not seeking reemployment. The pandemic increased the demand for some occupations and reduced the demand for others.

Unemployment rates for both men and women in STEM and non-STEM occupations increased over this 3-year period. Within each occupation group, the amount of increase for men and women was comparable. However, for both years (2019 and 2021) and both occupation groups (STEM and non-STEM), the unemployment rates for men were higher than those for women.

Within the racial and ethnic groups, increases in STEM unemployment rates between 2019 and 2021 were statistically significant for Whites, Hispanics, and Blacks but not for Asians. However, when comparing among groups, the increase was greater for Blacks (3.4 percentage points from 3.2% to 6.6%) and Hispanics (2.9 percentage points from 2.8% to 5.7%) than it was for Whites (1.0 percentage point from 1.9% to 2.9%). For non-STEM occupations, the unemployment rates in 2021 were higher than the rates in 2019 for White, Asian, Hispanic, and Black workers. The increases over the 3-year period for Asians (4.6 percentage points from 2.9% to 7.6%), Hispanics (3.9 percentage points from 4.3% to 8.2%), and Blacks (3.8 percentage points from 6.7% to 10.5%) were higher than those for Whites (2.0 percentage points 3.2% to 5.2%). Both in 2019 and 2021 and for all racial and ethnic groups, those working in STEM occupations had lower unemployment rates than those in non-STEM occupations.

Overall, the unemployment rates of workers with or without one or more disabilities were lower if they worked in STEM rather than in non-STEM occupations. However, both groups experienced comparable increases in unemployment in STEM and non-STEM occupations from 2019 to 2021. The unemployment rate for workers in non-STEM occupations with at least one disability was higher in 2021 (11.2%) than in 2019 (8.5%). In contrast, unemployment rates were comparable in 2021 (5.3%) and in 2019 (3.8%) for workers with at least one disability in STEM occupations. For those without a disability, the unemployment rates for both occupation groups increased over the 3-year period.

**Figure 5-1**

**Unemployment rate of the workforce ages 18–74 in STEM and non-STEM occupations, by sex, ethnicity, race and disability status: 2019 and 2021**

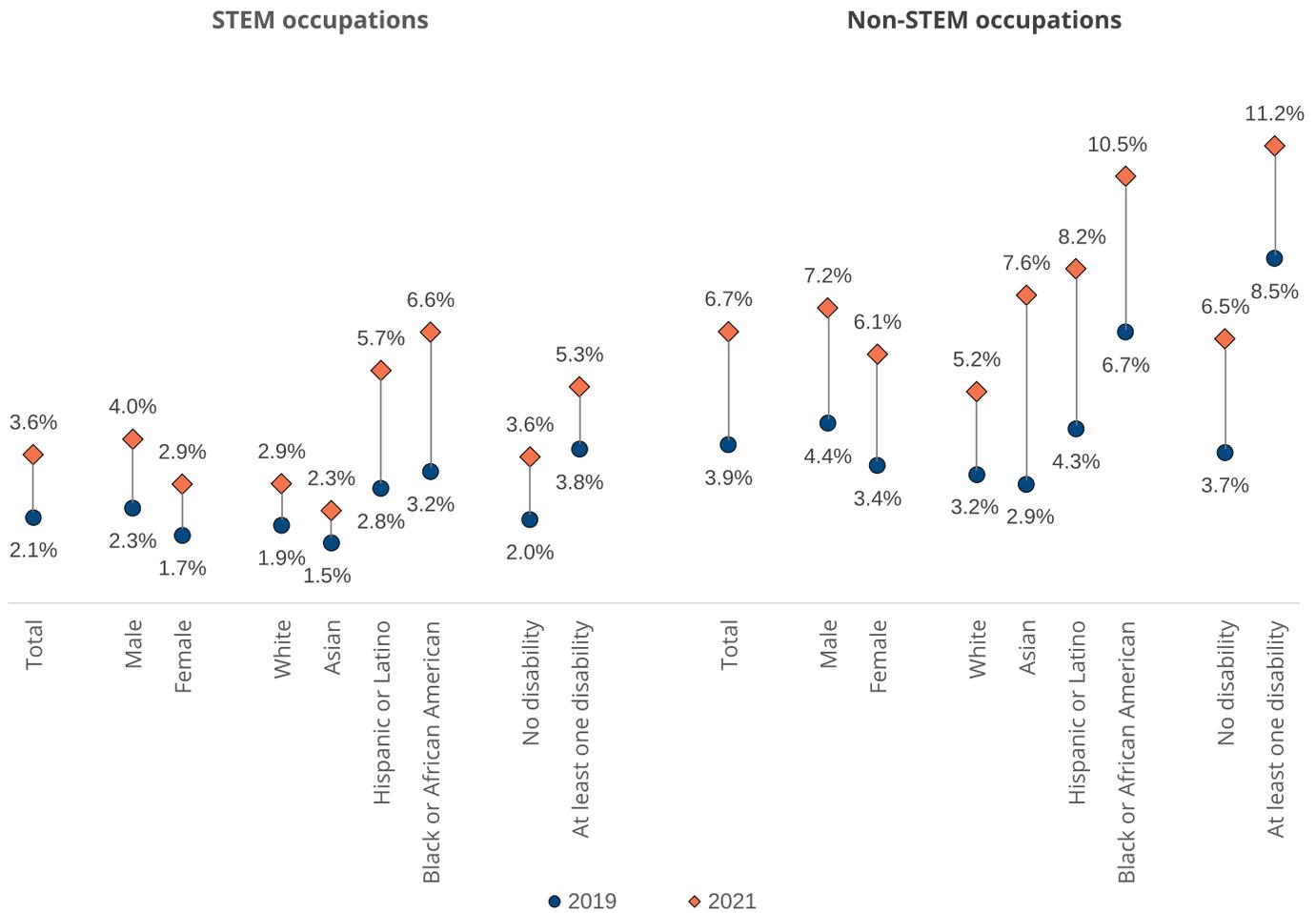

STEM = science, technology, engineering, and mathematics.

**Note(s):**
Civilian noninstitutionalized population plus armed forces living off post or with their families on post. Hispanic or Latino may be any race; race categories exclude Hispanic origin. Respondents can report more than one disability. Those who reported difficulty with one or more functionalities were classified as having a disability. Estimates suppressed for reliability for American Indian or Alaska Native and other (Native Hawaiian and Other Pacific Islander and more than one race); does not meet unweighted cell size requirements.

**Source(s):**
Census Bureau, Current Population Survey, Annual Social and Economic Supplement.

# Unemployment Rates in S&E, S&E-Related, and Middle-Skill Occupations

## STEM workers in middle-skill occupations have a higher unemployment rate than those in S&E or S&E-related occupations.

When the STEM workforce is divided by broad occupation type, those in middle-skill occupations had the highest unemployment rate in 2021 (5.9%), which was about twice that of S&E (2.5%) and over twice that of S&E-related (2.1%) occupations (figure 5-2). Higher unemployment rates for middle-skill occupations occurred across all groups of workers except for those with at least one disability.

### Figure 5-2

**Unemployment rate of the STEM workforce ages 18–74, by occupation, sex, ethnicity, race, and disability status: 2021**

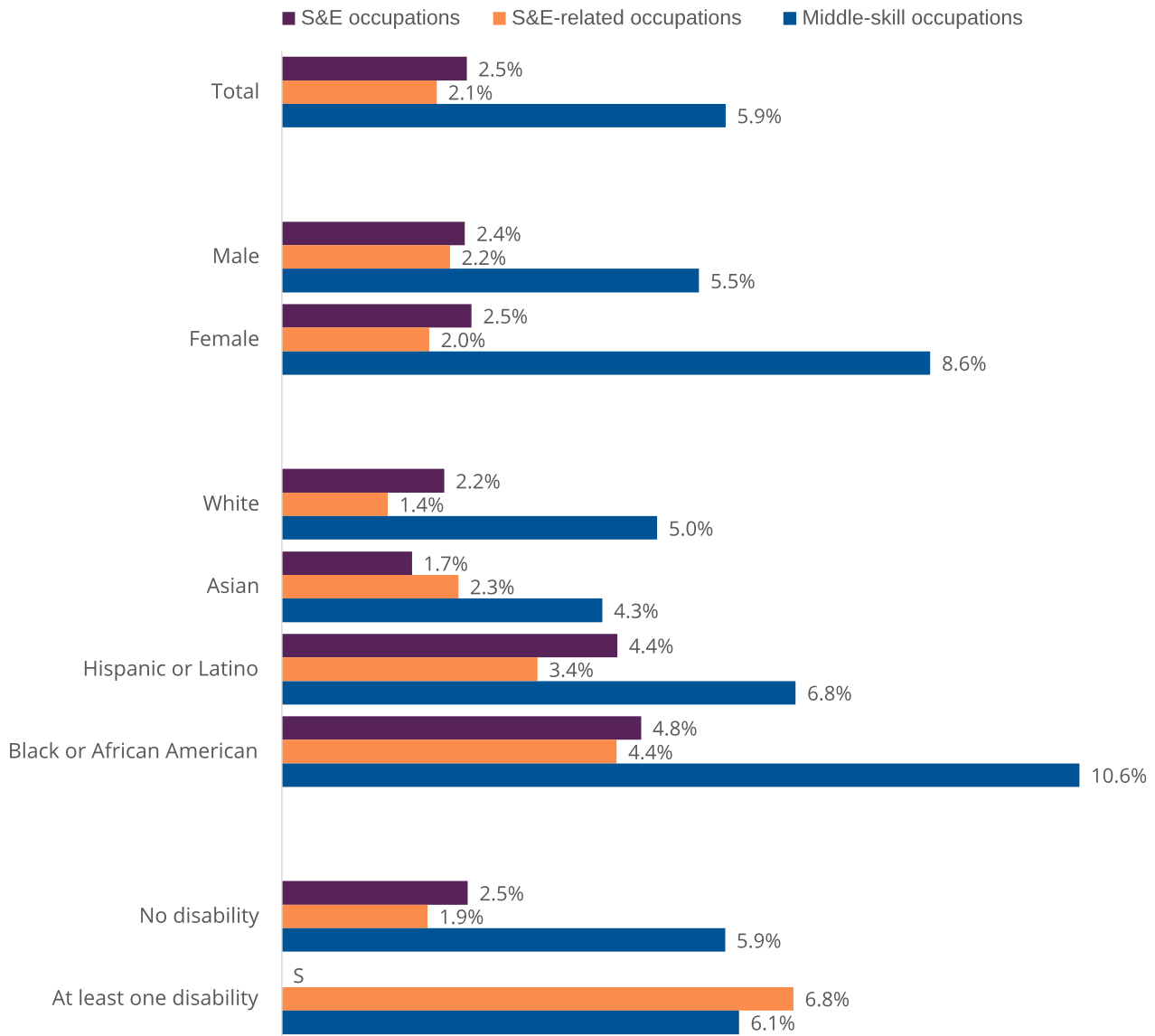

S = suppressed for reliability; coefficient of variation exceeds publication standards.

S&E = science and engineering; STEM = science, technology, engineering, and mathematics.

**Note(s):**
Civilian noninstitutionalized population plus armed forces living off post or with their families on post. Hispanic or Latino may be any race; race categories exclude Hispanic origin. Respondents can report more than one disability. Those who reported difficulty with one or more functionalities were classified as having a disability. Estimates suppressed for reliability for American Indian or Alaska Native and other (Native Hawaiian and Other Pacific Islander and more than one race); does not meet unweighted cell size requirements.

**Source(s):**
Census Bureau, Current Population Survey, Annual Social and Economic Supplement, 2021.

Although women in middle-skill occupations had a higher unemployment rate than men in these occupations, the unemployment rates for men and women in S&E or S&E-related occupations were similar.

In the middle-skill occupations, Blacks experienced the highest unemployment rate (10.6%) out of all the racial and ethnic groups. Unemployment rates for S&E workers who are Black (4.8%) or Hispanic (4.4%) were higher than those who are Asian (1.7%) or White (2.2%). In S&E-related occupations, Whites had lower unemployment rates (1.4%) than did Blacks (4.4%) and Hispanics (3.4%), but they had rates that were comparable to Asians (2.3%).

Although STEM workers with no disabilities had similar unemployment rates as those with disabilities in middle-skill occupations (5.9% and 6.1%, respectively), their rate was notably lower than that for those with a disability in S&E-related occupations (1.9% and 6.8%, respectively).

## Unemployment Rates by Educational Attainment in the STEM Labor Force

### The skilled technical workforce has a higher unemployment rate than STEM workers with a bachelor's degree or higher.

Overall, among all STEM workers in 2021, higher educational attainment is associated with lower unemployment (figure 5-3). STEM workers with a bachelor's degree or higher had a lower unemployment rate than those employed in the skilled technical workforce (2.4% vs. 4.9%).

For both men and women, unemployment was higher if they worked in the skilled technical workforce than if they were STEM workers with at least a bachelor's degree. However, there was no difference in the unemployment rates of men and women in the STEM workforce by education: 2.6% of men and 2.2% of women with at least a bachelor's degree were unemployed, as were 5.1% men and 4.2% of women without a bachelor's degree.

The pattern of higher unemployment rates for workers in the skilled technical workforce than for STEM workers with at least a bachelor's degree persists across all racial and ethnic groups. However, in both educational groups, Blacks had higher unemployment rates than Whites or Asians, but their rates were not statistically different from rates of Hispanic unemployment. In the skilled technical workforce, Hispanic unemployment was higher than White unemployment, but it was not statistically different from the unemployment rates of Asians and Blacks.

For STEM workers with at least one disability, educational attainment had little effect on unemployment. For those with a disability, the unemployment rate among the skilled technical workforce (5.2%) was about the same as the rate among the workforce with a bachelor's degree or higher (5.4%). Likewise, when comparing the unemployment rates of each educational group by disability status, the differences between the estimates are not statistically significant. That is, unemployment rates are comparable between those with at least a bachelor's degree who have one or more disabilities (5.4%) and those who have no disability (2.4%). Additionally, there is no statistical difference between the unemployment rates for those without a bachelor's degree by disability status (5.2% with at least one disability vs. 4.8% with no disability).

**Figure 5-3**

**Unemployment rate of the STEM workforce ages 18–74, by education, sex, ethnicity, race, and disability status: 2021**

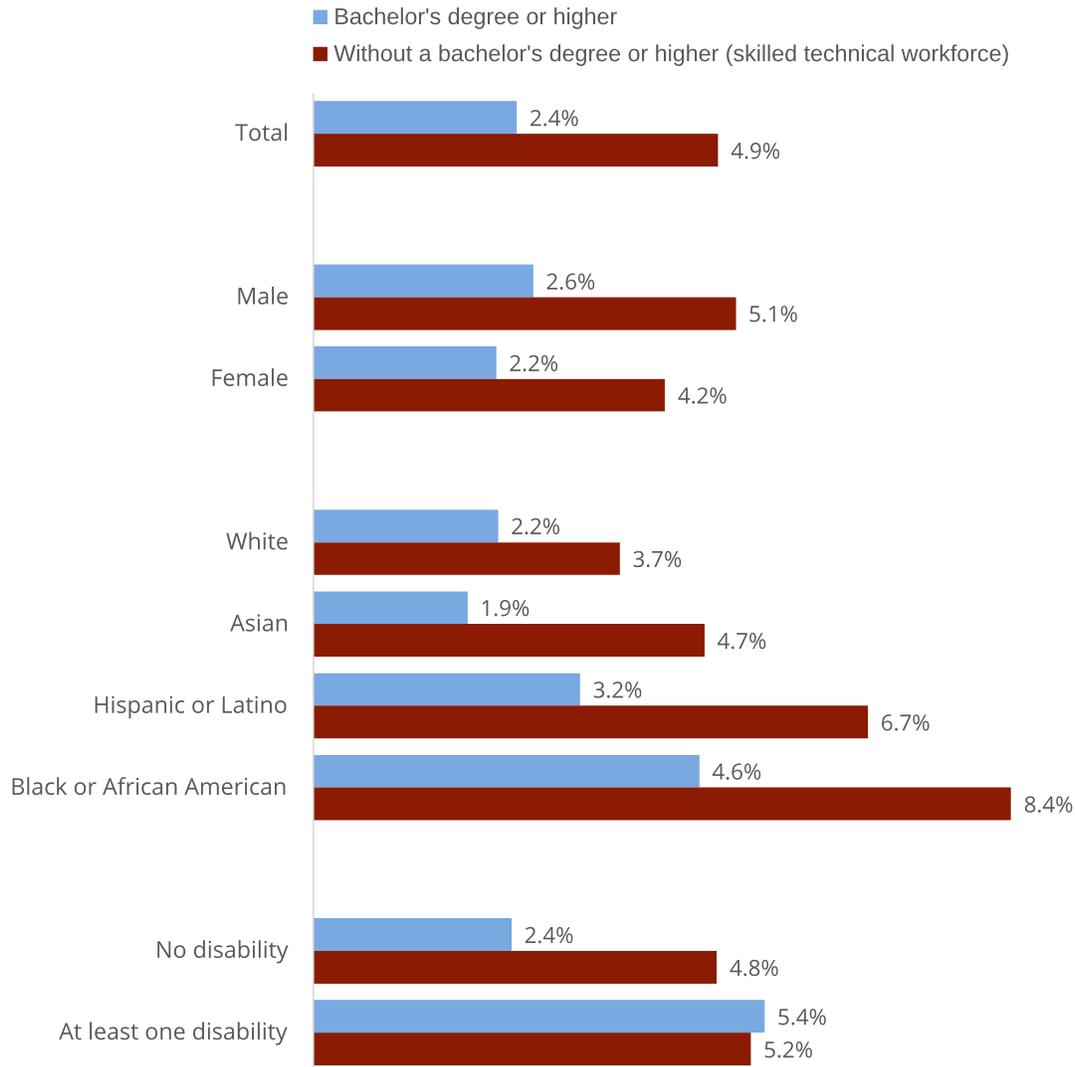

STEM = science, technology, engineering, and mathematics.

**Note(s):**
Civilian noninstitutionalized population plus armed forces living off post or with their families on post. Hispanic or Latino may be any race; race categories exclude Hispanic origin. Respondents can report more than one disability. Those who reported difficulty with one or more functionalities were classified as having a disability. Estimates suppressed for reliability for American Indian or Alaska Native and other (Native Hawaiian and Other Pacific Islander and more than one race); does not meet unweighted cell size requirements.

**Source(s):**
Census Bureau, Current Population Survey, Annual Social and Economic Supplement, 2021.

# STEM Workforce with at Least a Bachelor's Degree

## Overview

A bachelor's degree is typically seen as the gateway to better jobs, higher wages, and for some, a pathway out of poverty. Using data from the National Center for Science and Engineering Statistics' (NCSES's) 2021 National Survey of College Graduates (NSCG), this report evaluates the distribution of college-educated women, racial and ethnic minority groups, and people with one or more disabilities among workers in science and engineering (S&E) and S&E-related occupations. The section also analyzes college graduates in non-S&E occupations for comparative purposes. The information in this section complements that in the previous sections of this report on the science, technology, engineering, and mathematics (STEM) workforce that use data from the Census Bureau's Current Population Survey (CPS). This section provides more details about the representation of these groups in specific S&E occupations (see https://ncses.nsf.gov/pubs/nsb20212/table/SLBR-4). Information is presented about full-time (35 hours or more per week) and part-time (less than 35 hours per week) employment of women, the group composed of underrepresented minorities (Hispanic or Latino, Black or African American, and American Indian or Alaska Native), and workers with at least one disability. Where sufficient data are available, specific S&E occupation details for each underrepresented minority group are provided. Because the data sources differ, there are some differences in the values between this section and the other sections on the STEM workforce. For details, see sidebar Defining Race and Ethnicity and sidebar Defining Persons with at Least One Disability.

Women with at least a bachelor's degree were unevenly represented in S&E occupations, with the highest representation among social and related scientists and the lowest representation among engineers. Workers with one or more disabilities were relatively evenly distributed across S&E occupations. Among those working in S&E occupations with at least a bachelor's degree, a greater proportion of women and those with one or more disabilities were employed part time than their counterparts (men and those without disabilities). Underrepresented minorities and their counterparts (collectively, Whites, Asians, and those in the "other" race group, which includes single race, Native Hawaiian or Other Pacific Islander, and multi-race groups, not Hispanic) had similar rates of part-time employment.

## Representation in the Workforce with at Least a Bachelor's Degree

### Women make up a smaller portion of the S&E workforce than they do of the college-educated workforce overall.

Women made up 51% of the total labor force with at least a bachelor's degree (college-educated labor force) in 2021 (figure 6-1). Similar to the overall STEM workforce (figure 2-3), women represent a smaller proportion than men of the college-educated workforce in S&E occupations—29% were women (figure 6-1). The distribution of women among the S&E occupations is uneven in the college-educated labor force. In 2021, 61% of social and related scientists were women, as were 46% of biological, agricultural, and other life scientists, 33% of physical and related scientists, 26% of computer and mathematical scientists, and 16% of engineers. Compared with men, a greater share of college-educated workers in S&E-related occupations and non-S&E occupations were women (58% and 54%, respectively) in 2021.

35**Figure 6-1**

**College-educated women, by occupation: 2021**

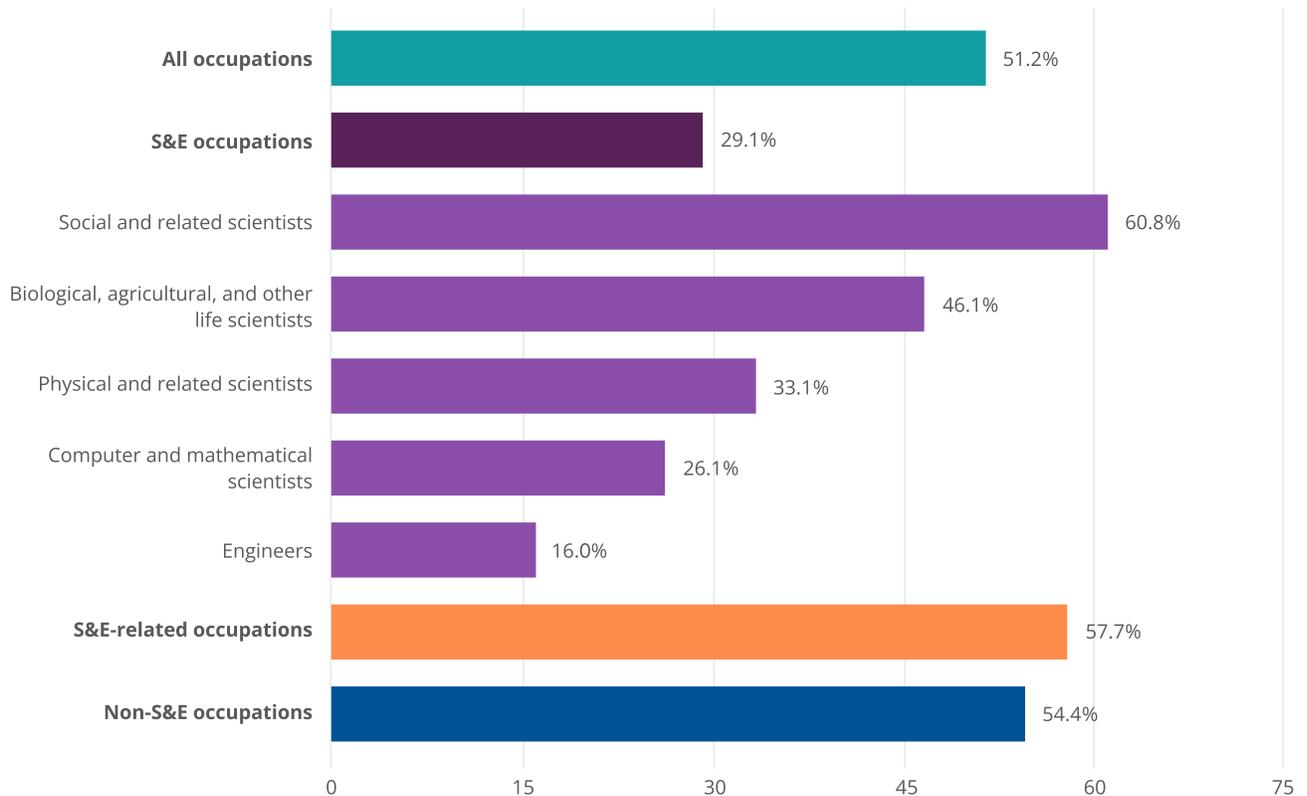

S&E = science and engineering.

**Note(s):**
Data include workers younger than age 75.

**Source(s):**
National Center for Science and Engineering Statistics, National Survey of College Graduates, 2021.

## Black Americans make up a smaller proportion of workers in S&E occupations than they do of workers in S&E-related occupations.

In 2021, 10% of the college-educated workforce were Hispanic, 8% were Black, and 0.2% were American Indian or Alaska Native (figure 6-2). Collectively, these underrepresented minorities constituted 14% of the college-educated workforce in S&E occupations, 17% of those in S&E-related occupations, and 20% of those in non-S&E occupations.

#### Figure 6-2

**Race and ethnicity of the college-educated workforce, by occupation: 2021**

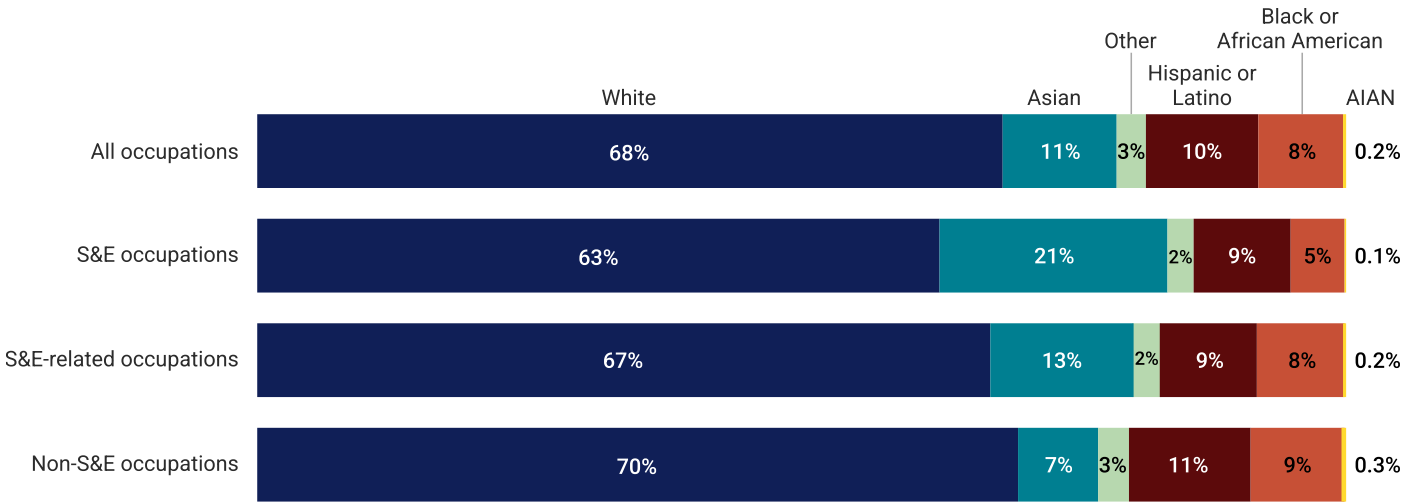

AIAN = American Indian or Alaska Native; S&E = science and engineering.

**Note(s):**
Hispanic or Latino may be any race; race categories exclude Hispanic origin. Other includes Native Hawaiian and Other Pacific Islander and more than one race. Data include workers younger than age 75.

**Source(s):**
National Center for Science and Engineering Statistics, National Survey of College Graduates, 2021.

## Representation of underrepresented minority groups as a whole in S&E occupations ranged from 12% to 18%.

Within S&E occupations, a total of 18% of social and related scientists were Black, Hispanic, or American Indian or Alaska Native, and 12% of physical and related scientists and engineers were made up of these underrepresented groups (figure 6-3). The representation of Black workers varied little among the S&E occupations. Hispanic workers were highly represented among social and related scientists, making up 11% of this group. American Indian or Alaska Native individuals were less than 0.5% of the workers in any S&E occupation.

**Figure 6-3**

**Race and ethnicity of the college-educated workforce, by S&E occupation: 2021**

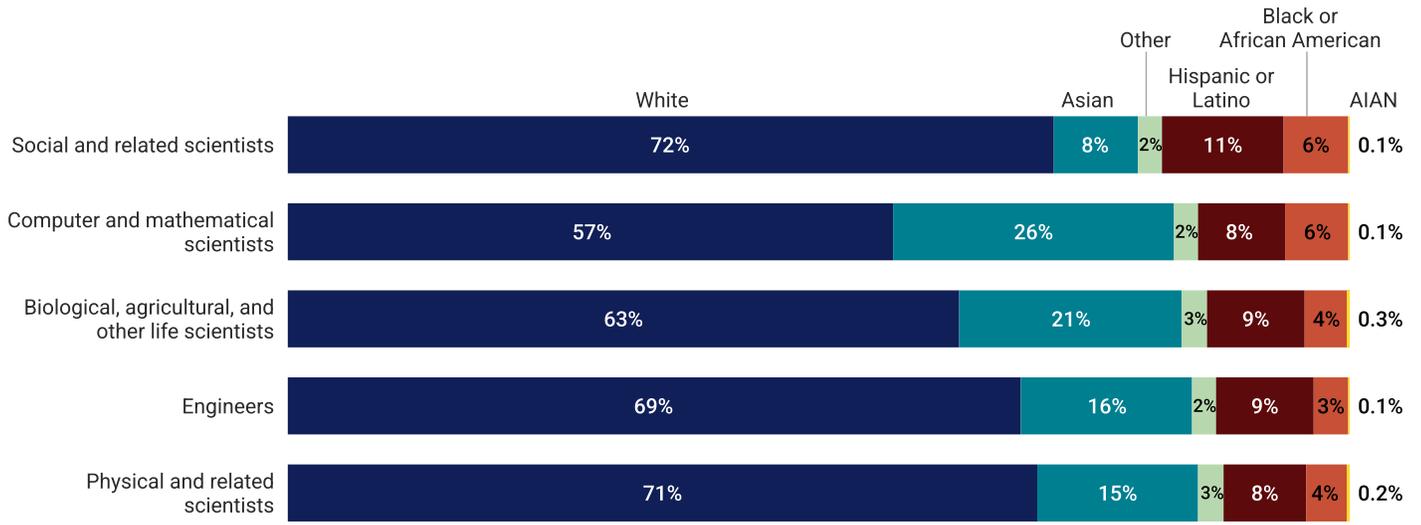

AIAN = American Indian or Alaska Native; S&E = science and engineering.

**Note(s):**
Hispanic or Latino may be any race; race categories exclude Hispanic origin. Other includes Native Hawaiian and Other Pacific Islander and more than one race. Data include workers younger than age 75.

**Source(s):**
National Center for Science and Engineering Statistics, National Survey of College Graduates, 2021.

## Representation of workers with at least one disability varied little among S&E and S&E-related occupations.

Among the employed college-educated workforce, 12% reported at least one disability in 2021 (figure 6-4). Although 13% of college-educated workers in non-S&E occupations had at least one disability, a smaller proportion of workers in S&E and S&E-related occupations reported having at least one disability (11% and 10%, respectively). Among those in S&E occupations, rates of disability were similar across the major occupation groups.

Figure 6-4

**College-educated persons with disabilities, by occupation: 2021**

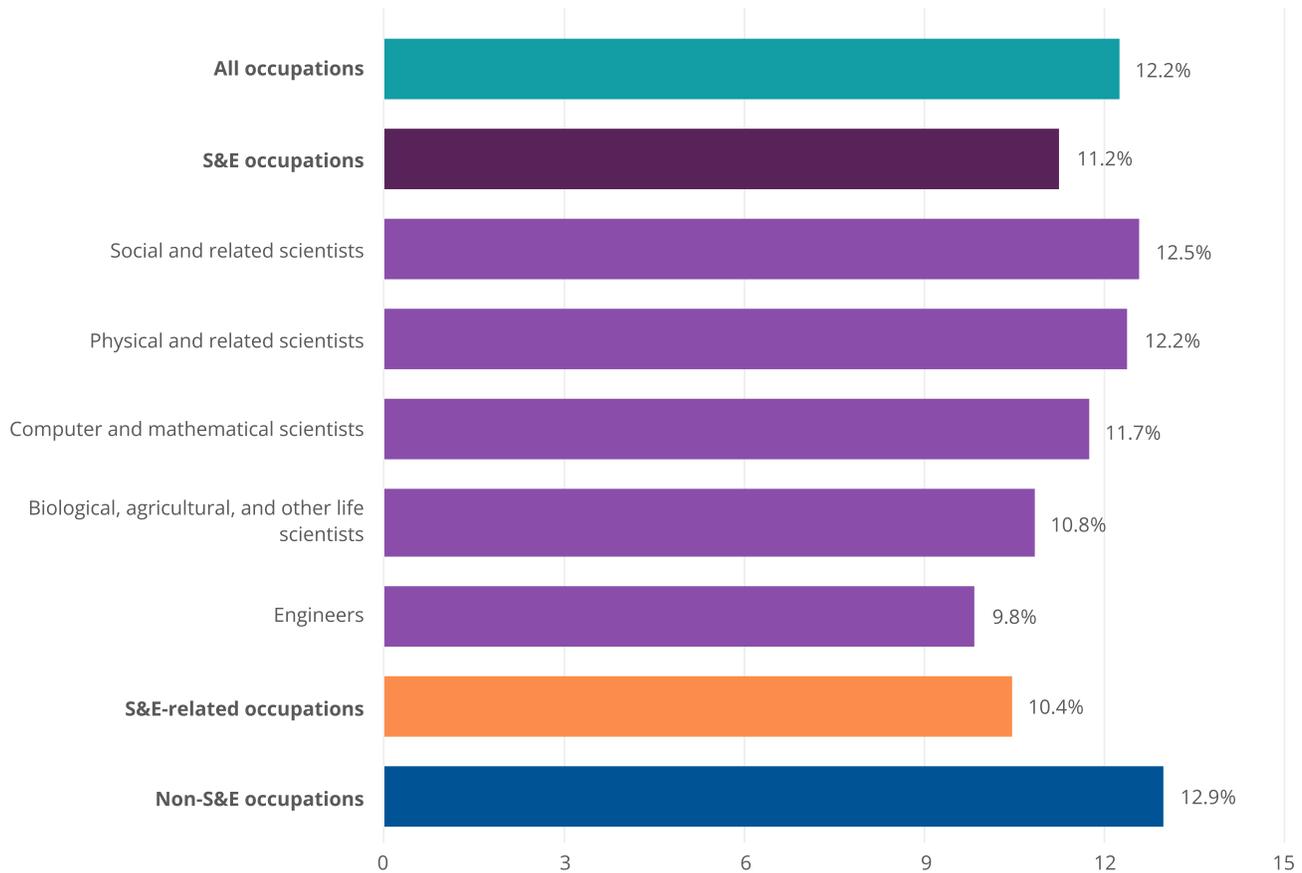

S&E = science and engineering.

**Note(s):**
Data include workers younger than age 75.

**Source(s):**
National Center for Science and Engineering Statistics, National Survey of College Graduates, 2021.

## Part-Time Employment in the Workforce with at Least a Bachelor's Degree

### The share of women working part time in the S&E workforce is nearly twice the share of men.

Among college-educated workers employed in S&E occupations in 2021, 13% of women and 7% of men worked part time (less than 35 hours per week) (figure 6-5). Among computer and mathematical scientists and engineers, women's part-time employment rate was higher than that of men. The distribution of part-time work across the S&E occupations is uneven. A greater proportion of social and related scientists worked part time (30% among women and 27% among men) than did any other occupation group. Fewer engineers and computer and mathematical scientists worked part time: 8%–9% among women and 4%–5% among men. The only significant difference between men and women was among engineers and computer and mathematical scientists. In S&E-related occupations, a greater proportion of college-educated women (20%) than men (11%) worked part time.

**Figure 6-5**

**College-educated S&E workforce employed part time, by occupation and sex: 2021**

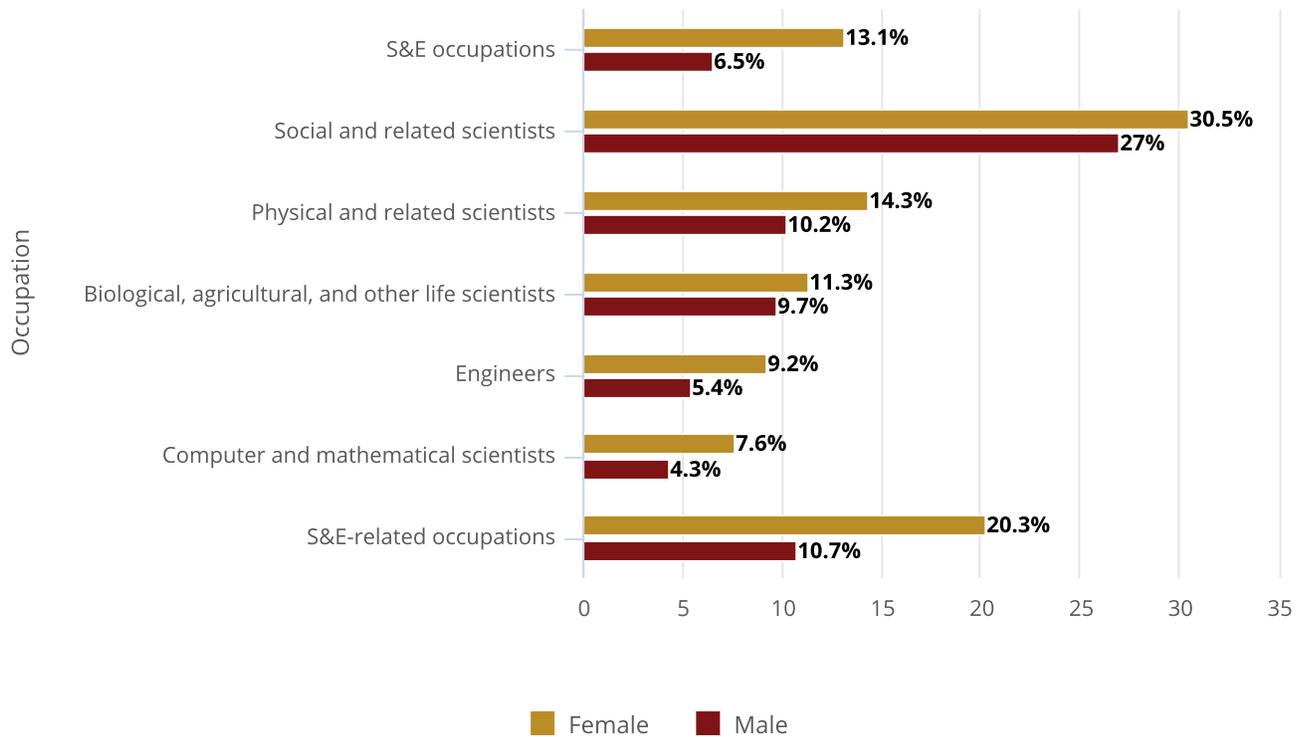

S&E = science and engineering.

**Note(s):**
Data include workers younger than age 75.

**Source(s):**
National Center for Science and Engineering Statistics, National Survey of College Graduates, 2021.

## The rate of part-time employment in the S&E workforce is independent of underrepresented minority status.

To make comparisons among racial and ethnic groups, part-time employment in S&E occupations among college-educated U.S. workers is presented based on two broad groups. The first group is the underrepresented minority group, which includes Hispanic, Black, and American Indian or Alaska Native workers. The second group consists of Whites, Asians, and those in the "other" category. Among college-educated workers employed in S&E occupations in 2021, 8% of each of these two broad groups were employed part time (figure 6-6). Furthermore, only among engineers was there a difference between these two broad groups among the S&E occupations: 3% of engineers in the underrepresented minority group worked part time, compared with 6% of engineers in the counterpart group. These two broad groups of college-educated workers also had similar rates of part-time employment in S&E-related occupations.

**Figure 6-6**

**College-educated S&E workforce employed part time, by occupation and underrepresented minority status: 2021**

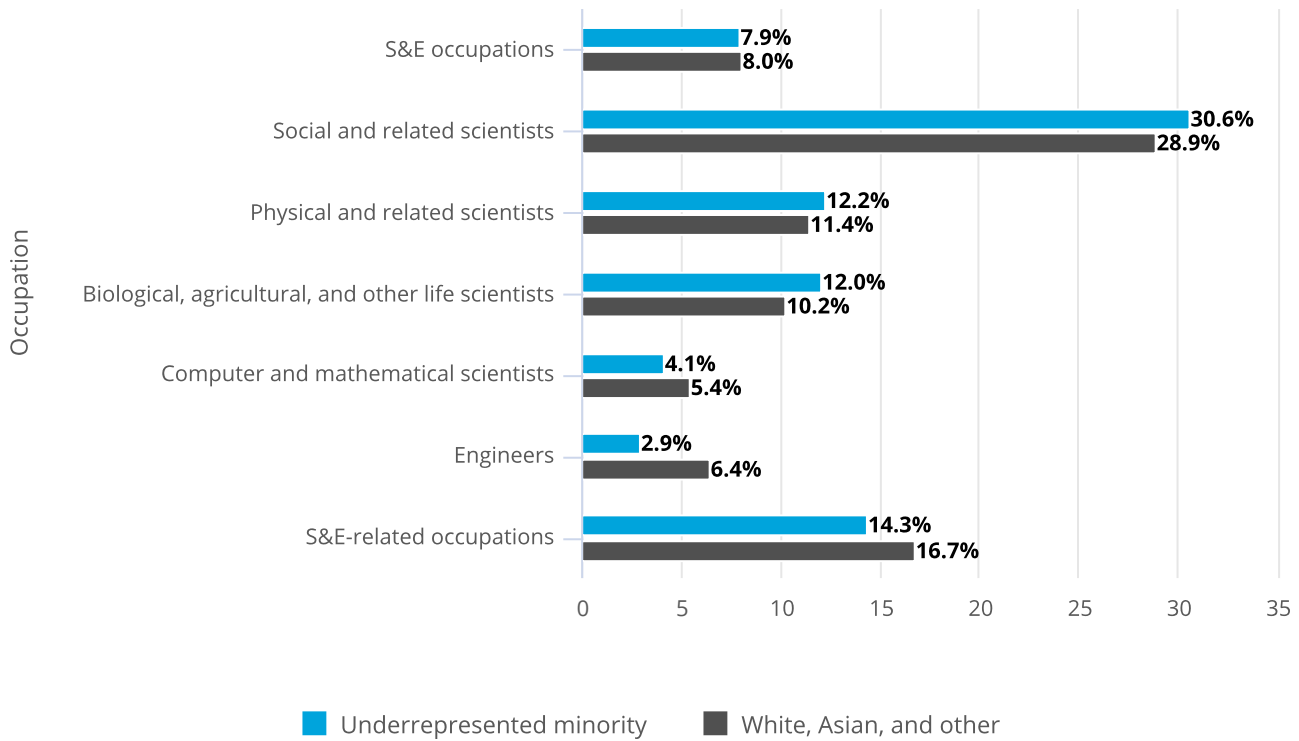

S&E = science and engineering.

**Note(s):**
Underrepresented minority includes Black or African American, Hispanic or Latino, and American Indian or Alaska Native. Hispanic or Latino may be any race; race categories exclude Hispanic origin. Other includes Native Hawaiian and Other Pacific Islander and more than one race. Data include workers younger than age 75.

**Source(s):**
National Center for Science and Engineering Statistics, National Survey of College Graduates, 2021.

## A greater proportion of workers with a disability than those without a disability are employed part time in S&E occupations.

Among college-educated workers employed in S&E occupations in 2021, 11% with at least one disability worked part time, whereas 8% with no disability worked part time (figure 6-7). Although a similar pattern occurred among workers in S&E-related occupations, this difference was not statistically significant. Part-time employment was higher among engineers with at least one disability (12%) than among those without a disability (5%), but there were no statistically significant differences by disability in the other S&E occupation groups.

**Figure 6-7**

**College-educated S&E workforce employed part time, by occupation and disability status: 2021**

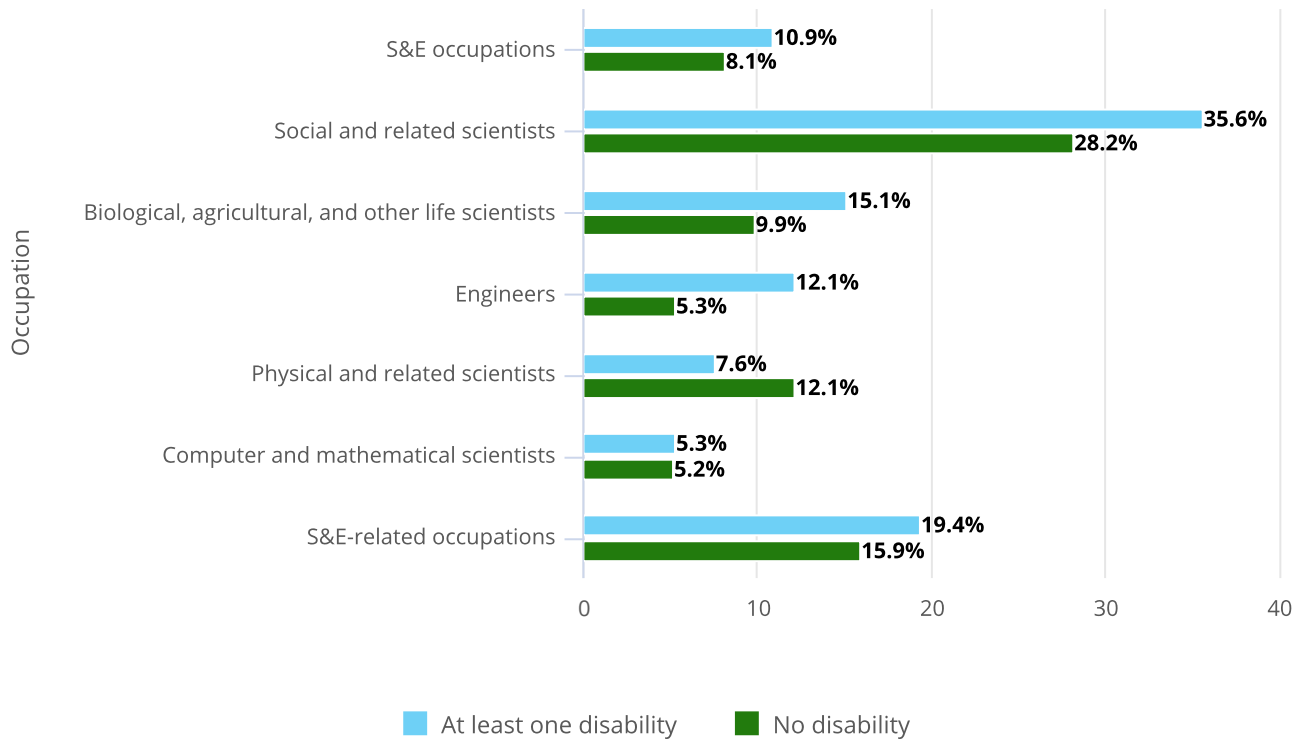

S&E = science and engineering.

**Note(s):**
Data include workers younger than age 75.

**Source(s):**
National Center for Science and Engineering Statistics, National Survey of College Graduates, 2021.

## Involuntary part-time employment falls disproportionately on persons with a disability for those in S&E-related occupations.

Among the college-educated workforce employed part time in S&E occupations in 2021, similar proportions of those with at least one disability (27%) and those without a disability (28%) reported wanting to work full time (figure 6-8). In contrast, among those in S&E-related occupations, persons with a disability who worked part time reported wanting to work full time at about twice the rate (28%) of those without a disability (15%).

**Figure 6-8**

**College-educated S&E workforce employed involuntarily part time, by occupation and disability status: 2021**

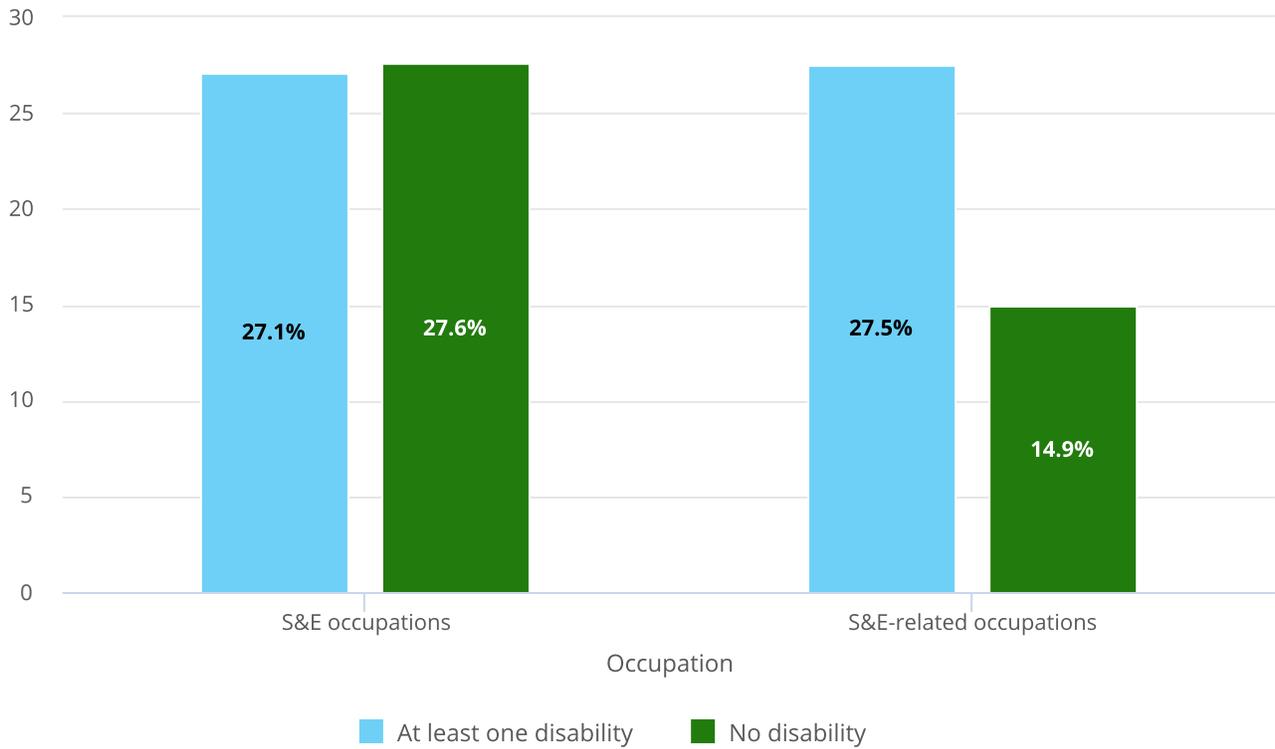

S&E = science and engineering.

**Note(s):**
Data include workers younger than age 75.

**Source(s):**
National Center for Science and Engineering Statistics, National Survey of College Graduates, 2021.

43# Science and Engineering Degrees Earned

## Overview

The representation of women, minorities, and persons with disabilities among postsecondary science and engineering (S&E) degree recipients may indicate their forthcoming participation in the science, technology, engineering, and mathematics (STEM) workforce. Most college graduates employed in S&E occupations work in the same broad field as their highest degree (see NSB, NSF 2021: table LBR-2). Demographic disparities in S&E education may thus signal future differences in employment in related occupations. Using data from 2011 through 2020 from the Integrated Postsecondary Education Data System (IPEDS), this section examines high-level patterns and demographic trends of S&E degree recipients. Information on the sex of the degree recipient is available for all individuals, whereas race and ethnicity information is available only for U.S. citizens and permanent residents. Baseline demographic information is derived from the Census Bureau's Current Population Survey (CPS) for individuals ages 18–34 years, who approximate the population receiving postsecondary education at the associate's, master's, and doctorate levels.

S&E degree data are presented according to five broad fields: social and behavioral sciences, agricultural and biological sciences, physical and earth sciences, mathematics and computer sciences, and engineering. Within these broad S&E fields, selected specific S&E fields are described with especially low or high proportions of bachelor's degrees earned by women or underrepresented minority groups (Hispanics or Latinos, Blacks or African Americans, and American Indians or Alaska Natives). One sidebar presents information about women earning associate's degrees in S&E technologies, a field that contributes to the pool of workers trained for most middle-skill and some S&E-related occupations. A second sidebar highlights the changing racial and ethnic composition of the college-age population in the United States. The final section uses data from the National Center for Science and Engineering Statistics' (NCSES's) Survey of Earned Doctorates (SED) to provide insight into S&E doctoral degrees earned by persons with disabilities. The data show that between 2011 and 2020, the greatest gains by women and underrepresented minorities were in S&E degrees earned at the associate's level. At higher degree levels and within S&E fields of study, representation of women and minorities is unevenly distributed. The proportion of S&E doctorate recipients with at least one disability was similar to the proportion of individuals with a disability among all doctoral recipients in 2021.

Comprehensive annual degree data by field, sex, race, and ethnicity are provided by degree level in the related detailed data tables (table 2-1, table 2-2, table 2-3, table 2-4) that accompany the report online.

## Overall S&E Degrees Earned by Women

### The number of S&E degrees earned by women increased at all degree levels between 2011 and 2020.

In 2020, women were awarded approximately 53,000 associate's degrees, 375,000 bachelor's degrees, 99,000 master's degrees, and 17,000 doctoral degrees in S&E fields (figure 7-1). From 2011 to 2020, the number of S&E degrees awarded to women increased the most at the bachelor's degree level (96,000 degrees), representing a 34% increase. The 63% increase in S&E associate's degrees awarded to women was the largest percentage increase over this period among all degree levels. S&E master's degrees awarded to women increased by 45%, and doctoral degrees awarded to women increased by 18% over this decade.

**Figure 7-1**

**S&E degrees awarded to women, by degree level: 2011 and 2020**

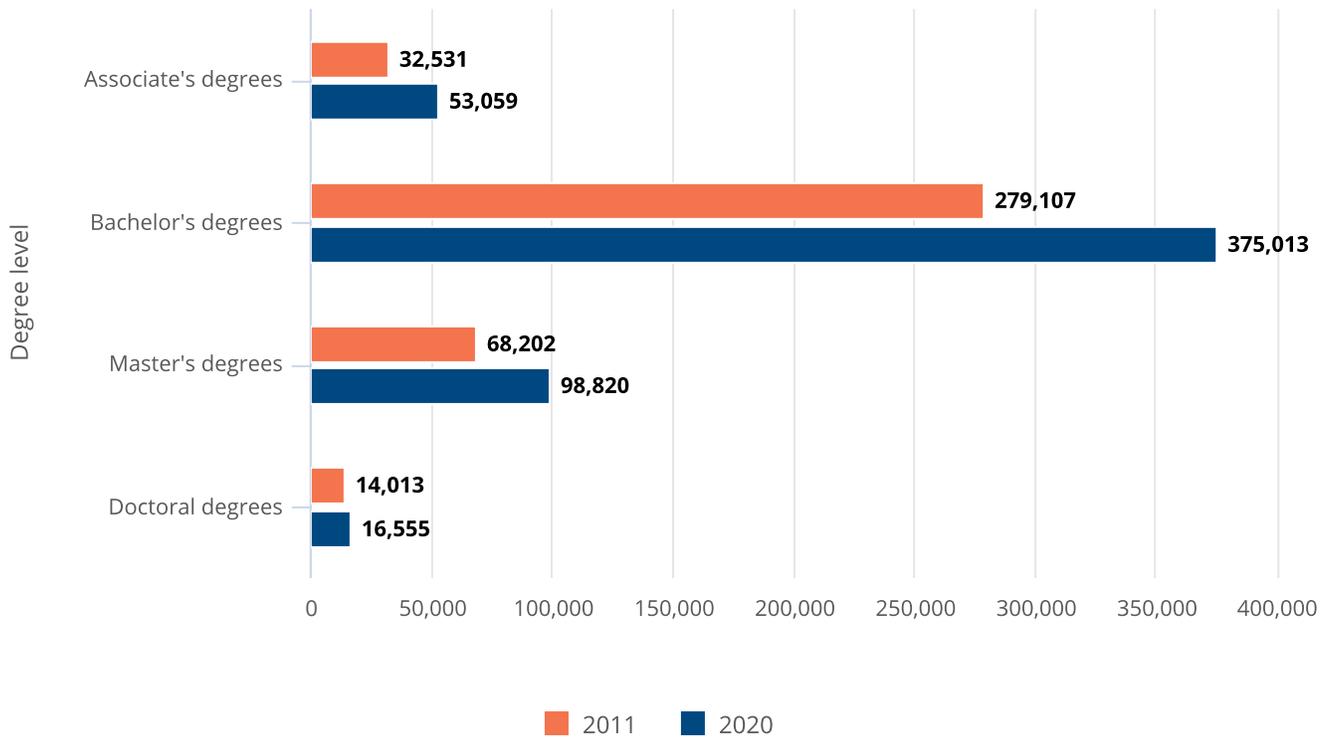

S&E = science and engineering.

**Note(s):**
Data are based on degree-granting, primarily postsecondary institutions eligible to participate in Title IV federal financial aid programs.

**Source(s):**
National Center for Science and Engineering Statistics, special tabulations (2022, Table Builder) of the National Center for Education Statistics, Integrated Postsecondary Education Data System, Completions Survey, provisional release data.

## The share of S&E degrees earned by women increased at the associate's degree level between 2011 and 2020.

Between 2011 and 2020, women's share of S&E degrees remained relatively consistent across degree levels, with the exception of associate's degrees (figure 7-2). The share of associate's degrees earned by women grew from 43% in 2011 to 49% in 2020. This increase mostly reflected a large number of Hispanic students earning associate's degrees in social and behavioral sciences (table 2-1). Women earn half of S&E bachelor's degrees, which has been a consistent pattern for more than a decade (NCSES 2017). Women were slightly underrepresented in S&E at the master's (46%) and doctoral (41%) levels in 2020.

**Figure 7-2**

**S&E degrees awarded to women, by degree level: 2011–20**

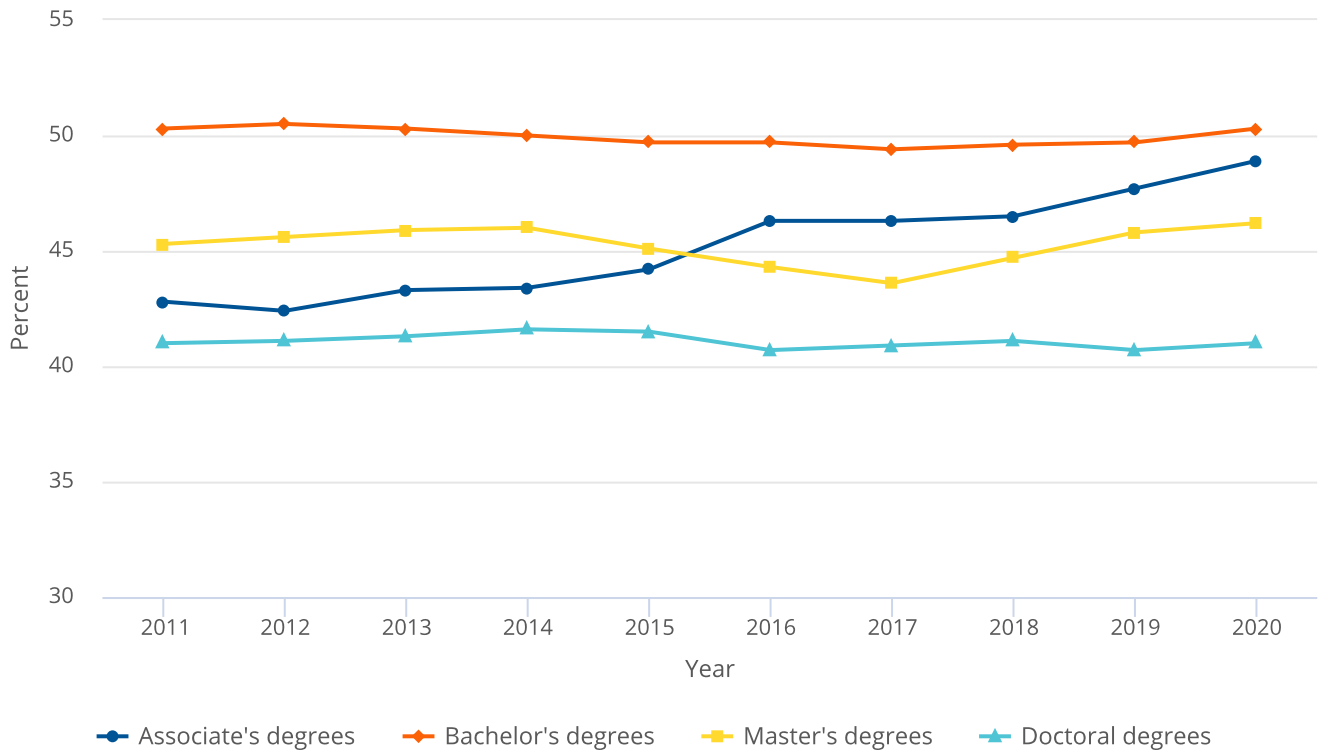

S&E = science and engineering.

**Note(s):**
Data are based on degree-granting, primarily postsecondary institutions eligible to participate in Title IV federal financial aid programs.

**Source(s):**
National Center for Science and Engineering Statistics, special tabulations (2022, Table Builder) of the National Center for Education Statistics, Integrated Postsecondary Education Data System, Completions Survey, provisional release data.

## Degrees Earned by Women in Broad S&E Fields

### Representation of women varies greatly by broad S&E field.

Women accounted for 66% of bachelor's degrees in social and behavioral sciences and 64% of degrees in agricultural and biological sciences in 2020 (figure 7-3). In contrast, they accounted for 26% of these degrees in mathematics and computer sciences and 24% in engineering. Comparable differences for these fields occur at the master's and doctoral degree levels between women and men (table 2-3, table 2-4).

#### Figure 7-3

**S&E degrees awarded to women, by field and degree level: 2011 and 2020**

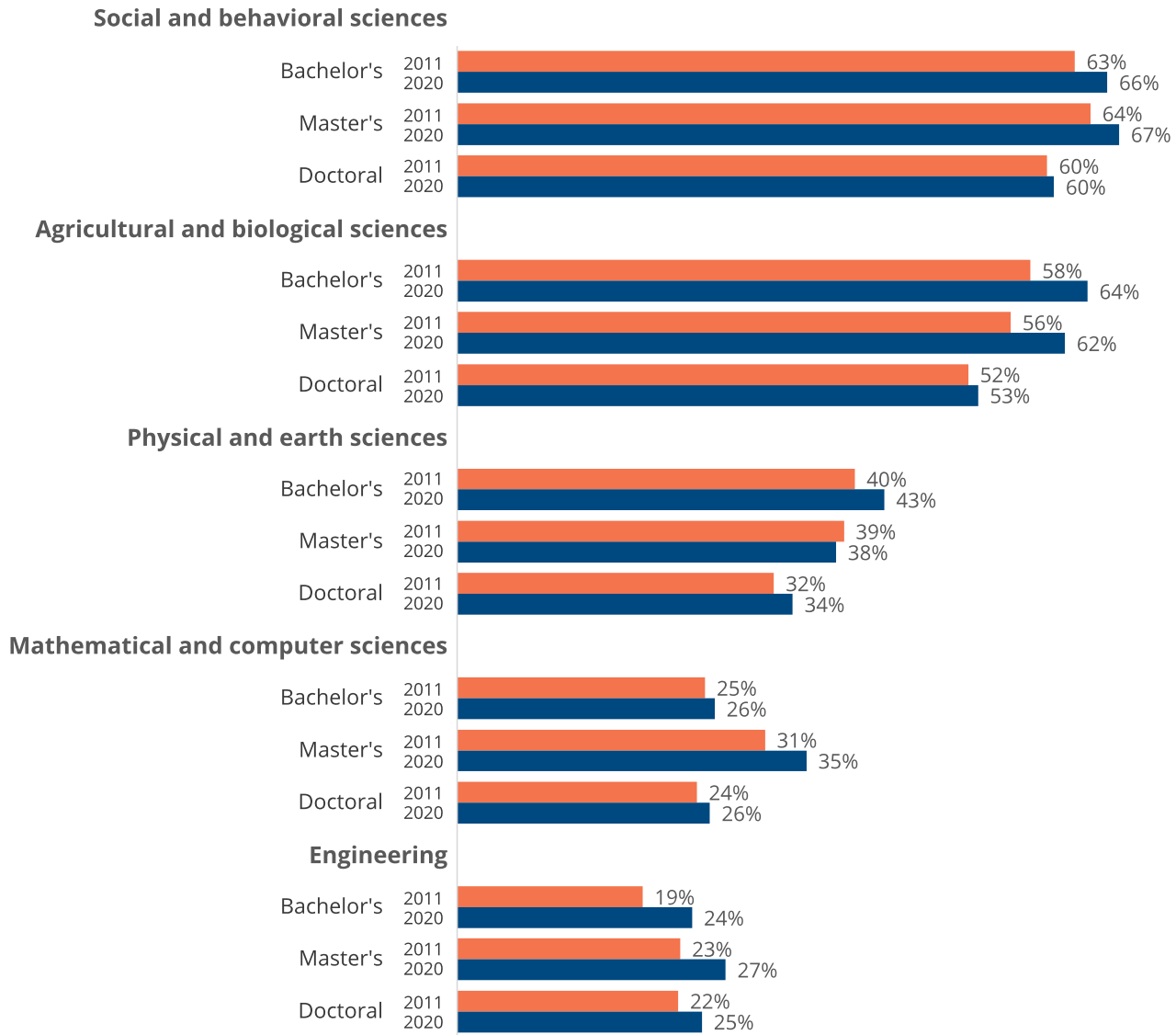

S&E = science and engineering.

**Note(s):**
Data are based on degree-granting, primarily postsecondary institutions eligible to participate in Title IV federal financial aid programs.

**Source(s):**
National Center for Science and Engineering Statistics, special tabulations (2022, Table Builder) of the National Center for Education Statistics, Integrated Postsecondary Education Data System, Completions Survey, provisional release data.

## Women earn the highest proportion of degrees in social and behavioral sciences overall but a much smaller share of economics degrees, compared with men.

In 2020, women earned 66% of bachelor's, 67% of master's, and 60% of doctoral degrees in social and behavioral sciences (figure 7-3). Women's share of degrees has increased in this broad field at all degree levels since 2011.


The distribution of degrees awarded to women within specific fields of social and behavioral sciences is uneven (table 2-2 to table 2-4). The overall high representation of women in this broad S&E field is partly due to the very high share of degrees earned by women in psychology (79% of bachelor's degrees in 2020). Although women earned a majority of the degrees across nearly all fields within social and behavioral sciences at all degree levels, economics remains a notable exception. Women earned 33% of bachelor's degrees in economics in 2020 and were underrepresented in economics at the master's and doctoral levels. This gap between the share of degrees earned by women in economics and in social and behavioral sciences has persisted over the past decade.

### Women earn more than half of degrees in agricultural and biological sciences.

The share of degrees in agricultural and biological sciences earned by women increased from 2011 to 2020 (figure 7-3). Similar to the overall pattern for S&E degrees, women earned lower shares of agricultural and biological sciences degrees at the doctorate level, earning 53% of doctoral degrees, 62% of master's degrees, and 64% of bachelor's degrees in 2020. Among all S&E broad fields, the proportion of degrees in agricultural and biological sciences earned by women is second only to those earned by women in social and behavioral sciences.

### Women are underrepresented in physical and earth sciences, especially physics.

At all degree levels, women earn less than half of degrees in physical and earth sciences (figure 7-3). The pattern of a smaller proportion of degrees at the higher levels occurs here. In 2020, women earned 34% of doctoral degrees, 38% of master's degrees, and 43% of bachelor's degrees in this broad S&E field.

Chemistry and physics are the most common fields within physical and earth sciences by number of degrees awarded at the bachelor's level (table 2-2). Compared to the overall share of all physical and earth sciences degrees earned by women, women earned higher shares of degrees in chemistry and much lower shares in physics at all degree levels (table 2-2 to table 2-4). The difference is greatest at the bachelor's level: women earn more than half (53%) of chemistry degrees and about a quarter (24%) of physics degrees. Despite modest gains in the proportion of physics degrees earned by women over the past decade, physics remains a key field where women are significantly underrepresented among postsecondary degree recipients.

### Women earn a low share of degrees in mathematics and computer sciences.

Women account for a low share of degree recipients in the broad field of mathematics and computer sciences. In 2020, 35% of master's degrees in this broad S&E field were awarded to women, whereas 26% of bachelor's and doctoral degrees were awarded to women (figure 7-3). Women's representation increased the most at the master's degree level, where the share of degrees earned by women grew from 31% to 35%.

Within this broad S&E field, the representation of women is particularly low in computer sciences. At the bachelor's level, the proportion of women receiving computer science degrees is lower than it was a generation ago. Women earned 21% of these degrees in 2020 compared with 29% in 1995 (NCSES 2017: figure 2-C). However, the 21% of computer science bachelor's degree earned by women in 2020 represented an increase from 2011, when 18% of these degrees were earned by women (table 2-2).

### The number and proportion of engineering degrees awarded to women has increased.

Although women remain underrepresented in engineering degree awards, their share of engineering degrees earned increased at all levels between 2011 and 2020 (figure 7-3). Growth was particularly pronounced at the bachelor's level, where the number of engineering degrees awarded to women more than doubled from nearly 15,000 in 2011 to over 31,000 in 2020 (table 2-2). As of 2020, women earned approximately a quarter of engineering degrees at the bachelor's, master's, and doctoral degree levels (figure 7-3). Women are also underrepresented in engineering technologies, where they earned 15% of associate's degrees in 2020 (see sidebar S&E Technology Associate's Degrees).

The distribution of degrees awarded to women within specific fields of social and behavioral sciences is uneven (table 2-2 to table 2-4). The overall high representation of women in this broad S&E field is partly due to the very high share of degrees earned by women in psychology (79% of bachelor's degrees in 2020). Although women earned a majority of the degrees across nearly all fields within social and behavioral sciences at all degree levels, economics remains a notable exception. Women earned 33% of bachelor's degrees in economics in 2020 and were underrepresented in economics at the master's and doctoral levels. This gap between the share of degrees earned by women in economics and in social and behavioral sciences has persisted over the past decade.

### Women earn more than half of degrees in agricultural and biological sciences.

The share of degrees in agricultural and biological sciences earned by women increased from 2011 to 2020 (figure 7-3). Similar to the overall pattern for S&E degrees, women earned lower shares of agricultural and biological sciences degrees at the doctorate level, earning 53% of doctoral degrees, 62% of master's degrees, and 64% of bachelor's degrees in 2020. Among all S&E broad fields, the proportion of degrees in agricultural and biological sciences earned by women is second only to those earned by women in social and behavioral sciences.

### Women are underrepresented in physical and earth sciences, especially physics.

At all degree levels, women earn less than half of degrees in physical and earth sciences (figure 7-3). The pattern of a smaller proportion of degrees at the higher levels occurs here. In 2020, women earned 34% of doctoral degrees, 38% of master's degrees, and 43% of bachelor's degrees in this broad S&E field.

Chemistry and physics are the most common fields within physical and earth sciences by number of degrees awarded at the bachelor's level (table 2-2). Compared to the overall share of all physical and earth sciences degrees earned by women, women earned higher shares of degrees in chemistry and much lower shares in physics at all degree levels (table 2-2 to table 2-4). The difference is greatest at the bachelor's level: women earn more than half (53%) of chemistry degrees and about a quarter (24%) of physics degrees. Despite modest gains in the proportion of physics degrees earned by women over the past decade, physics remains a key field where women are significantly underrepresented among postsecondary degree recipients.

### Women earn a low share of degrees in mathematics and computer sciences.

Women account for a low share of degree recipients in the broad field of mathematics and computer sciences. In 2020, 35% of master's degrees in this broad S&E field were awarded to women, whereas 26% of bachelor's and doctoral degrees were awarded to women (figure 7-3). Women's representation increased the most at the master's degree level, where the share of degrees earned by women grew from 31% to 35%.

Within this broad S&E field, the representation of women is particularly low in computer sciences. At the bachelor's level, the proportion of women receiving computer science degrees is lower than it was a generation ago. Women earned 21% of these degrees in 2020 compared with 29% in 1995 (NCSES 2017: figure 2-C). However, the 21% of computer science bachelor's degree earned by women in 2020 represented an increase from 2011, when 18% of these degrees were earned by women (table 2-2).

### The number and proportion of engineering degrees awarded to women has increased.

Although women remain underrepresented in engineering degree awards, their share of engineering degrees earned increased at all levels between 2011 and 2020 (figure 7-3). Growth was particularly pronounced at the bachelor's level, where the number of engineering degrees awarded to women more than doubled from nearly 15,000 in 2011 to over 31,000 in 2020 (table 2-2). As of 2020, women earned approximately a quarter of engineering degrees at the bachelor's, master's, and doctoral degree levels (figure 7-3). Women are also underrepresented in engineering technologies, where they earned 15% of associate's degrees in 2020 (see sidebar S&E Technology Associate's Degrees).

SIDEBAR

**S&E Technology Associate's Degrees**

The fields that make up S&E technologies are distinct from but related to the five broad S&E fields. Fields within S&E technologies focus on the application of a defined set of technical skills, with most degrees awarded at the associate's level (table 2-1 in the online data tables). These S&E technologies include technician programs in health, engineering, and other fields. Associate's degrees in these fields provide professional preparation for entering middle-skill and some S&E-related occupations that do not require bachelor's degrees (see sidebar The STEM Workforce of the United States).

Substantial differences in men's and women's shares of S&E technologies associate's degrees are apparent by degree field (figure 7-A). In 2020, women earned the majority (82%) of associate's degrees in health technologies and a small share (15%) of degrees in engineering technologies. Because associate's degrees in health technologies are by far the most common, representing 77,000 out of 110,000 degrees in S&E technologies, women also earned most (63%) of the S&E technologies degrees. Women's share of degrees in engineering technologies and health technologies has changed little over the past decade, although the total number of degrees in these fields awarded to both women and men has declined (table 2-1). Women earned less than half of the degrees in science technologies and other S&E technologies.

**Figure 7-A**

S&E technologies associate's degrees awarded, by field and sex: 2020

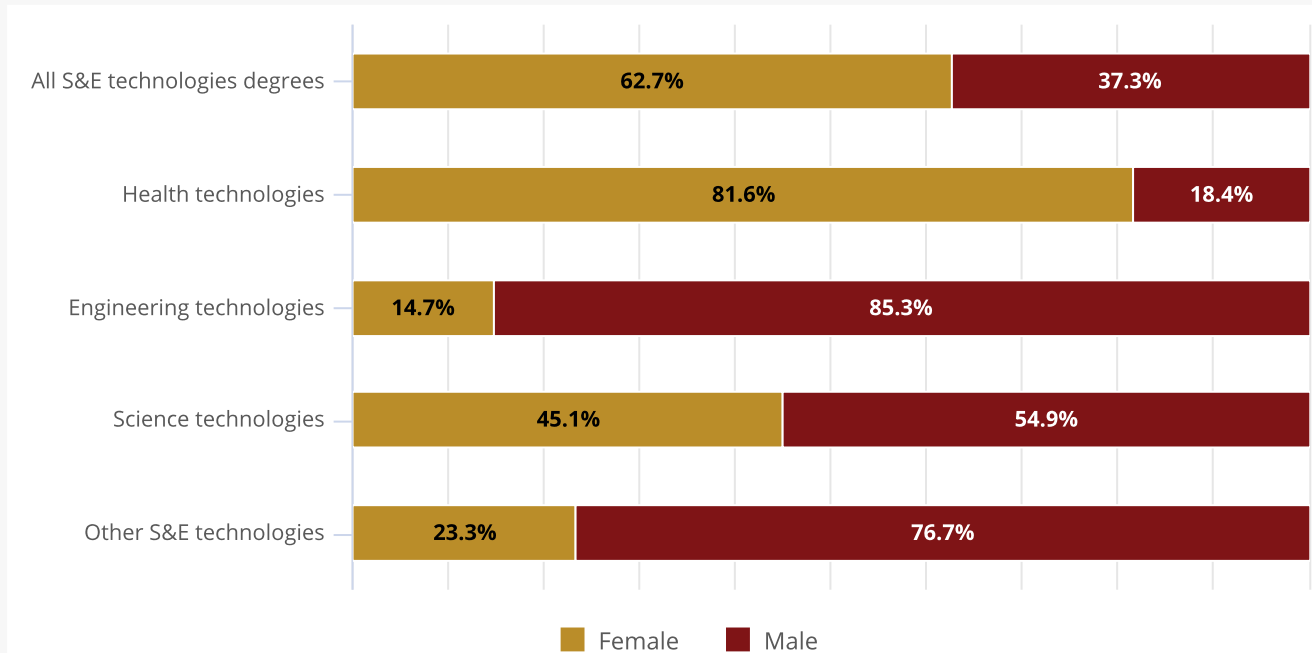

S&E = science and engineering.

**Note(s):**
Data are based on degree-granting, primarily postsecondary institutions eligible to participate in Title IV federal financial aid programs.

**Source(s):**
National Center for Science and Engineering Statistics, special tabulations (2022, Table Builder) of the National Center for Education Statistics, Integrated Postsecondary Education Data System, Completions Survey, provisional release data.

## Overall S&E Degrees Earned by Underrepresented Minorities

### Among S&E degrees earned by underrepresented minorities, the highest proportion was associate's degrees.

In 2020, persons from underrepresented minority groups—Hispanic, Black, and American Indian or Alaska Native—collectively earned 43% of associate's degrees, 26% of bachelor's degrees, 24% of master's degrees, and 16% of doctoral degrees in the five broad S&E fields of study (figure 7-4). At the associate's level, Hispanic individuals earned a higher share of S&E degrees (32%) relative to their share of the college-age population (22%) (figure 7-4). However, Hispanic students and those from other underrepresented minority groups account for much lower shares of degree recipients at the bachelor's degree level and above. For example, the percentage of S&E associate's degrees awarded to American Indian or Alaska Native students is twice as large as the percentage of S&E degrees awarded to this group at any other level. In contrast, White students and Asian students each make up a disproportionately large share of S&E degree recipients at the bachelor's level and above. In 2020, White students accounted for 70% of S&E doctoral degrees and Asian students accounted for 11% (figure 7-4).

**Figure 7-4**

**U.S. population ages 18–34 and S&E degree recipients, by degree level and race and ethnicity: 2020**

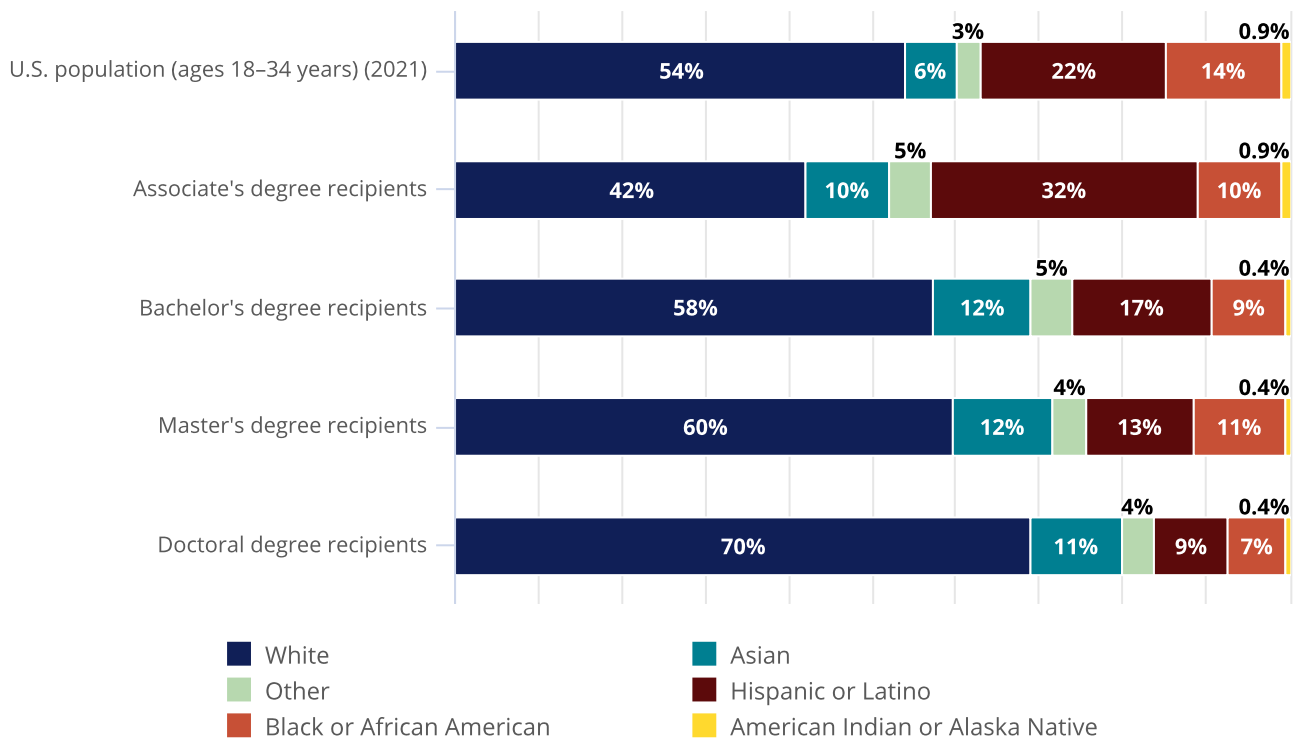

S&E = science and engineering.

**Note(s):**
Hispanic or Latino may be any race; race categories exclude Hispanic origin. Other includes Native Hawaiian and Other Pacific Islander and more than one race. Race and ethnicity data for S&E degree recipients are available only for U.S. citizens and permanent residents. Data are based on degree-granting, primarily postsecondary institutions eligible to participate in Title IV federal financial aid programs.

**Source(s):**
National Center for Science and Engineering Statistics, special tabulations (2022, Table Builder) of the National Center for Education Statistics, Integrated Postsecondary Education Data System, Completions Survey, provisional release data. Population data from Census Bureau, Current Population Survey, Annual Social and Economic Supplement, 2021.

## The proportion of S&E degrees earned by underrepresented minorities increased between 2011 and 2020.

The number and share of S&E degrees earned by underrepresented minorities increased at all degree levels over the past decade (table 2-1 to table 2-4). S&E associate's degrees earned by underrepresented minorities grew at the highest rate, increasing by 13 percentage points from 31% in 2011 to 43% in 2020 (figure 7-5). The increase is much less pronounced at higher degree levels, especially for S&E doctorates, where the share of degrees earned by underrepresented minorities increased from 13% to 16%.

**Figure 7-5**

**S&E degrees awarded to persons from underrepresented minority groups, by degree level: 2011 and 2020**

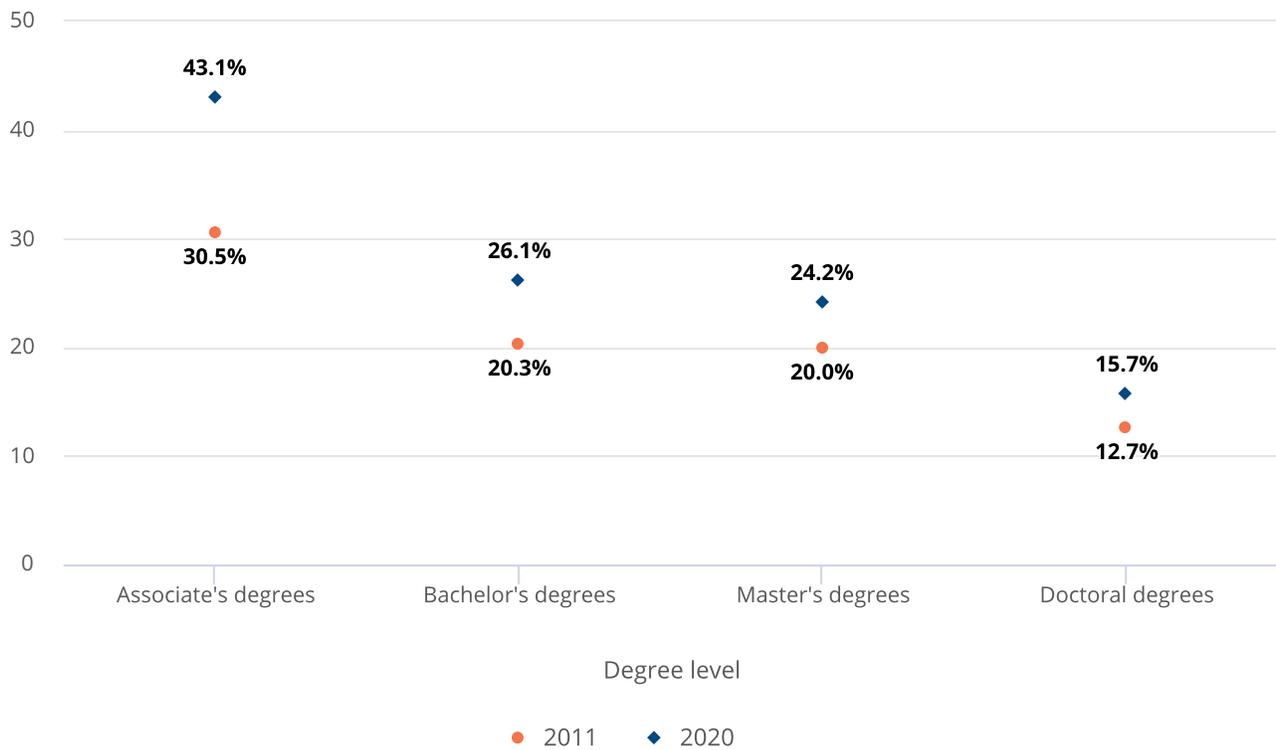

S&E = science and engineering.

**Note(s):**
Underrepresented minority groups include Black or African American, Hispanic or Latino, and American Indian or Alaska Native. Hispanic or Latino may be any race; race categories exclude Hispanic origin. Race and ethnicity data are available only for U.S. citizens and permanent residents. Data are based on degree-granting, primarily postsecondary institutions eligible to participate in Title IV federal financial aid programs.

**Source(s):**
National Center for Science and Engineering Statistics, special tabulations (2022, Table Builder) of the National Center for Education Statistics, Integrated Postsecondary Education Data System, Completions Survey, provisional release data.

## S&E associate's degrees awarded to Hispanic students more than tripled between 2011 and 2020.

Virtually all of the increase in S&E associate's degrees earned by underrepresented minority students in the past decade is due to an increase in degrees earned by Hispanic students. The number of associate's degrees earned by Hispanic students more than tripled, going from 10,000 degrees in 2011 to 33,000 degrees in 2020 (figure 7-6). In contrast, the number of associate's degrees earned by Black students and American Indian or Alaska Native students were similar in 2011 and 2020. Relative to all associate's degree recipients, Hispanic students are concentrated in social and behavioral sciences (table 2-1).

**Figure 7-6**

**S&E associate's degrees awarded, by race and ethnicity: 2011 and 2020**

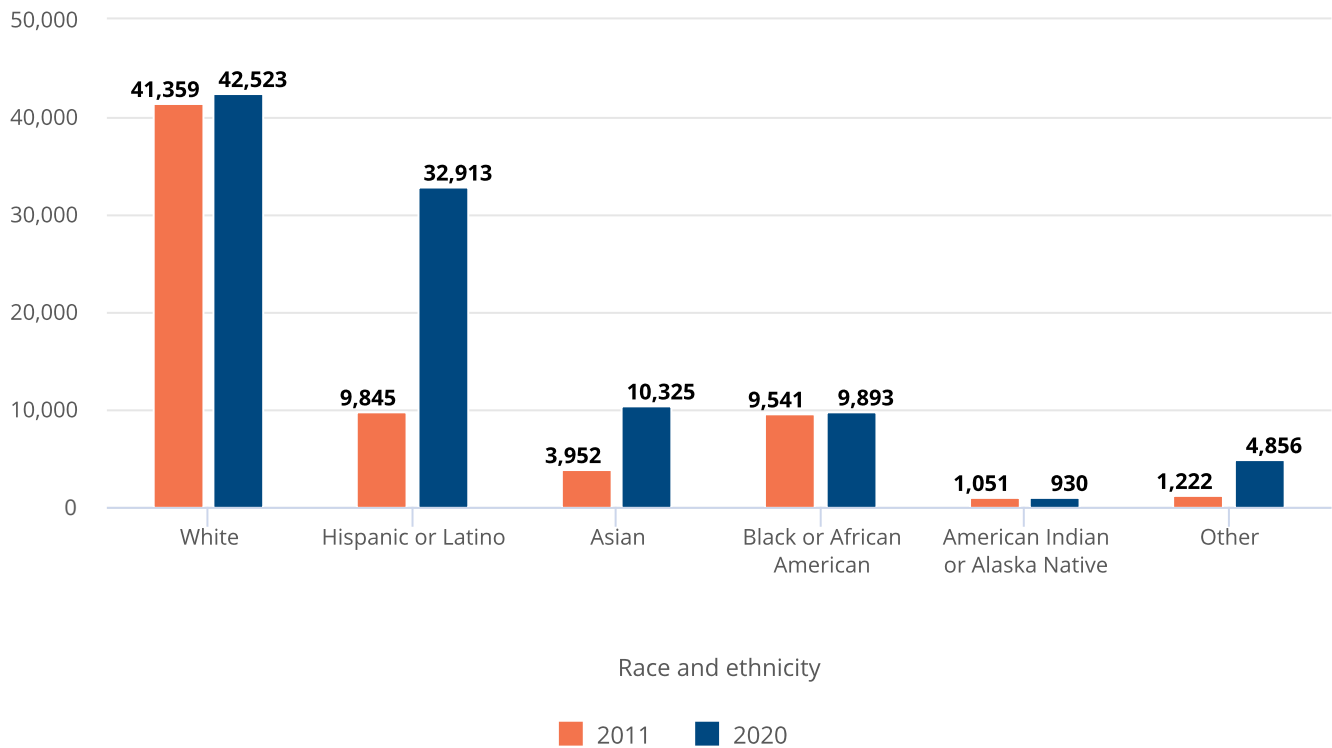

S&E = science and engineering.

**Note(s):**
Hispanic or Latino may be any race; race categories exclude Hispanic origin. Other includes Native Hawaiian and Other Pacific Islander and more than one race. Race and ethnicity data are available only for U.S. citizens and permanent residents. Data are based on degree-granting, primarily postsecondary institutions eligible to participate in Title IV federal financial aid programs.

**Source(s):**
National Center for Science and Engineering Statistics, special tabulations (2022, Table Builder) of the National Center for Education Statistics, Integrated Postsecondary Education Data System, Completions Survey, provisional release data.

## Bachelor's Degrees Earned by Hispanic or Latino Students

### The percentage of S&E bachelor's degrees earned by Hispanic students increased from 2011 to 2020.

Over the past decade, Hispanic students have earned increasing shares of S&E bachelor's degrees across all broad S&E fields (figure 7-7). This growth has occurred in tandem with a significant increase in the Hispanic share of the population of typical degree-seeking age (see sidebar The Changing Racial and Ethnic Composition of the U.S. College-Age Population). The share of social and behavioral sciences degrees earned by Hispanic students increased rapidly, from 12% in 2011 to 21% in 2020. Consequently, the Hispanic share of bachelor's degrees in social and behavioral sciences is nearly identical to their share of the 18- to 34-year-old population, which was 22% in 2021 (figure 7-B).

**Figure 7-7**

**S&E bachelor's degrees awarded to Hispanic or Latino students, by field: 2011 and 2020**

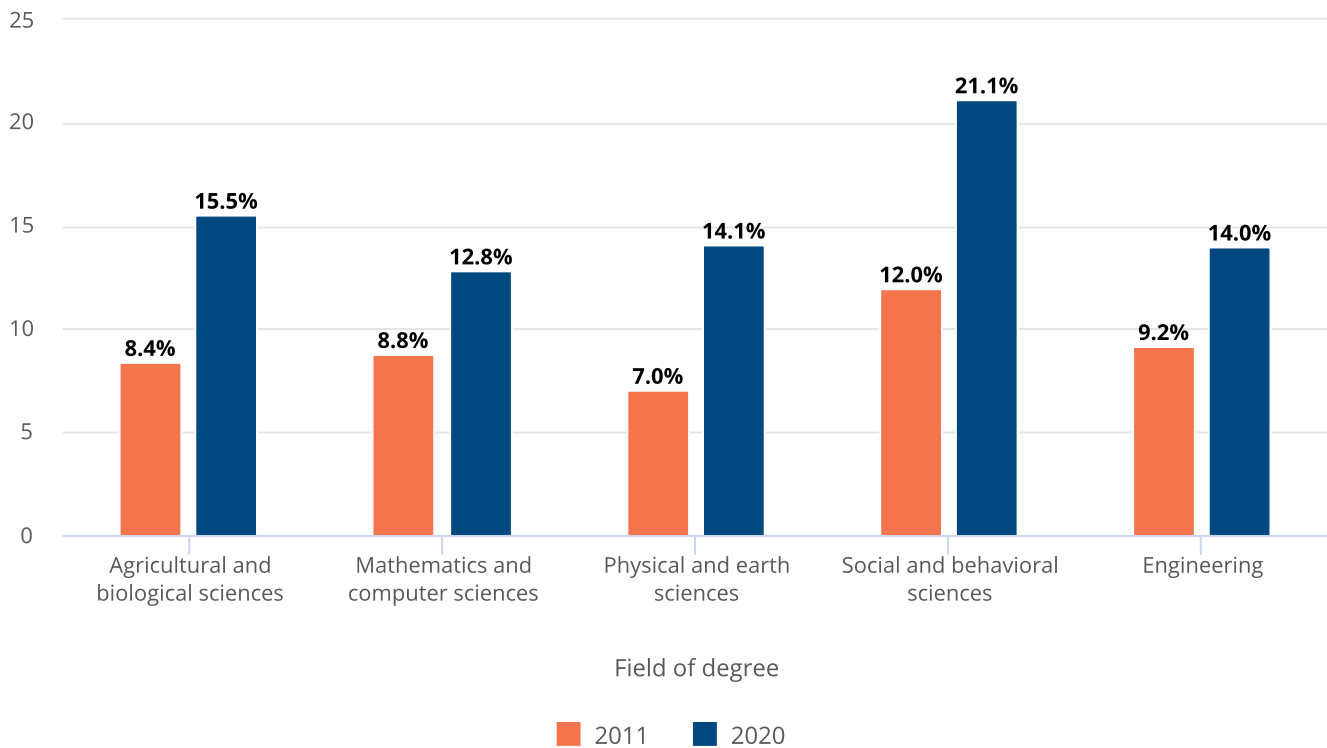

S&E = science and engineering.

**Note(s):**
Hispanic or Latino may be any race; race categories exclude Hispanic origin. Race and ethnicity data are available only for U.S. citizens and permanent residents. Data are based on degree-granting, primarily postsecondary institutions eligible to participate in Title IV federal financial aid programs.

**Source(s):**
National Center for Science and Engineering Statistics, special tabulations (2022, Table Builder) of the National Center for Education Statistics, Integrated Postsecondary Education Data System, Completions Survey, provisional release data.

SIDEBAR

**The Changing Racial and Ethnic Composition of the U.S. College-Age Population**

Data from the Census Bureau show that the racial and ethnic composition of the U.S. college-age population changed from 2011 to 2021 (figure 7-B). Underrepresented minorities collectively accounted for a larger share of the 18- to 34-year-old population in 2021 than 2011, and this increase occurred primarily among Hispanic individuals. The proportion of the U.S. college-age population that is Hispanic increased from 20% in 2011 to 22% in 2021, but the proportions of this population that are Black or that are American Indian or Alaska Native were similar.

**Figure 7-B**

U.S. population ages 18–34, by race and ethnicity: 2011 and 2021

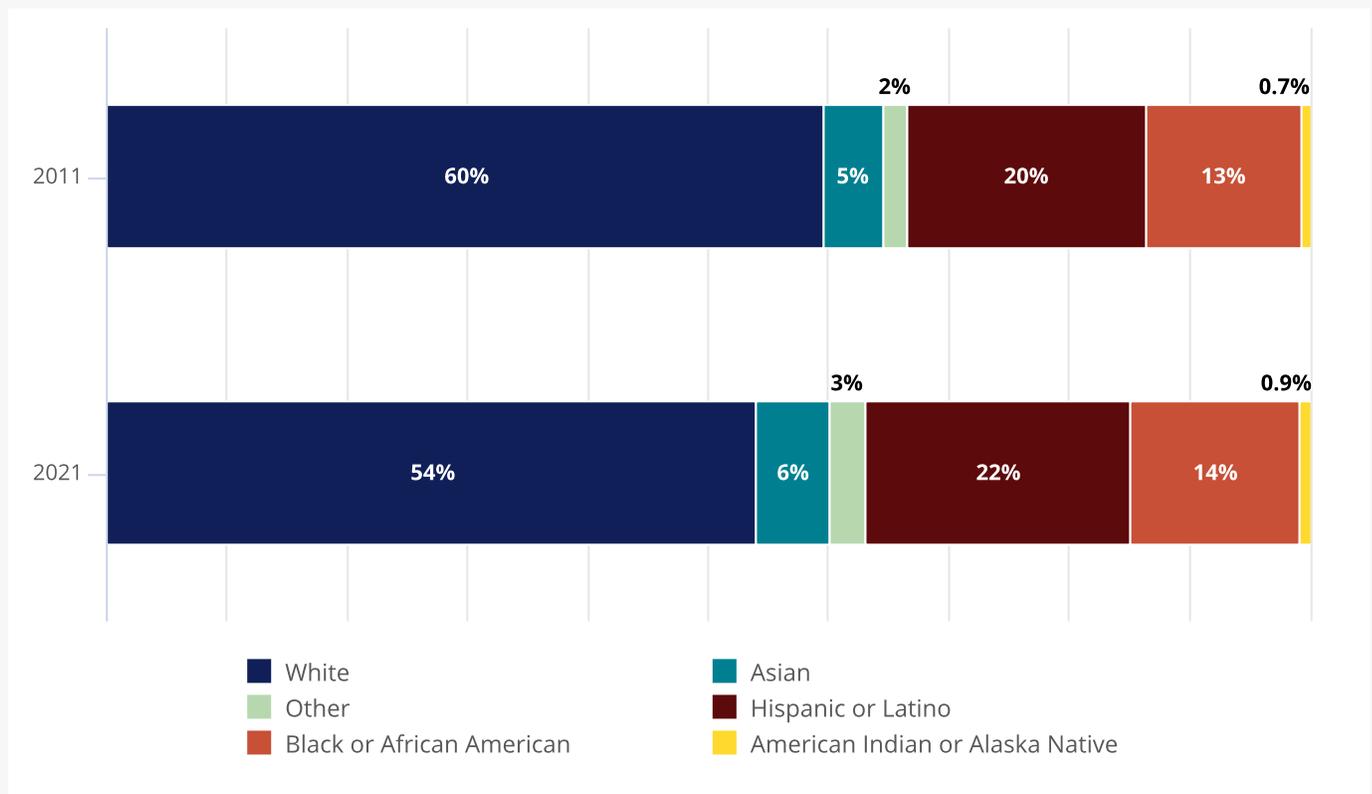

**Note(s):**
Population data includes the civilian noninstitutionalized population as well as individuals in the armed forces living off post or with their families on post. Hispanic or Latino may be any race; race categories exclude Hispanic origin. Other includes Native Hawaiian and Other Pacific Islander and more than one race.

**Source(s):**
Census Bureau, Current Population Survey, Annual Social and Economic Supplement.

The growth in the share of S&E degrees earned by underrepresented minorities (figure 7-5) in the last decade may be driven at least in part by the increasing share of these groups among the U.S. college-age population. If the rate of increase in S&E degrees earned by members of a historically underrepresented group is not commensurate with their simultaneous growth in the college-age population, then their degree of underrepresentation in S&E is not necessarily diminishing. For example, if the Hispanic share of S&E degrees increased by 1% over the same time period that the Hispanic share of college-age individuals increased by 10%, then Hispanic individuals would have become proportionately less represented among S&E degree recipients.

Within social and behavioral sciences, Hispanic students earned 30% of the bachelor's degrees in sociology and 13% in economics (table 2-2). The proportion of degrees earned by Hispanic students varies only slightly across the remaining broad fields of S&E. In 2020, the share of bachelor's degrees earned by Hispanic students in physical and earth sciences, which in 2011 had the lowest share of Hispanic degree recipients, was comparable to their share of these degrees in agricultural and biological sciences, mathematics and computer sciences, and engineering. Hispanic students earned between 13% and 16% of the bachelor's degrees in these four broad fields.

## Bachelor's Degrees Earned by Black or African American Students

### Black students are underrepresented among S&E bachelor's degree recipients.

Representation of Black students among S&E bachelor's degree recipients varies by S&E field (figure 7-8). Black students have the highest representation in social and behavioral sciences, earning 12% of bachelor's degrees in 2020. Representation is lowest in engineering, where Black students earned 5% of bachelor's degrees. From 2011 to 2020, this share remained nearly unchanged, even though the number of engineering degrees earned by Black students steadily increased (table 2-2). This pattern holds true across all fields of S&E. The share of degrees earned by Black students has not changed substantially over the past decade, despite an increase in the overall numbers of S&E bachelor's degrees earned. Across all broad fields of S&E, Black individuals are underrepresented at the bachelor's degree level relative to their share of the 18- to 34-year-old population (14% in 2021) (figure 7-B).

**Figure 7-8**

**S&E bachelor's degrees awarded to Black or African American students, by field: 2011 and 2020**

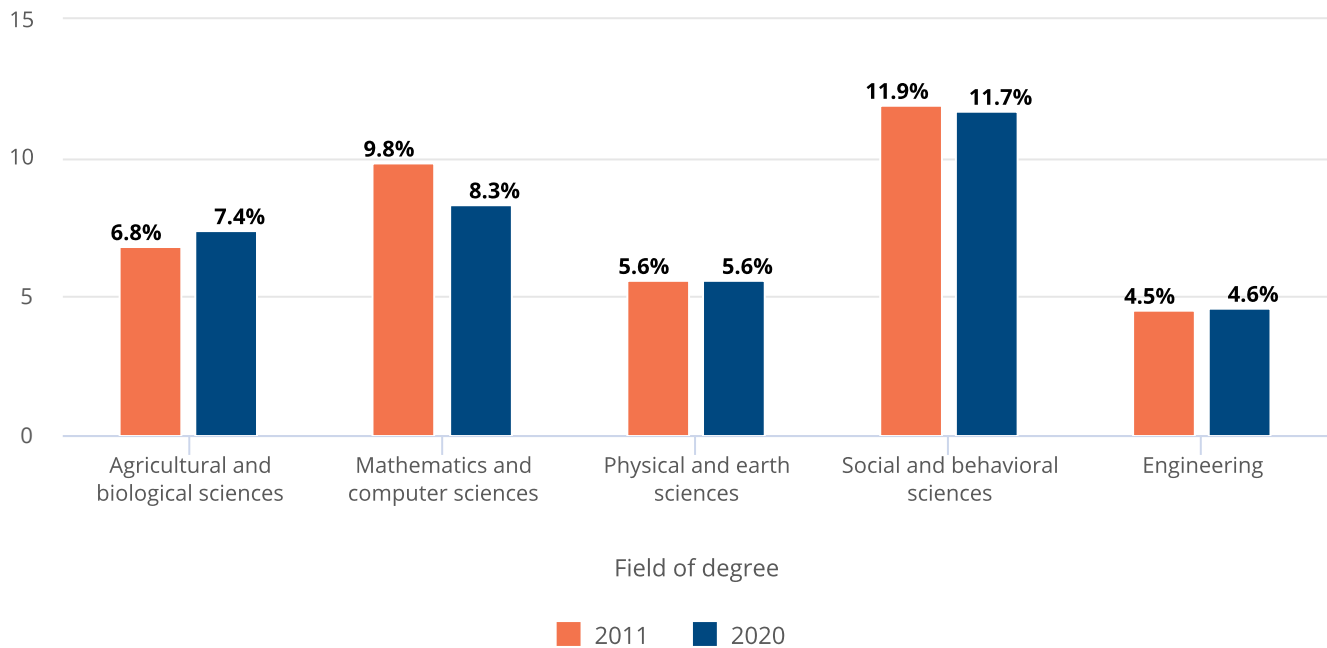

S&E = science and engineering.

**Note(s):**
Race and ethnicity data are available only for U.S. citizens and permanent residents. Data are based on degree-granting, primarily postsecondary institutions eligible to participate in Title IV federal financial aid programs.

**Source(s):**
National Center for Science and Engineering Statistics, special tabulations (2022, Table Builder) of the National Center for Education Statistics, Integrated Postsecondary Education Data System, Completions Survey, provisional release data.

# Bachelor's Degrees Earned by American Indian or Alaska Native Students

## S&E bachelor's degrees earned by American Indian or Alaska Native students decreased between 2011 and 2020.

American Indian or Alaska Native students account for a low and declining share of S&E degree recipients (figure 7-9). Both the number (table 2-2) and share of bachelor's degrees earned by American Indian or Alaska Native students across all fields of S&E declined from 2011 to 2020 (figure 7-9). The share of degrees earned by American Indian or Alaska Native students is lowest in mathematics and computer sciences and in engineering. About 0.3% of bachelor's degrees in these two broad S&E fields were earned by American Indian or Alaska Native students. Like students from other underrepresented minority groups, American Indian or Alaska Native students account for a larger share of degrees in social and behavioral sciences (0.5% in 2020) relative to other fields of S&E; however, this difference is not as pronounced as it is for Hispanic and Black students.

Total S&E bachelor's degrees awarded to American Indian or Alaska Native students dropped by 22% from 2011 to 2020 (table 2-2). Declines also occurred at the master's and doctoral levels (table 2-3, table 2-4). American Indian or Alaska Native students are substantially underrepresented among S&E degree recipients at the bachelor's level and above. American Indian or Alaska Native individuals accounted for 0.9% of the 18- to 34-year-old population in 2021 (figure 7-B) and received 0.4% of S&E bachelor's, master's, and doctoral degrees in 2020 (figure 7-4).

**Figure 7-9**

**S&E bachelor's degrees awarded to American Indian or Alaska Native students, by field: 2011 and 2020**

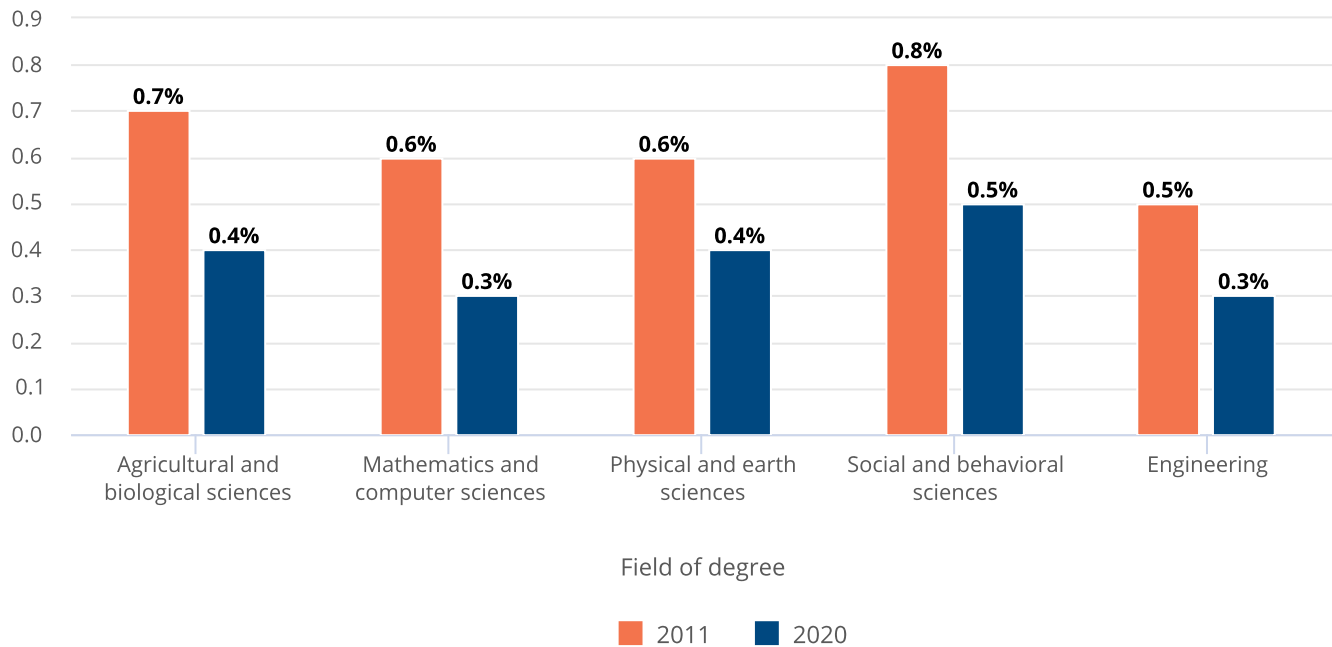

S&E = science and engineering.

**Note(s):**
Race and ethnicity data are available only for U.S. citizens and permanent residents. Data are based on degree-granting, primarily postsecondary institutions eligible to participate in Title IV federal financial aid programs.

**Source(s):**
National Center for Science and Engineering Statistics, special tabulations (2022, Table Builder) of the National Center for Education Statistics, Integrated Postsecondary Education Data System, Completions Survey, provisional release data.

## Doctorates Earned by Persons with Disabilities

### Individuals with at least one disability earned 11% of S&E doctorate degrees.

Compared with data for other groups, data on postsecondary degrees earned by persons with disabilities are limited. However, data from NCSES's Survey of Earned Doctorates are available for doctorate recipients with one or more disabilities. Slight variations are observable in the share of doctorate recipients reporting a disability between S&E and non-S&E fields, as well as across different S&E fields (figure 7-10). In 2021, among all doctorate recipients, 11% reported at least one disability, as did a similar proportion (11%) of S&E doctorate recipients. A slightly higher proportion (13%) of doctoral degree recipients in non-S&E fields had at least one disability. Among S&E doctorate recipients, individuals earning degrees in psychology and social sciences had the highest rate of disability (13%) and those in engineering had the lowest rate (8%). Disability is defined as an individual reporting at least moderate difficulty on at least one of several tasks (see the sidebar Defining Persons with at Least One Disability in the Introduction for details).

**Figure 7-10**

**Doctorate recipients with disability, by selected field: 2021**

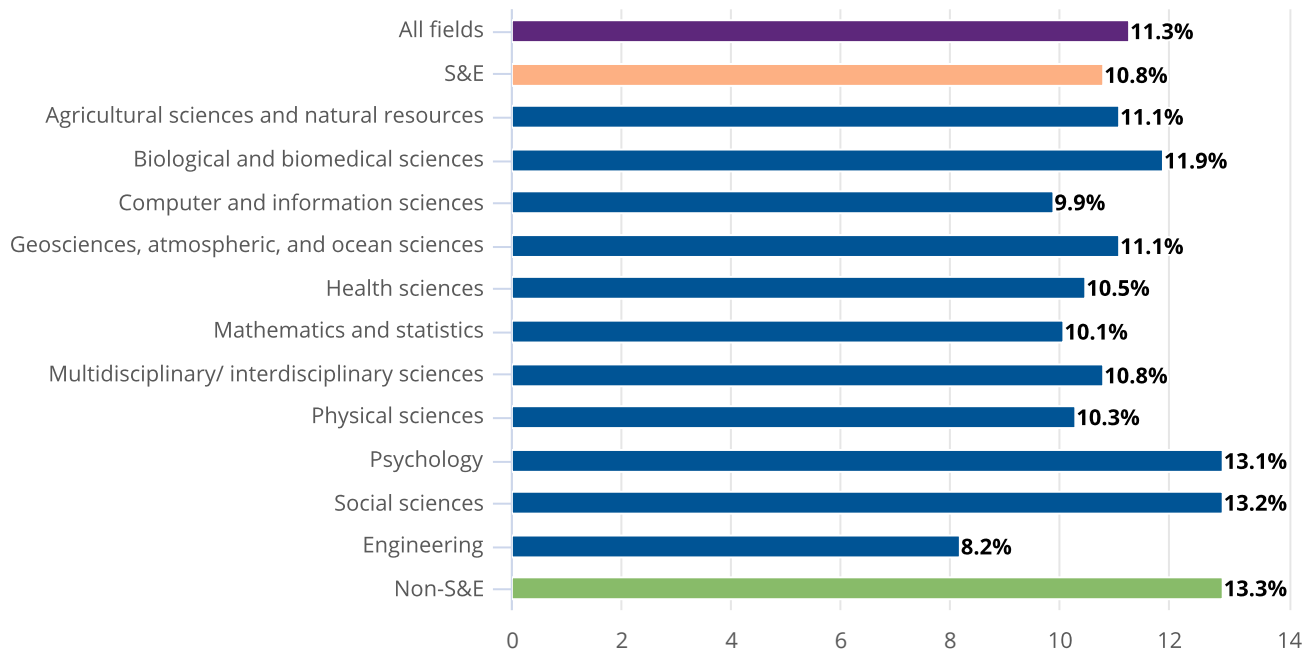

S&E = science and engineering.

**Note(s):**
S&E fields listed here differ from those used in other figures that used data from the Integrated Postsecondary Education Data System, Completions Survey. In previous years, health sciences was classified as non-S&E. Doctorate recipients could report more than one disability. Survey asks degree of difficulty—none, slight, moderate, severe, or unable to do—an individual has in seeing (with glasses), hearing (with hearing aid), walking without assistance, lifting 10 pounds, or concentrating, remembering, or making decisions. Those respondents who answered "moderate," "severe," or "unable to do" for any activity were classified as having a disability. Data in this table are based on research doctorates only. Percentages were calculated by taking the number of persons who indicated that they had a disability and dividing it by the number of persons who responded to the disability question.

**Source(s):**
National Center for Science and Engineering Statistics, special tabulations (2022) of the 2021 Survey of Earned Doctorates.

# Graduate Enrollment in Science and Engineering

## Overview

Although a bachelor's degree in a science and engineering (S&E) field can open doors to a variety of occupations, a master's or doctoral degree is required for employment or advancement in some S&E professions. Using data from the National Center for Science and Engineering Statistics' (NCSES's) Survey of Graduate Students and Postdoctorates in Science and Engineering (GSS), this section reports on the enrollment of women and underrepresented racial and ethnic minorities in S&E master's and doctoral programs from fall 2017 through fall 2021.[6] The data show that the gap between men and women in S&E graduate enrollment is shrinking. Women are concentrated in social and behavioral sciences, and the lower rate of federal funding for graduate students in these fields contributes to a slightly lower share of female graduate students receiving federal financial support. Students who are Hispanic or Latino, Black or African American, and American Indian or Alaska Native are underrepresented among S&E graduate students, and students who are Hispanic or Black are especially underrepresented at the doctoral level. These students from underrepresented minority groups have higher enrollment rates in graduate programs in social and behavioral sciences compared with other S&E fields. A significantly lower share of Black students are enrolled full time, compared with students from other races and ethnicities. During the COVID-19 pandemic, overall S&E graduate enrollment experienced slowing growth in 2020 and accelerating growth in 2021.

For the GSS, the analysis of graduate student enrollment and funding by sex pertains to all individuals, whereas analysis by race and ethnicity applies only to U.S. citizens and permanent residents. Student disability status is not available in the survey data. Starting in 2017, the fields of study in the GSS were updated, limiting comparability with earlier data (see the "Technical Notes" section for details). The year of the data refers to the fall semester of the academic year; for example, data reported for 2021 are for fall 2021.

## Overall Enrollment by Women

### The number of women enrolled in S&E graduate programs increased from 2017 to 2021.

Between 2017 and 2021, the number of women enrolled in S&E master's programs increased by 37% (from 137,000 to 188,000), and the number of women in doctoral programs grew by 16% (from 104,000 to 122,000) (figure 8-1). Men outnumber women at both degree levels, but the enrollment gap has decreased due to faster growth in enrollment of women. In 2021, 47% of S&E master's students and 44% of doctoral students were female, compared with 42% and 41% in 2017. Female enrollment exhibited small growth from 2019 to 2020 but especially large growth from 2020 to 2021, suggesting that the COVID-19 pandemic did not have long-term adverse effects on women's overall enrollment in S&E graduate programs.

## Enrollment of Women in S&E Fields

### Women account for the highest share of S&E graduate student enrollment in social and behavioral sciences and the lowest share in engineering and in mathematics and computer sciences.

In 2021, the largest number of women (70,000) were enrolled in programs at the master's level in social and behavioral sciences (figure 8-2). About 42,000 women were master's degree students in mathematics and computer sciences, which was the second-largest field for female students. At the doctoral level, enrollment of women was highest in agricultural and biological sciences and natural resources (38,000), followed by social and behavioral sciences (34,000).

**Figure 8-1**

**S&E graduate students, by sex and level of enrollment: 2017–21**

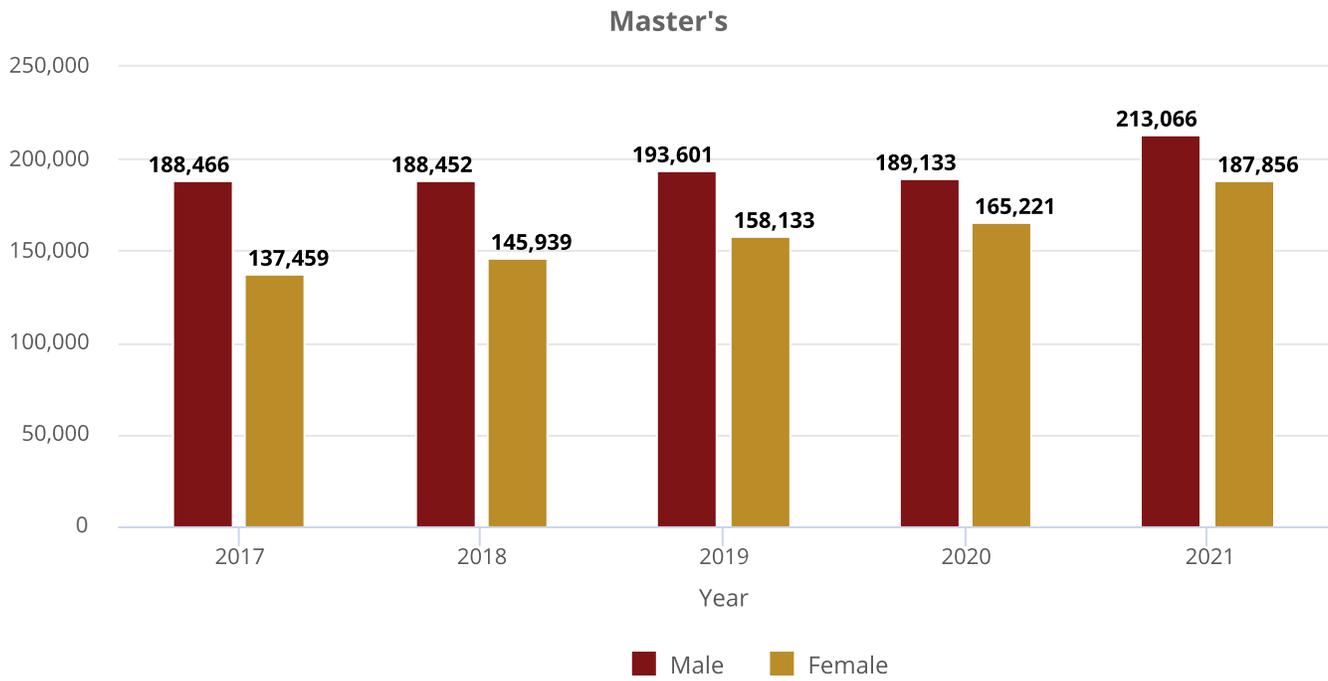

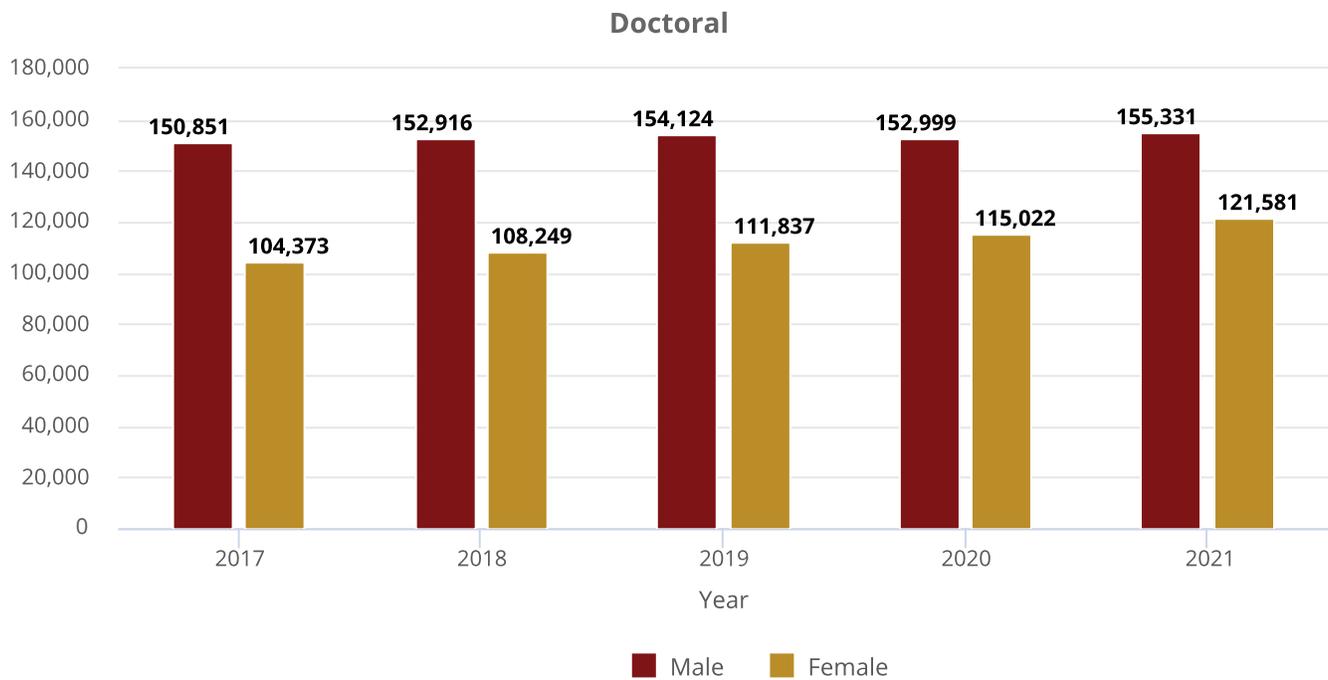

S&E = science and engineering.

**Source(s):**
National Center for Science and Engineering Statistics, Survey of Graduate Students and Postdoctorates in Science and Engineering, 2021.

**Figure 8-2**

**S&E graduate student enrollment of women, by field and level of enrollment: 2021**

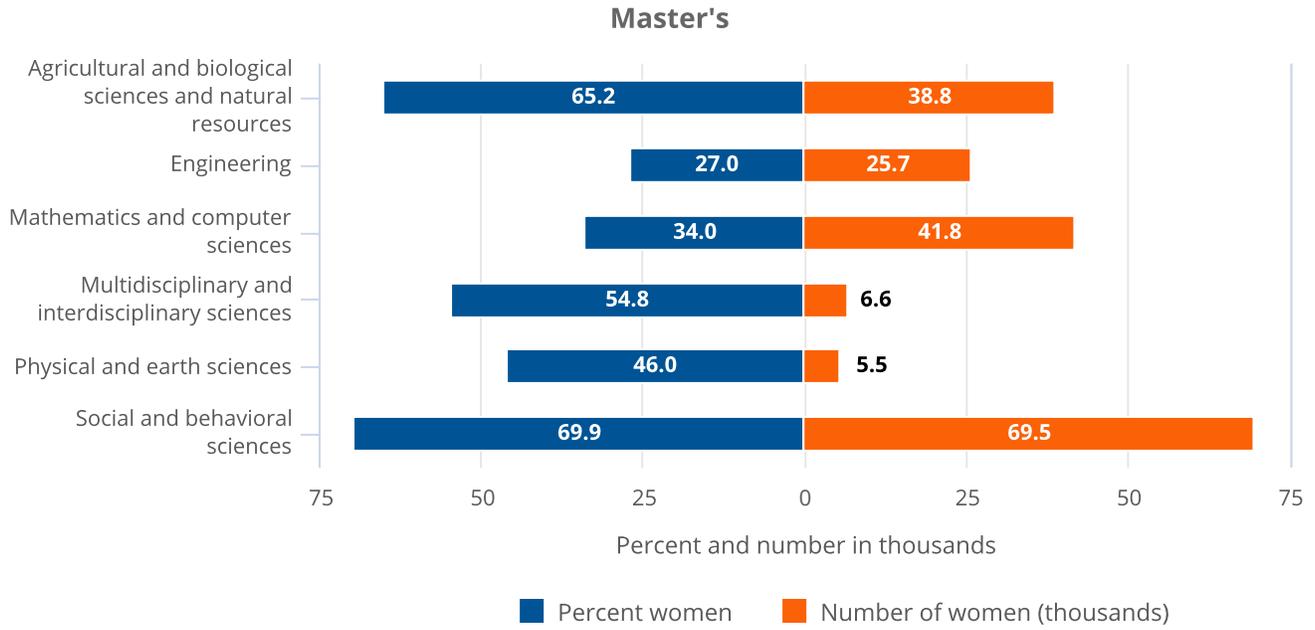

Master's

| Field | Percent women | Number of women (thousands) |
|---|---|---|
| Agricultural and biological sciences and natural resources | 65.2 | 38.8 |
| Engineering | 27.0 | 25.7 |
| Mathematics and computer sciences | 34.0 | 41.8 |
| Multidisciplinary and interdisciplinary sciences | 54.8 | 6.6 |
| Physical and earth sciences | 46.0 | 5.5 |
| Social and behavioral sciences | 69.9 | 69.5 |

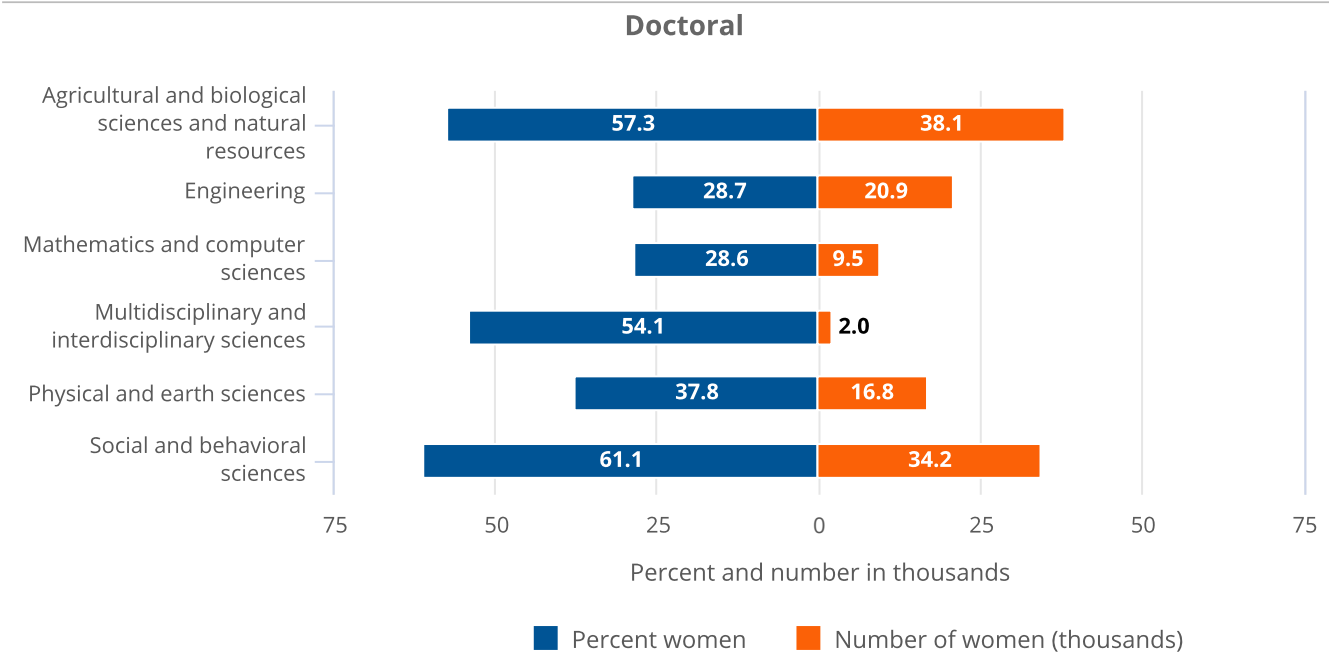

Doctoral

| Field | Percent women | Number of women (thousands) |
|---|---|---|
| Agricultural and biological sciences and natural resources | 57.3 | 38.1 |
| Engineering | 28.7 | 20.9 |
| Mathematics and computer sciences | 28.6 | 9.5 |
| Multidisciplinary and interdisciplinary sciences | 54.1 | 2.0 |
| Physical and earth sciences | 37.8 | 16.8 |
| Social and behavioral sciences | 61.1 | 34.2 |

S&E = science and engineering.

**Note(s):**
Degree fields in this figure differ from fields presented in other tables and figures in this report that are based on the National Center Education Statistics Integrated Postsecondary Education Data System, Completions Survey.

**Source(s):**
National Center for Science and Engineering Statistics, Survey of Graduate Students and Postdoctorates in Science and Engineering, 2021.

Women accounted for the highest share of S&E graduate enrollment in social and behavioral sciences, where they were 70% of master's and 61% of doctoral students (figure 8-2). This concentration of women in social and behavioral sciences contributes to the disparity in federal support between male and female graduate students (see sidebar Sources of Support for S&E Graduate Education of Women). Women were also a majority of graduate students in agricultural and biological sciences and natural resources at both the master's (65%) and doctoral (57%) levels, and slightly more than half of master's and doctoral students in multidisciplinary and interdisciplinary sciences. However, women's share of enrollment was 27% in engineering and 34% in mathematics and computer sciences at the master's level, and their share was 29% for each of these fields at the doctoral level.

### SIDEBAR

**Sources of Support for S&E Graduate Education of Women**

**A slightly lower share of women than men who are full-time S&E graduate students receive federal financial support for their education.**

Three main sources of graduate funding for tuition, living expenses, and other education-related costs are self-support, which includes loans and family sources; federal funding; and institutional support. In 2021, the majority (70%) of full-time science and engineering (S&E) master's students primarily relied on their own funds, including loans, to support their studies. More than 9 out of 10 full-time S&E doctoral students, however, receive substantial financial support for their studies through assistantships, scholarships, or other means.* The federal government provides some of this support, and slightly lower percentages of women than men receive federal funding at both the master's and doctoral levels (figure 8-A).

Federal support for graduate students varies according to degree level and field. Women receive slightly lower rates of federal funding than men within some fields of S&E, but funding rates for both women and men follow the same broad pattern by field. Federal funding supports larger percentages of students in agricultural and biological sciences, physical and earth sciences, and engineering than in other fields at both the master's and doctoral levels (table 3-3). In contrast, the percentage of full-time students in social and behavioral sciences supported by federal funding in 2021 was low at both the master's (3% of women and 5% of men) and doctoral (9% and 7%) levels. Because women are concentrated in behavioral and social sciences, the low rate of federal funding for graduate students in this field contributes to the lower rates of funding for female compared to male graduate students (figure 8-2).

**Figure 8-A**

Primary source of support for full-time S&E graduate students, by level of enrollment and sex: 2021

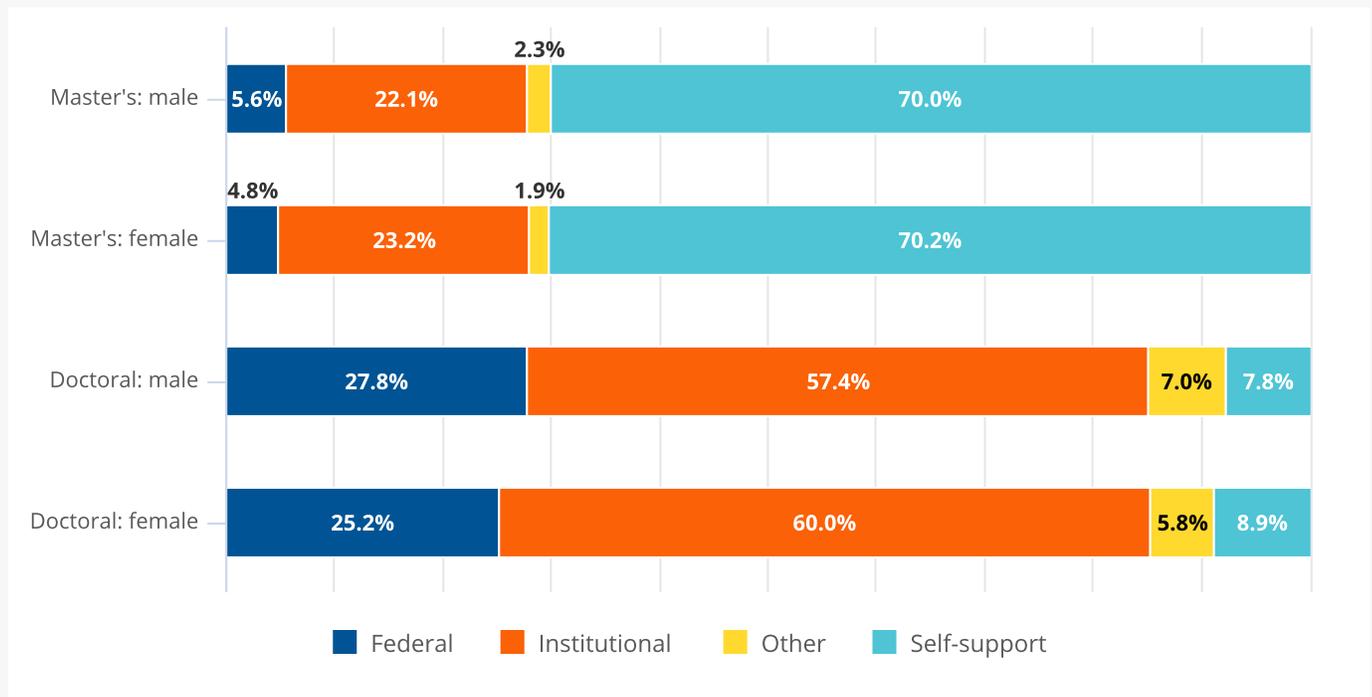

S&E = science and engineering.

**Note(s):**
Other includes support from nonfederal domestic and foreign sources.

**Source(s):**
National Center for Science and Engineering Statistics, Survey of Graduate Students and Postdoctorates in Science and Engineering, 2021.

* From NCSES's Survey of Graduate Students and Postdoctorates in Science and Engineering.

## Overall Minority Enrollment

### Hispanic, Black, and American Indian or Alaska Native students are underrepresented among S&E graduate students.

Racial and ethnic disparities in S&E graduate enrollment parallel those disparities among S&E degree recipients. White and Asian students are overrepresented among S&E graduate students, and Hispanic, Black, and American Indian or Alaska Native students are underrepresented (figure 8-3). The disparity is more pronounced at the doctoral level, where each underrepresented minority group accounts for a share of enrolled students that is roughly half its share of the 18- to 34-year-old population (figure 7-B). Collectively, underrepresented minorities were 37% of the college-age population but 25% of S&E master's students and 19% of S&E doctoral students in 2021.

**Figure 8-3**

**S&E graduate students, by level of enrollment, race, and ethnicity: 2021**

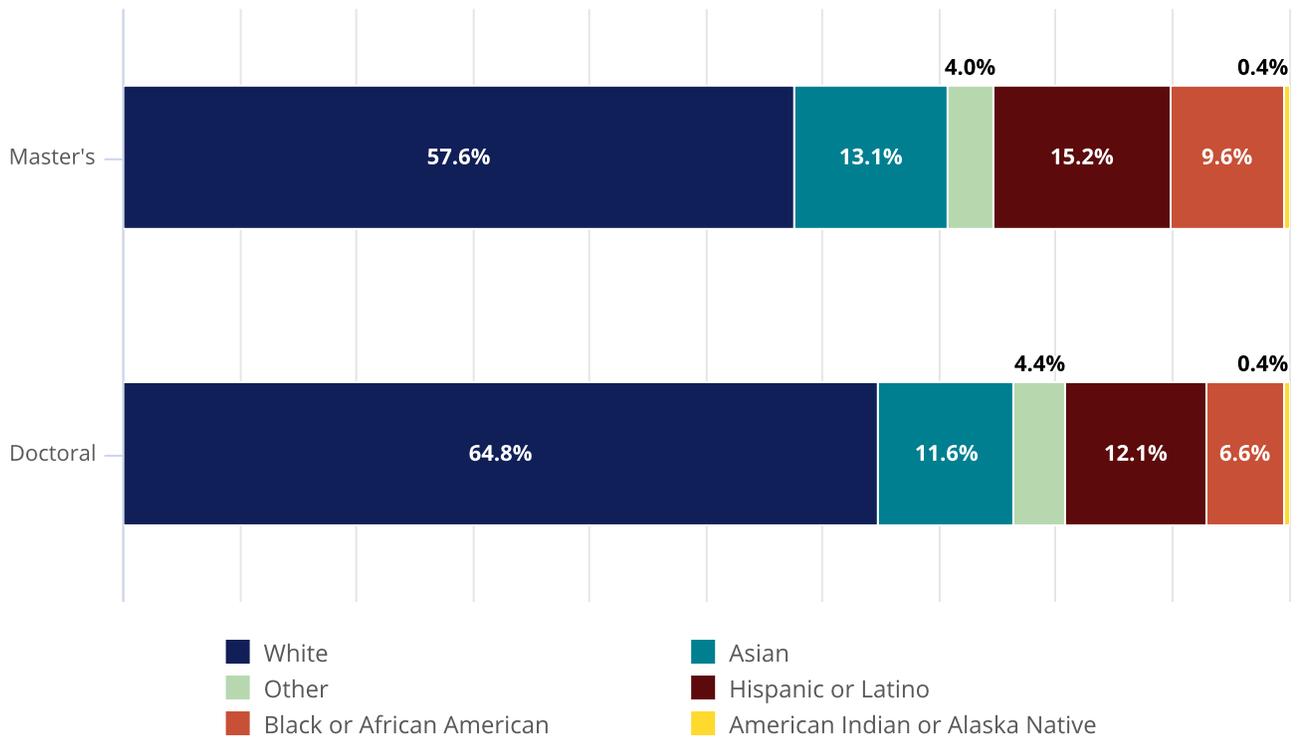

S&E = science and engineering.

**Note(s):**
Hispanic or Latino may be any race; race categories exclude Hispanic origin. Other includes Native Hawaiian and Other Pacific Islander and more than one race.

**Source(s):**
National Center for Science and Engineering Statistics, Survey of Graduate Students and Postdoctorates in Science and Engineering, 2021.

## Overall S&E graduate enrollment of underrepresented minorities increased in recent years, primarily due to expanding enrollment of Hispanic students.

Total S&E graduate enrollment from underrepresented minorities grew from 66,000 in 2017 to 97,000 in 2021 (figure 8-4). Hispanic enrollment grew most rapidly during this period, increasing by 56%, compared with 36% growth in Black enrollment and 11% growth in enrollment of American Indian or Alaska Native students.

**Figure 8-4**

**S&E graduate students from underrepresented minority groups, by race and ethnicity: 2017–21**

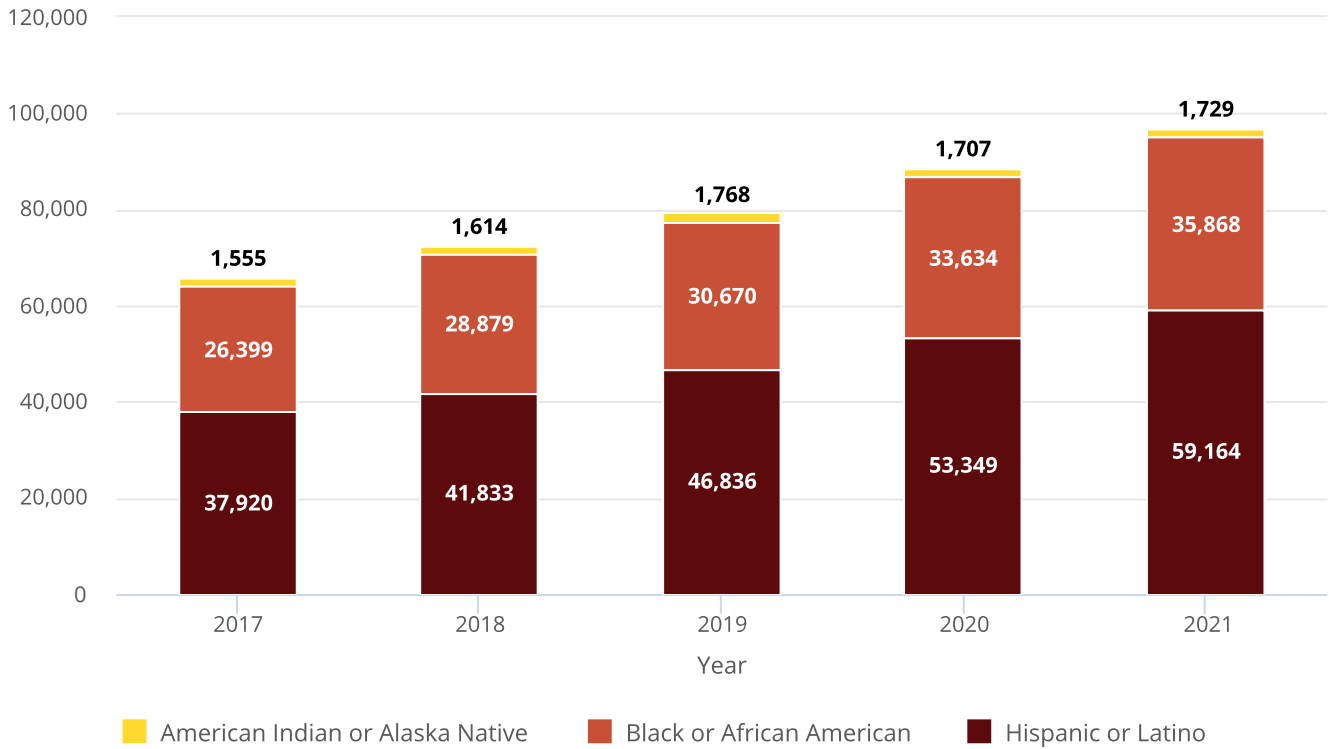

S&E = science and engineering.

**Note(s):**
Hispanic or Latino may be any race; race categories exclude Hispanic origin.

**Source(s):**
National Center for Science and Engineering Statistics, Survey of Graduate Students and Postdoctorates in Science and Engineering.

Total S&E graduate enrollment among persons from underrepresented minority groups increased by 12% from 2019 to 2020, a higher rate of growth than the preceding years, suggesting that the first year of the COVID-19 pandemic did not negatively affect overall minority enrollment. However, S&E graduate enrollment of underrepresented minority students increased by a smaller amount (9%) from 2020 to 2021.

## Enrollment of Hispanic or Latino Students by S&E Field

### Representation of Hispanic graduate students is highest in social and behavioral sciences and lowest in mathematics and computer sciences.

Like students from other underrepresented groups, Hispanic graduate students are concentrated in social and behavioral sciences at both the master's (20% of students in 2021) and doctoral (14%) levels (figure 8-5). Hispanic representation was lowest in mathematics and computer sciences: 11% of master's students and 9% of doctoral students were Hispanic.

**Figure 8-5**

**Hispanic or Latino S&E graduate student enrollment, by field and level of enrollment: 2021**

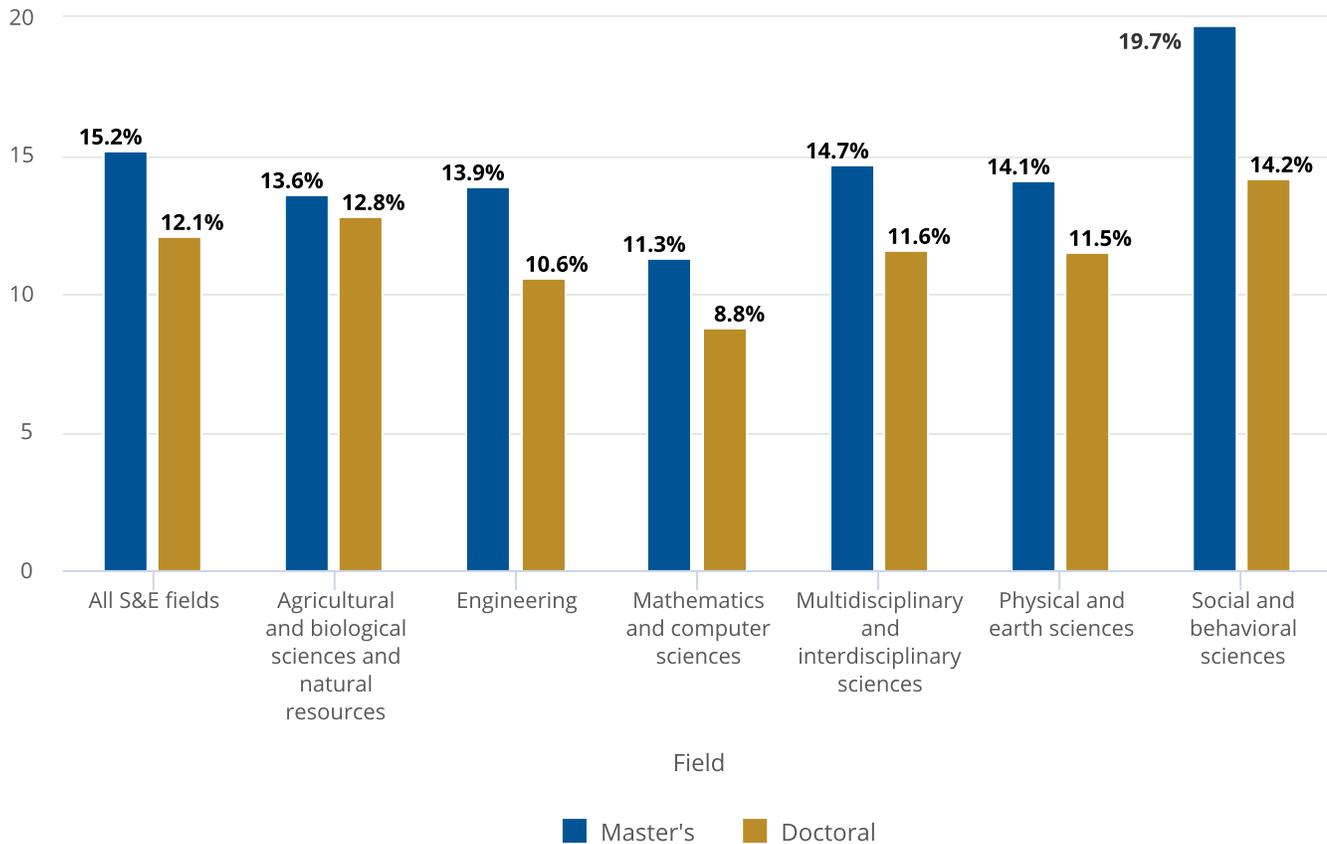

S&E = science and engineering.

**Note(s):**
Hispanic or Latino may be any race; race categories exclude Hispanic origin. Degree fields in this figure differ from fields presented in other tables and figures in this report that are based on the National Center Education Statistics Integrated Postsecondary Education Data System Completions Survey.

**Source(s):**
National Center for Science and Engineering Statistics, Survey of Graduate Students and Postdoctorates in Science and Engineering, 2021.

## Enrollment of Black or African American Students by S&E Field

### Black graduate students are concentrated in social and behavioral sciences at both degree levels and in mathematics and computer sciences at the master's level only.

At the master's level, the proportion of graduate students who are Black in social or behavioral sciences (12%) and mathematics and computer sciences (11%) was higher than the average Black representation across all S&E fields (figure 8-6). Black representation among master's students was lowest in engineering (6%) and physical and earth sciences (5%). Patterns of Black enrollment are slightly different at the doctoral level. Black students were most prevalent in social or behavioral sciences (10%) and multidisciplinary sciences (9%) and had the lowest representation in physical and earth sciences (4%). In terms of enrollment status, Black graduate students in S&E fields are less frequently enrolled on a full-time basis compared with students from other racial and ethnic groups (see sidebar Minority S&E Graduate Student Enrollment Status).

**Figure 8-6**

**Black or African American S&E graduate student enrollment, by field and level of enrollment: 2021**

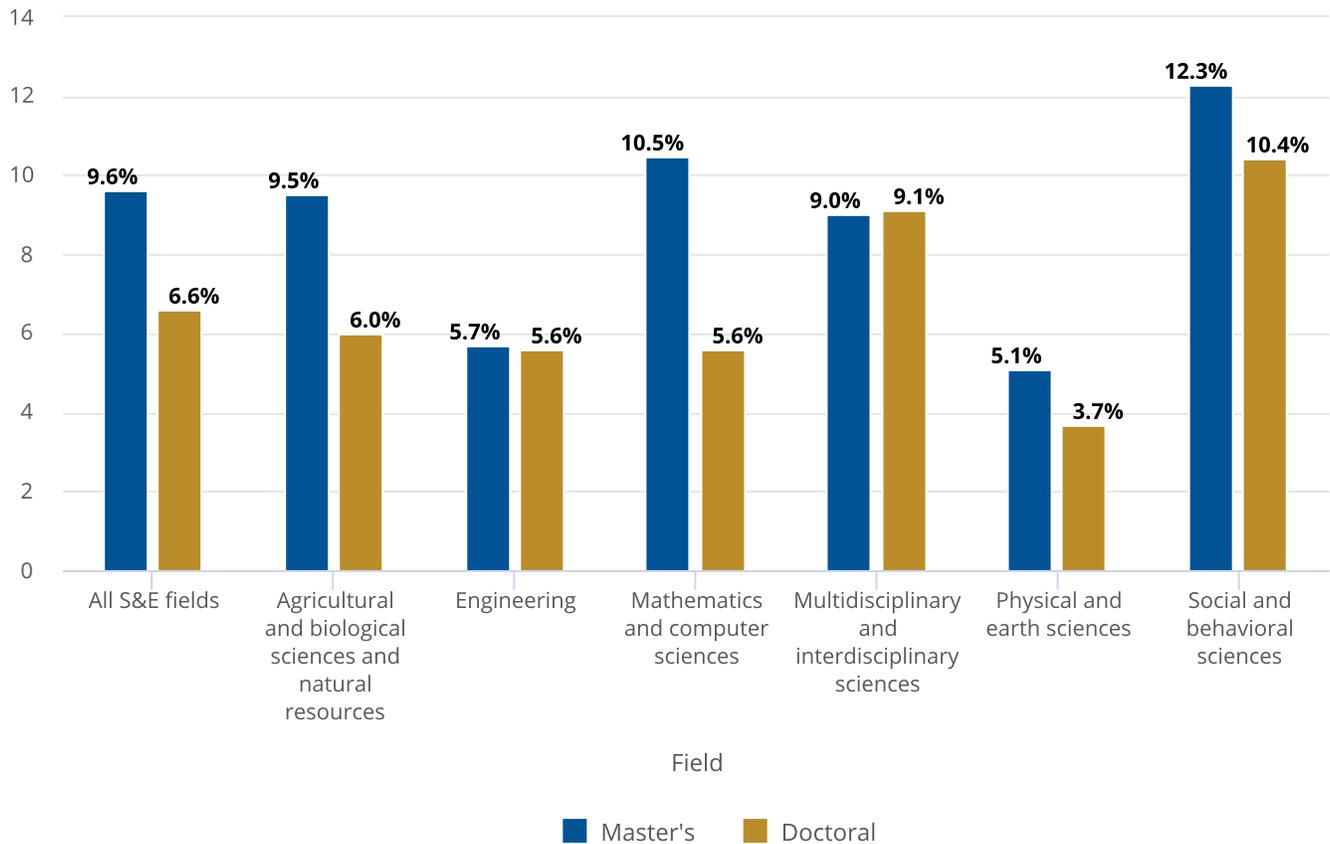

Field

■ Master's   ■ Doctoral

S&E = science and engineering.

**Note(s):**
Degree fields in this figure differ from fields presented in other tables and figures in this report that are based on the National Center Education Statistics Integrated Postsecondary Education Data System Completions Survey.

**Source(s):**
National Center for Science and Engineering Statistics, Survey of Graduate Students and Postdoctorates in Science and Engineering, 2021.

**SIDEBAR**

**Minority S&E Graduate Student Enrollment Status**

**A relatively low proportion of Black S&E graduate students are enrolled full time.**

Many graduate programs allow students to enroll on a part-time or full-time basis. Enrollment status may be indicative of students' level of professional or family commitments outside of the classroom, as well as financial barriers to full-time study. From 2017 to 2021, the share of graduate students in science and engineering with full-time status declined slightly for most racial and ethnic groups (figure 8-B). A significantly lower share of Black students were enrolled full time, compared with students from other racial and ethnic groups. For example, 58% of Black students had full-time status in 2021, whereas 64% of White students were enrolled full time. Given the minor changes in full-time enrollment shares from 2019 to 2021, the data do not indicate that the COVID-19 pandemic had a significant impact on enrollment status during this period.

**Figure 8-B**

S&E graduate students with full-time status, by race and ethnicity: 2017–21

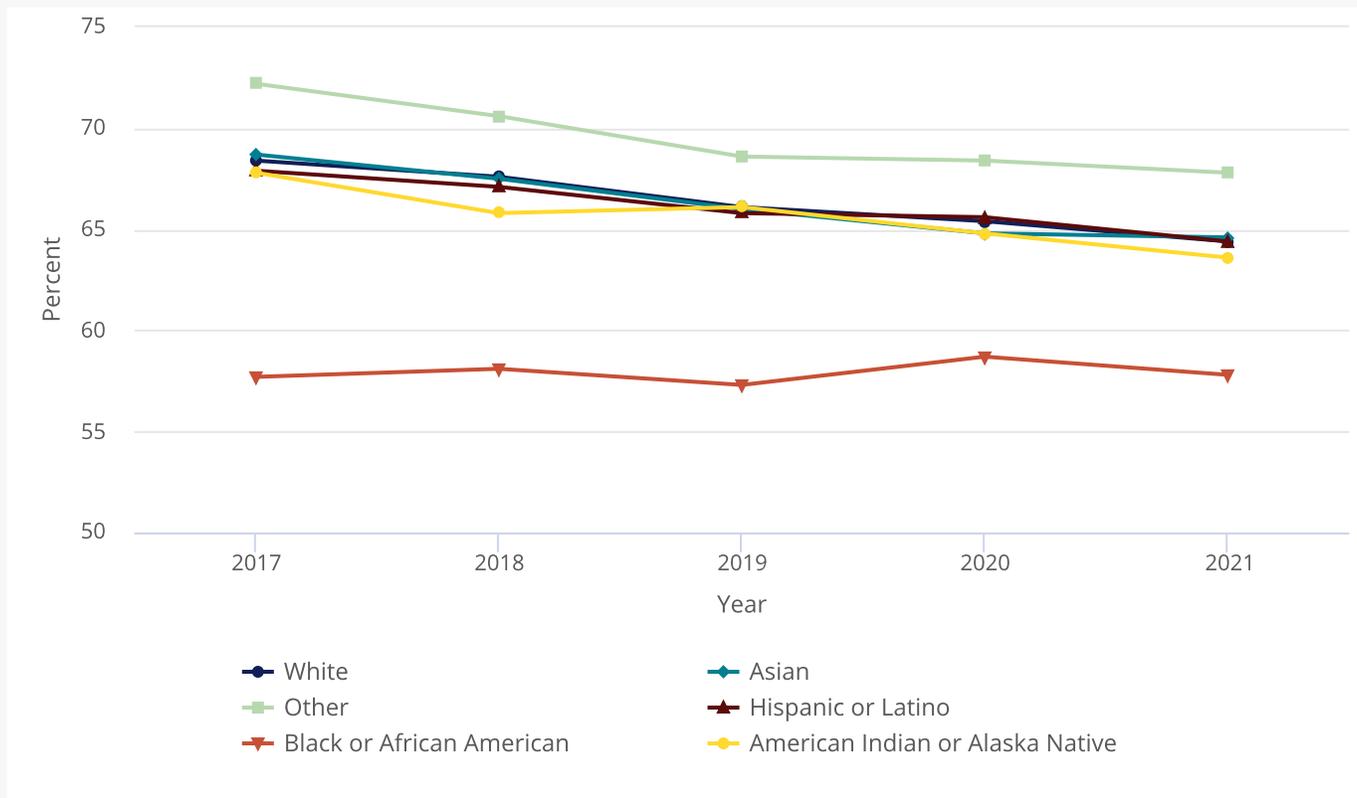

S&E = science and engineering.

**Note(s):**
Hispanic or Latino may be any race; race categories exclude Hispanic origin. Other includes Native Hawaiian and Other Pacific Islander and more than one race.

**Source(s):**
National Center for Science and Engineering Statistics, Survey of Graduate Students and Postdoctorates in Science and Engineering.

## Enrollment of American Indian or Alaska Native Students by S&E Field

### S&E graduate enrollment of American Indian or Alaska Native students is low overall but proportionately high in natural resources and conservation.

American Indian or Alaska Native graduate students accounted for 0.4% of master's and doctoral students across all S&E fields in 2021 (figure 8-7). At the broad field of degree level, they had highest representation in social and behavioral sciences at both the master's (0.5%) and doctoral (0.7%) levels. American Indian or Alaska Native students accounted for 0.4% of graduate students in agricultural and biological sciences and natural resources, but within this field they accounted for a relatively higher share of enrollment in natural resources and conservation: 0.9% of master's students and 1.8% of doctoral students (table 3-1).

**Figure 8-7**

**American Indian or Alaska Native S&E graduate student enrollment, by field and level of enrollment: 2021**

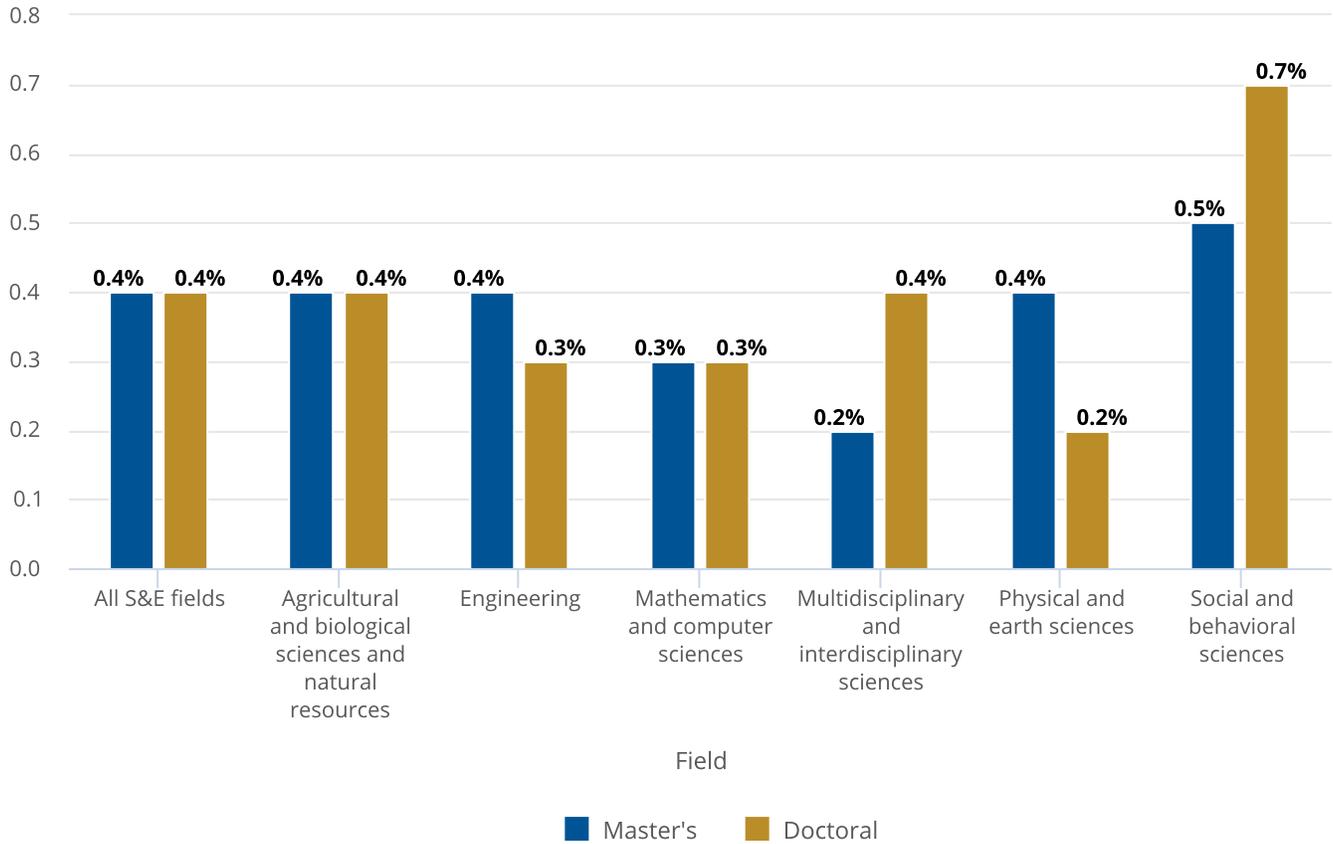

S&E = science and engineering.

**Note(s):**
Degree fields in this figure differ from fields presented in other tables and figures in this report that are based on the National Center Education Statistics Integrated Postsecondary Education Data System Completions Survey.

**Source(s):**
National Center for Science and Engineering Statistics, Survey of Graduate Students and Postdoctorates in Science and Engineering, 2021.

# Glossary

## Definitions

**Labor force:** A subset of the population that includes only those who are employed and those who are not working but actively seeking work (unemployed).

**Middle-skill occupations:** As a subset of science, technology, engineering, and mathematics (STEM) occupations, middle-skill occupations require a high level of STEM skills and expertise but do not typically require a bachelor's degree for entry. These positions are primarily in the areas of construction trades, installation, maintenance, and production.

**S&E occupations:** As a subset of STEM occupations, science and engineering (S&E) occupations typically require a bachelor's degree for entry and include five major categories of workers: (1) computer and mathematical scientists, (2) biological, agricultural, and environmental life scientists, (3) physical scientists, (4) social scientists, and (5) engineers.

**S&E-related occupations:** As a subset of STEM occupations, S&E-related occupations require STEM skills and expertise, but they do not fall into the five main S&E categories. The main occupational categories and positions that make up this group include health care workers, S&E managers, S&E precollege teachers, and technologists and technicians.

**S&E technologies:** This group of degree fields includes science, engineering, health, and other technologies that prepare students for the STEM workforce.

**Skilled technical workforce:** This subset of the STEM workforce is workers in S&E, S&E-related, and middle-skill occupations who do not have a bachelor's degree or higher.

**STEM workforce:** This subset of the U.S. workforce is comprised of workers in S&E, S&E-related, and middle-skill occupations.

**Underrepresented minorities:** Races or ethnicities whose representation in STEM employment and S&E education is smaller than their representation in the U.S. population. This includes Blacks or African Americans, Hispanics or Latinos, and American Indians or Alaska Natives.

**Workforce:** A subset of the labor force that includes only employed individuals.

# Key to acronyms

**COVID-19:** coronavirus disease 2019

**CPS:** Current Population Survey

**GSS:** Survey of Graduate Students and Postdoctorates in Science and Engineering

**IPEDS**: Integrated Postsecondary Education Data System

**NCSES:** National Center for Science and Engineering Statistics

**NSCG:** National Survey of College Graduates

**S&E:** science and engineering

**SDR:** Survey of Doctorate Recipients

**SED:** Survey of Earned Doctorates

**STEM:** science, technology, engineering, and mathematics

# Data Sources

The report highlights key statistics drawn from several data sources. Accompanying tables provide additional details. The data sources are as follows:

- Census Bureau's Current Population Survey (CPS) Annual Social and Economic Supplement (2011, 2019, and 2021)
- NCSES's National Survey of College Graduates (NSCG) (2021)
- NCSES's Survey of Earned Doctorates (SED) (2021)
- NCSES's Survey of Doctorate Recipients (SDR) (2021) (only presented in the tables)
- National Center for Education Statistics' Integrated Postsecondary Education Data System (IPEDS) (2011–20)
- NCSES's Survey of Graduate Students and Postdoctorates in Science and Engineering (GSS) (2017–21)

# References

Akinyooye L, Nezamis E. 2021. As the COVID-19 pandemic affects the nation, hires and turnover reach record highs in 2020. *Monthly Labor Review,* Bureau of Labor Statistics, June 2021. Available at https://doi.org/10.21916/mlr.2021.11.

National Center for Science and Engineering Statistics (NCSES). 2017. *Women, Minorities, and Persons with Disabilities in Science and Engineering: 2017.* Special Report NSF 17-310. Arlington, VA. Available at https://www.nsf.gov/statistics/2017/nsf17310/.

National Center for Science and Engineering Statistics (NCSES). 2022. *Workforce Statistics.* Infographic NCSES 22-203. Alexandria, VA. Available at https://ncses.nsf.gov/136/assets/0/files/ncses_workforcestatistics_onepager.pdf.

National Science Board, National Science Foundation (NSB, NSF). 2021. The STEM Labor Force of Today: Scientists, Engineers and Skilled Technical Workers. *Science and Engineering Indicators 2022*. NSB-2021-2. Alexandria, VA. Available at https://ncses.nsf.gov/pubs/nsb20212.

# Notes

**1**  Data from the National Center for Education Statistics (NCES) indicate that in fall 2019, 81% of undergraduate students and 71% of graduate students in degree-granting postsecondary institutions were within this age range (NCES *Digest of Education Statistics 2021*: table 303.50).

**2**  The Census Bureau's Current Population Survey defines money wages and salary as the "total money earnings received from work performed as an employee during the income year. It includes wages, salary, Armed Forces pay, commissions, tips, piece-rate payments, and cash bonuses earned, before deductions are made for taxes, bonds, pensions, union dues, etc. Earnings for self-employed incorporated businesses are considered wage and salary." Census Bureau, March 2021 Annual Social and Economic Supplement. Available at https://www.census.gov/data/datasets/time-series/demo/cps/cps-asec.html.

**3**  See S&E labor market conditions in *NSB Indicators 2020*: Science and Engineering Labor Force.

**4**  The question on total wage and salary earnings in the Current Population Survey asks about earnings from the longest job held during the previous year (e.g., longest job held in 2020 for the 2021 Current Population Survey).

**5**  See https://www.bls.gov/opub/reports/race-and-ethnicity/2020/home.htm.

**6**  Previous editions of this report examined undergraduate enrollment using data from the Integrated Postsecondary Education Data System (IPEDS), administered by the National Center for Education Statistics, which lacks data on students' fields of study. For analysis of total undergraduate enrollment, see the Undergraduate Enrollment chapter of the NCES Condition of Education at https://nces.ed.gov/programs/coe/indicator/cha/undergrad-enrollment.

# Acknowledgments and Citation

## Acknowledgments

Elizabeth Grieco, Program Director, Science, Technology, and Innovation—Public Information, National Center for Science and Engineering Statistics (NCSES), developed and wrote this report with co-author Steven Deitz (NCSES). Overall guidance and direction of the content and presentation of the report were provided by Amy Burke, Program Director, Science, Technology, and Innovation Analysis, NCSES; Emilda B. Rivers, Director, NCSES; Vipin Arora, former Deputy Director, NCSES; and John Finamore, Chief Statistician, NCSES.

Writing consultation was provided by Nancy Gough (Betah Associates) and analytical support provided by Laura Pomerance (ProSource 360). Statistical review was provided by Wan-Ying Chang (NCSES). Quality assurance of data and statistical accuracy was provided by Devi Mishra (NCSES) and staff from RTI International. Communication and outreach support, including social media guidance, was provided by Julia Milton (NCSES). The report was edited and produced by Christine Hamel and Tanya Gore, with guidance from Catherine Corlies, all at NCSES. Development of the Web version was guided by Rajinder Raut (NCSES).

## Suggested citation

National Center for Science and Engineering Statistics (NCSES). 2023. *Diversity and STEM: Women, Minorities, and Persons with Disabilities 2023*. Special Report NSF 23-315. Alexandria, VA: National Science Foundation. Available at https://ncses.nsf.gov/wmpd.

# Contact NCSES

**NCSES**

National Center for Science and Engineering Statistics
Directorate for Social, Behavioral and Economic Sciences
National Science Foundation
2415 Eisenhower Avenue, Suite W14200
Alexandria, VA 22314
Tel: (703) 292-8780
FIRS: (800) 877-8339
TDD: (800) 281-8749
E-mail: ncsesweb@nsf.gov

https://ncses.nsf.gov/wmpd

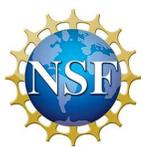 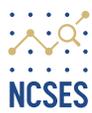 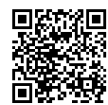

